\documentclass{aa}  

\usepackage{graphicx}
\usepackage{txfonts}
\usepackage{amsmath}    
\usepackage{amssymb}    
\usepackage{xspace}
\usepackage{xcolor}
\usepackage{natbib}
\bibpunct{(}{)}{;}{a}{}{,} 

\newcommand{\Ha}{H$\alpha$\xspace}

\newcommand{\hdue}{$\rm H_{2}$\xspace}
\newcommand{\Msun}{$\rm M_\odot$\xspace}

\newcommand{\kms}{$\rm km \, s^{-1}$\xspace}

\newcommand{\aco}{$\alpha_{CO}$\xspace}

\usepackage{hyperref}

\begin{document}

   \title{Molecular gas and star formation in GASP jellyfish galaxies}

   \author{A. Moretti
          \inst{1}
          \and
          R. Paladino \inst{2,3}
          \and
          C. Bacchini \inst{4}
          \and
          A. Marasco \inst{1}
          \and
          E. Giunchi \inst{5}
          \and
          B. M. Poggianti \inst{1}
          \and
          L. K. Hunt \inst{6}
          \and
          T. Deb \inst{7}
          \and
          B. Vulcani \inst{1}
          \and
          M. Gullieuszik \inst{1}
          \and
          A. Lassen \inst{1}
          \and
          A. Wolter \inst{8}
          \and
          M. Gitti \inst{5,2}
          \and
          M. Radovich \inst{1}
          \and
          J. Fritz \inst{9}
          \and
          N. Tomicic \inst{10}
          }

   \institute{INAF - Padova Astronomical Observatory, Vicolo dell'Osservatorio 5, I-35122 Padova, Italy\\
              \email{alessia.moretti@inaf.it}
              \and
              INAF - Istituto di Radioastronomia, via P. Gobetti 101, 40129 Bologna, Italy
              \and
              European Southern Observatory, Karl-Schwarzschild-Strasse 2, 85748 Garching, Germany
              \and
              DARK, Niels Bohr Institute, University of Copenhagen, Jagtvej 155, 2200 Copenhagen, Denmark
              \and
              Dipartimento di Fisica e Astronomia, Università di Bologna, Via Piero Gobetti 93/2, 40129 Bologna, Italy
              \and
              INAF - Osservatorio Astrofisico di Arcetri, Largo E. Fermi 5, 50125, Firenze, Italy
              \and
              Center for Astrophysics - Harvard \& Smithsonian, 60 Garden St, Cambridge, MA 02138, USA
              \and
              INAF - Osservatorio Astronomico di Brera, Via Brera, 28, 20121 Milano, Italy
              \and
              Instituto de Radioastronomia y Astrofisica, UNAM, Campus Morelia, AP 3-72, Morelia, CP 58089, Mexico   
              \and
              Department of Physics, Faculty of Science, University of Zagreb, Bijenicka 32, 10 000 Zagreb, Croatia
             }

   \date{Received November 18, 2025; accepted February 13, 2026}

  \abstract
   {Several studies have reported a nearly linear correlation between the molecular gas and star formation rate surface density, the so-called Kennicutt-Schmidt (KS) law.
}
   {We aim to retrieve the KS relation for a sample of four star-forming galaxies located in nearby clusters, disturbed by the effects of the ram pressure stripping, as testing this law in galaxies subject to different environmental conditions can provide key information on the physics of star formation.}
   {To perform our analysis, we used ALMA band 6 and band 3 data coupled with MUSE data at spatial resolution of $\sim$1 kpc. Moreover, we analyzed data of star-forming complexes detected through their \Ha ionized gas emission. We also derived the star formation efficiencies of the star-forming regions nested in these big complexes using the star formation rates derived from spatially resolved HST images and various recipes for the corresponding cold gas phase.}
   {We find that ram-pressure-stripped galaxies show normal-to-low star formation efficiencies, depending on the position within the galaxy and on the local gas density: the inner dense regions in the disk show higher efficiencies with respect to the outer regions, including the gaseous tails. The global relation between the star formation rate density and the molecular gas surface density is superlinear, likely suggesting the shortening of the depletion times at high gas mass densities caused by the ram pressure. Within the star-forming complexes, the star formation efficiency is very similar to the one observed at 1 kpc scale in undisturbed star-forming disks. Interestingly, this result holds also for the star-forming complexes located in the stripped gas tails.
   The analysis of HST resolved clumps suggests that the molecular gas is not uniformly distributed within the star-forming complexes, but its density distribution follows a steeper profile.
   }
   {}


   \maketitle

\section{Introduction} \label{sec:intro}
The process by which diffuse gas within galaxies coalesces to form new stars is one of the most fundamental and complex phenomena in astrophysics. Understanding the factors that govern the rate and efficiency of star formation is crucial for comprehending galaxy evolution across cosmic time. A cornerstone in this endeavor is the Kennicutt-Schmidt relation (often referred to as the KS relation), an empirical power-law relationship that links the surface density of gas to the surface density of star formation in galaxies \citep{Schmidt1959,Kennicutt1989}.
When considering the molecular gas only, the KS relation, i.e., the one relating the star formation rate (SFR) surface density to the \hdue surface mass density, $\Sigma_{SFR} \propto \Sigma_{H_2}^n$, has been found to be linear (n$\sim$1) or slightly sublinear \citep{Bacchini+2019_MW,Bacchini+2019_vol,Delosreyes+2019}.
This result has been confirmed at many scales, from the global scale \citep{KE2012} to the $\sim$1 kpc scale \citep{Bigiel2008,Bigiel2011,Leroy2013}, to the small scale of Fornax cluster galaxies \citep[$\sim$300 pc, see][]{Zabel+2020}, to the even smaller scales ($\sim$100 pc) studied in PHANGS \citep{Pessa+2021} for nearby star-forming disks.
Despite this, whether the molecular clouds properties are universal or not, i.e., whether they depend on the conditions in which they are located, is still debated \citep{HeyerDame2015}.
Recent works on nearby galaxies, including those produced by the PHANGS collaboration thanks to the exquisite resolution of ALMA \citep[][and references therein]{Sun+2020b}, show variations with the galactic environment, and, within each galaxy, with the position of the cloud with respect to the galaxy center, resulting in an increased scatter of the molecular KS law.
The radial dependence of the resolved star formation efficiency (i.e., the ratio between the SFR surface density and the \hdue surface density, SFE) has been recently studied in a sample of 81 nearby galaxies \citep{Villanueva+2021} that possess both IFU observations from the CALIFA survey \citep{Sanchez+2012} and interferometric molecular gas estimates from the EDGE-CALIFA survey \citep{Bolatto+2017}.
It has been suggested that the measured variations are linked to the pressure in the ambient medium, which can be increased in galaxy centers, bars, or where galaxies interact with each other.
\cite{Bacchini+2019_MW,Bacchini+2019_vol} showed that the projection effect due to the gas disk thickness also contributes significantly to local variations in the molecular KS law observed in nearby spirals.
In general, whenever the pressure of the interstellar medium (ISM) is increased, there the properties of the SFE can greatly vary.

As an example, \citet{Zabel+2020} have shown that Fornax galaxies close to the virial radius of the cluster show shorter depletion times with respect to more central galaxies, albeit this result can be biased by the limited size of the sample, which is mostly dominated by dwarf galaxies.
More recently, the analysis of Virgo cluster galaxies from the VERTICO survey \citep{jimenez+2023} has found a KS relation with a unitary slope broadly consistent with the one found for field galaxies, but with a larger scatter ($\sim 0.4$ dex).
In particular, HI-deficient galaxies, probably subject to cluster environmental effects, show a steeper KS relation and lower molecular gas efficiencies compared to HI-normal galaxies.
In particular, galaxies showing evidence of ram pressure stripping (RPS) seem to be slightly enriched in H$_2$\citep{Zabel+2022}.

In view of this, it is particularly interesting to show how the star-forming properties vary in disturbed galaxies, such as the jellyfish galaxies affected by the RPS. 
Molecular gas has already been observed in local ram-pressure-stripped galaxies \citep{Jachym2013,Jachym2014,Jachym2017} with single dish telescopes (APEX, IRAM) but also with the ALMA interferometry \citep{Jachym+2019}, revealing detectable amounts of molecular gas both in the galaxy disks and in the tails.
In particular, the tail of ESO137-001 observed with ALMA at 350 pc resolution shows the presence of molecular compact CO emitting regions embedded in the stripped gas tail.
A few molecular gas clumps have also been detected in the stripped tail of NGC4858 at $\sim$ 500 pc resolution \citep{Souchereau+2025}.

Moving to larger distances, the GASP sample \citep{GASPI} offers the first statistically significant sample of the RPS phenomenology \citep{Poggianti+2025}, highlighting the possibility of tracking this physical mechanism using the ionized gas emission.
As for the molecular gas content, it has already been shown that galaxies at the peak of RPS, i.e., galaxies with long ionized gas tails (the so-called jellyfish galaxies) exhibit an anomalously high molecular gas content with respect to undisturbed field galaxies with the same stellar mass \citep{Moretti+2020}, and further analysis on a larger sample has demonstrated that this happens at the expense of the galaxy HI content as galaxies travel through the cluster potential approaching and passing by the pericenter \citep{Moretti+2023}.
In fact, although the GASP survey is mainly based on the analysis of the resolved properties of the ionized gas emission detected with the MUSE integral field spectrograph, many follow-up observations have subsequently explored other gas phases, from the cold atomic gas \citep{GASPXVII, GASPXXV, GASPXXVI, Luber+2022} to the radio continuum \citep{Ignesti+2022}, to the X-ray emission of the hot ionized gas \citep{GASPXXIII,GASPXXXIV, Bartolini+2022}.
The cold molecular gas properties have benefited from both single dish \citep{GASPX, Moretti+2023} and interferometric data \citep{GASPXXII, Moretti+2020} of a subsample of jellyfish galaxies. 

A clear characterization of the SFE in ram-pressure-stripped galaxies is nevertheless still missing, as well as a clear understanding of how the star formation is taking place in the star-forming regions observed in the stripped gas tails.
In fact, both conditions, i.e., galaxy disks subject to an increased pressure due to the interaction with the hot intra-cluster medium (ICM) and galaxy tails where the molecular gas is exposed to the hot gas surrounding galaxies constitute a peculiar environment where star formation takes place, and understanding whether its efficiency is comparable or not with the one exhibited in disks is the topic we want to address in this paper, by studying four galaxies from the GASP sample for which we have ALMA data.

We describe our sample in Sec.~\ref{sec:data}, in which we also give a characterization of ALMA data used in what follows.
We then derive the KS relation at a scale of $\sim 1$ kpc in Sec.~\ref{sec:KS1} using both the ALMA CO(2-1) and CO(1-0) line emission.
Sec.~\ref{sec:KSc} contains the KS relation derived for the star-forming regions identified by the \Ha MUSE emission and an evaluation of their virial parameter.
Finally, we explore different possibilities to track the star formation at the scale of the HST-resolved clumps in Sec.~\ref{sec:hst}.
We summarize our results in Sec.~\ref{sec:discussion}.

\section{Data}\label{sec:data}
We base our results on the analysis of ALMA cycle 5 data taken in Band 3 and Band 6, targeting four galaxies of the GASP sample; namely, JO201\citep{GASPII}, JO204 \citep{GASPIV}, JO206 \citep{GASPI}, and JW100 \citep{GASPXXIII}. The GASP survey \citep{GASPI,Poggianti+2025} is devoted to the study of the mechanisms that favor the gaseous interplay between galaxies and their surrounding medium, with specific interest in RPS \citep{GG72} taking place in dense environments, but also in mechanisms shaping galaxy properties in less dense environments \citep{Vulcani+2021}.
The GASP cluster target galaxies are members of the WINGS/OmegaWINGS sample \citep{Fasano+2006,Gullieuszik+2015,Moretti+2017} and have a typical redshift of $z=0.05$. At this distance, 1\arcsec corresponds to $\sim1$ kpc.
We give in Table \ref{tab:galaxies} the main details on the target galaxies, i.e., the distance from the brightest cluster galaxy (BCG) in kiloparsecs, stellar mass, redshift, total SFR, and disk SFR derived from the MUSE \Ha emission \citep{Vulcani+2018}. These galaxies have been widely studied within the GASP project, and references to the relevant papers are given in \citep{Poggianti+2025}.
\begin{table*}[]
\caption[]{Target galaxies' properties.}
\label{tab:galaxies}
\centering
\begin{tabular}{|l|l|l|l|l|l|l|l|l|}
\hline
Galaxy & RA&DEC&Cluster &d$_{BCG}$& M$_\star$       & z & SFR(total) & SFR(disk) \\
       &&&         &[kpc]& [$10^{10}M_{\odot}$] &         &  [$M_{\odot}/yr$]&  [$M_{\odot}/yr$] \\      
\hline
JO201   &00:41:30.29 &-09:15:45.900& A85    &358& 3.55               &  0.045  &   6$\pm$1 & 5$\pm$1 \\ 
JO204   &10:13:46.83&-00:54:51.056& A957   &119& 4                  &  0.042  &   1.7$\pm$0.3 & 1.5$\pm$0.3 \\
JO206  &21:13:47.41&+02:28:34.383& IIZW108 &331& 9                  &  0.051  &   5$\pm$1 & 4.8$\pm$0.9 \\
JW100  &23:36:25.06&+21:09:02.529& A2626   &83& 30                 &  0.061  &   4$\pm$0.8 & 2.6$\pm$0.5 \\
\hline
\end{tabular}
\tablefoot{Coordinates, host cluster, distance from the BCG,  stellar mass, redshift, total SFR, and disk SFR for the four target galaxies.}
\end{table*}

\begin{table}[]
\caption[]{Properties of the CO(1-0) and CO(2-1) line images.}

    \label{tab:data}
    \centering
    \begin{tabular}{|c|c|c|c|c|c|c|}
    \hline
    Galaxy & $\nu_{obs}$ & $\theta_{maj}$ & $\theta_{min}$ & PA & rms & MRS  \\
           & GHz         & \arcsec        & \arcsec        & deg & mJy b$^{-1}$ & \arcsec\\
    \hline
    \multicolumn{7}{|c|}{CO(1-0)}\\
    \hline
    JO201  & 110.307 & 1.99 & 1.57 & -84.6 & 0.5 & 20 \\
    JO204  & 110.625 & 1.62 & 1.36 &  81.5 & 0.5 & 20 \\
    JO206  & 109.677 & 1.60 & 1.30 & -87.4 & 0.7 & 23 \\
    JW100  & 108.644 & 2.00 & 1.70 & 8.3   & 0.9 & 24 \\
    \hline
    \multicolumn{7}{|c|}{CO(2-1)}\\
    \hline
    JO201  & 220.680 & 1.38 & 1.02 & 78.5 & 0.7 & 18.7 \\
    JO204  & 221.152 & 0.98 & 0.86 & 70.8 & 0.6 & 18.6 \\
    JO206  & 219.288 & 1.23 & 1.01 & -72.7 & 0.6 & 18.8 \\
    JW100  & 217.258 & 1.40 & 1.12 & 33   & 0.8  & 19\\
    \hline
    \end{tabular}
\tablefoot{Observed frequency, synthesized beam, channel rms of the 20 \kms velocity resolution cubes, and MRS.}
\end{table}

The four galaxies that we analyze here are all in the peak stripping phase \citep{Poggianti+2025}, as is demonstrated by their position in the phase-space diagram \citep{GASPIX,GASPXXI} and confirmed by their extremely long tails of ionized gas. Their stellar masses span the range $10.0 <  \log\,(M/M_{\odot}) < 11.5$, and their SFR is in excess with respect to the GASP control sample of galaxies \citep{Vulcani+2018}, but for JW100, which shows a lower global SFR. Interestingly enough, all these galaxies possess a central active galactic nucleus \citep[AGN;][]{poggianti2017,GASPXIX}, which is quite common in strongly ram-pressure-stripped galaxies \citep{Peluso+2021}.
Recently, these galaxies have been observed with HST \citep{Gullieuszik+2023, Giunchi+2023}, revealing the presence of small ($\sim$70-500 pc) star-forming clumps distributed in the disk and in the gaseous tails, with star formation rate densities ($\Sigma_{SFR}$) that are intermediate between those of starbursting and normal galaxies. 

\begin{figure*}
    \centering
    \includegraphics[width=0.24\textwidth]{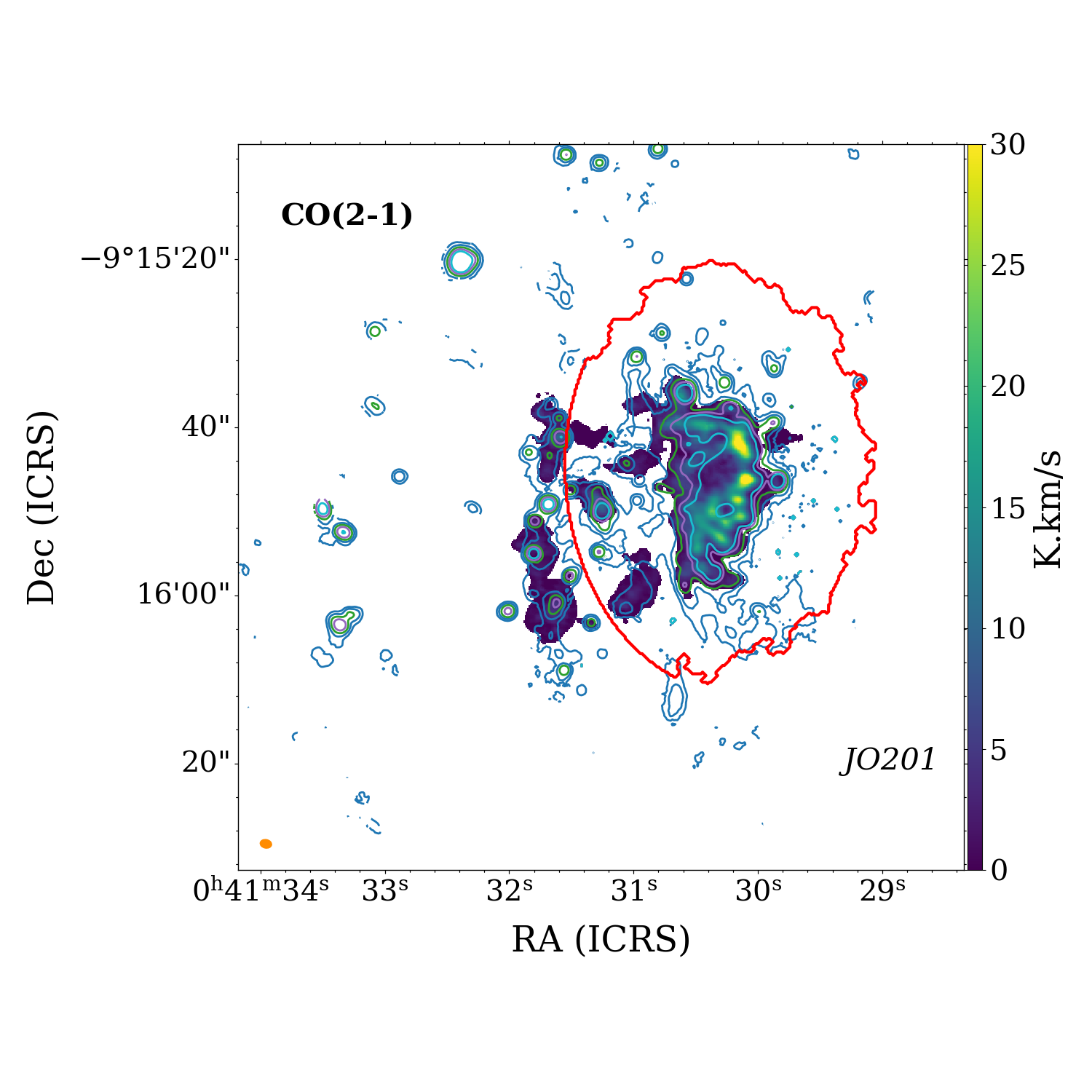}
    \includegraphics[width=0.24\textwidth]{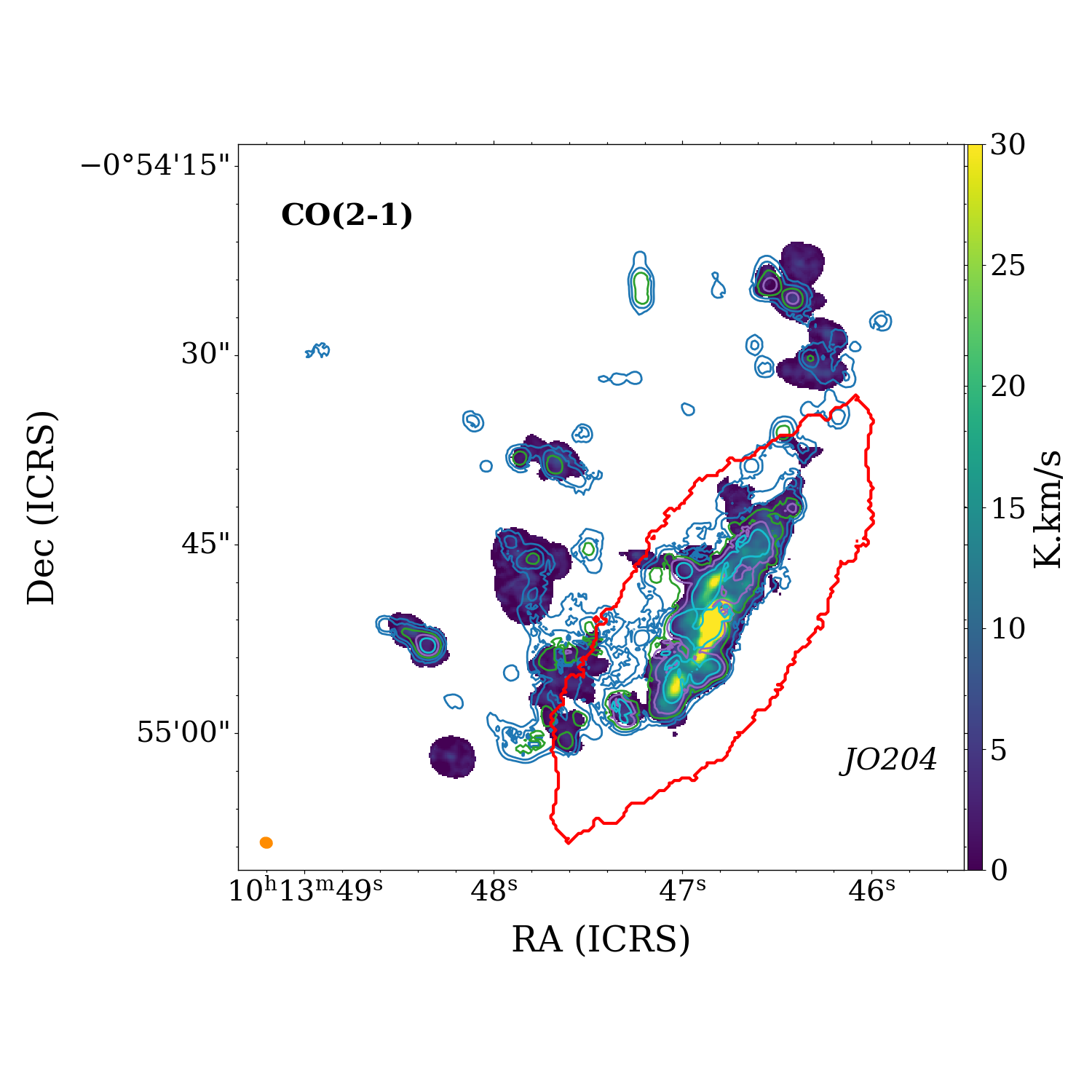}
    \includegraphics[width=0.24\textwidth]{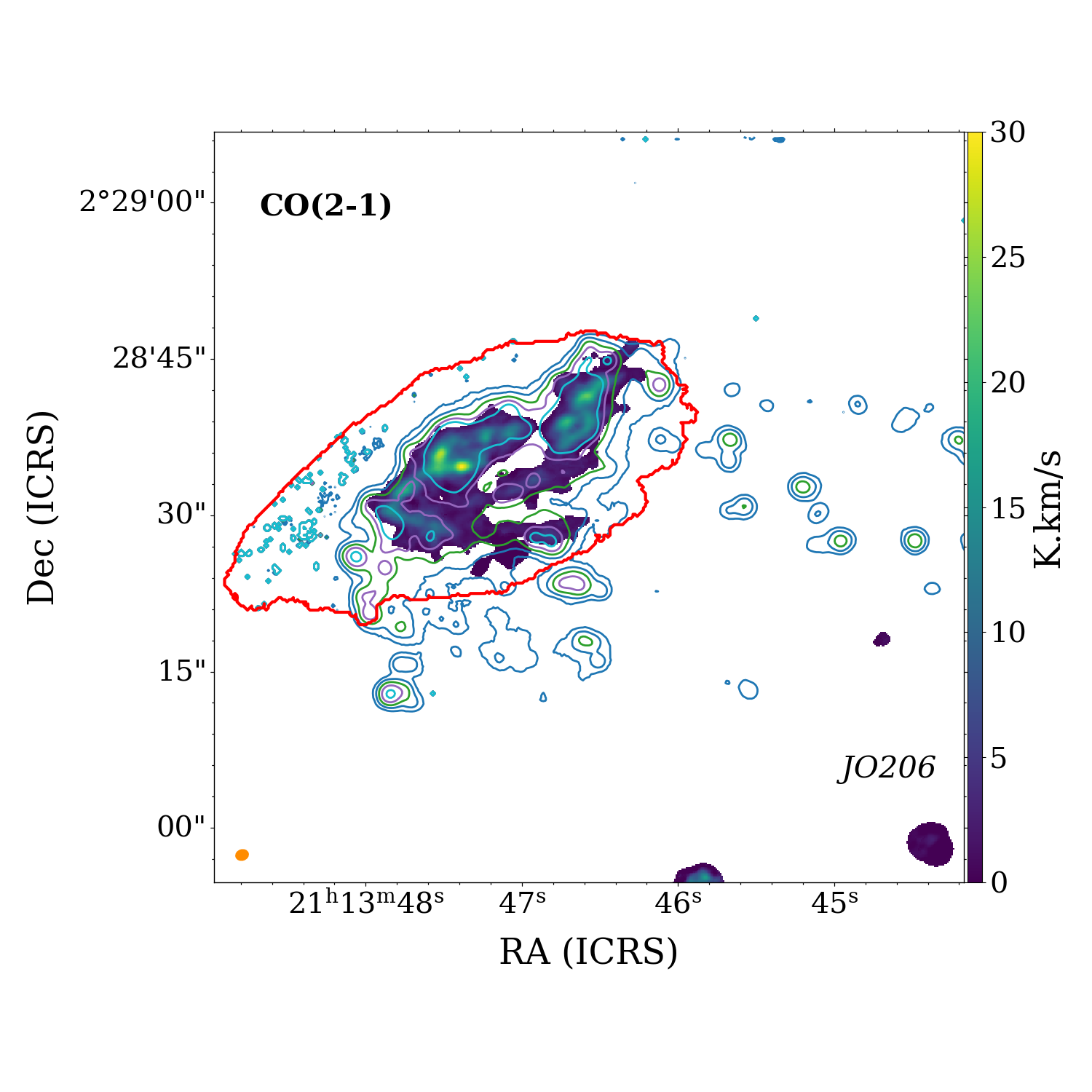}
    \includegraphics[width=0.24\textwidth]{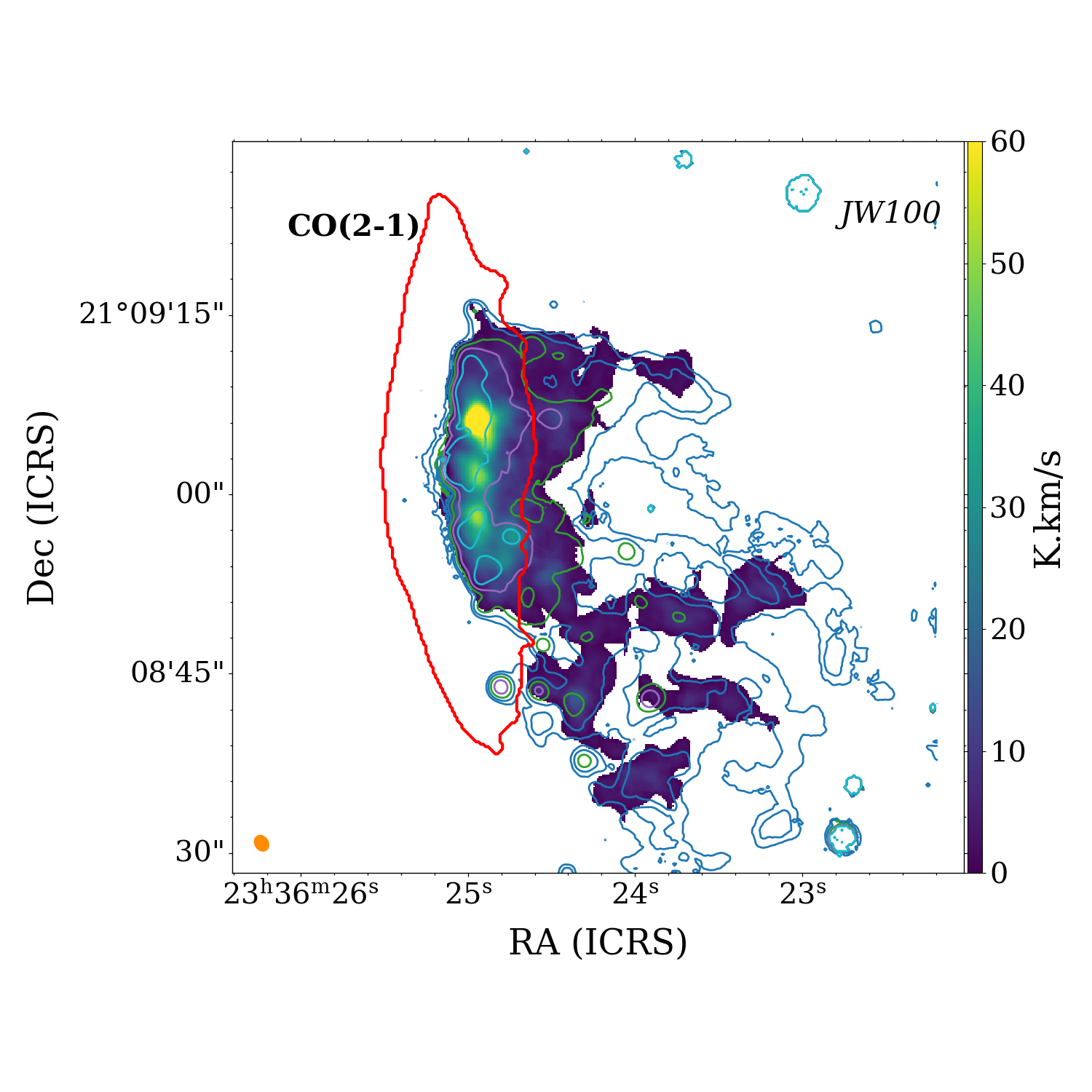}
    \includegraphics[width=0.24\textwidth]{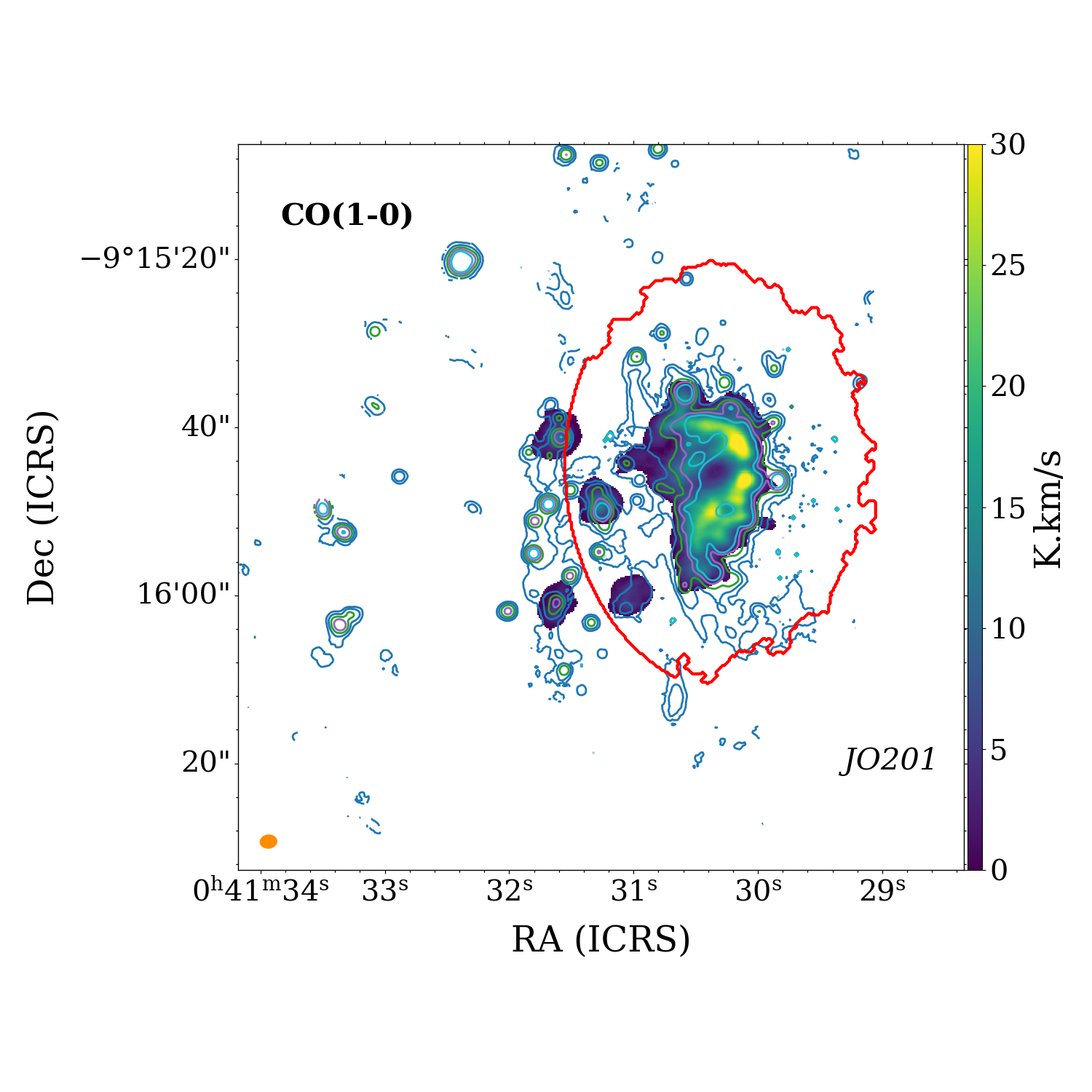}
    \includegraphics[width=0.24\textwidth]{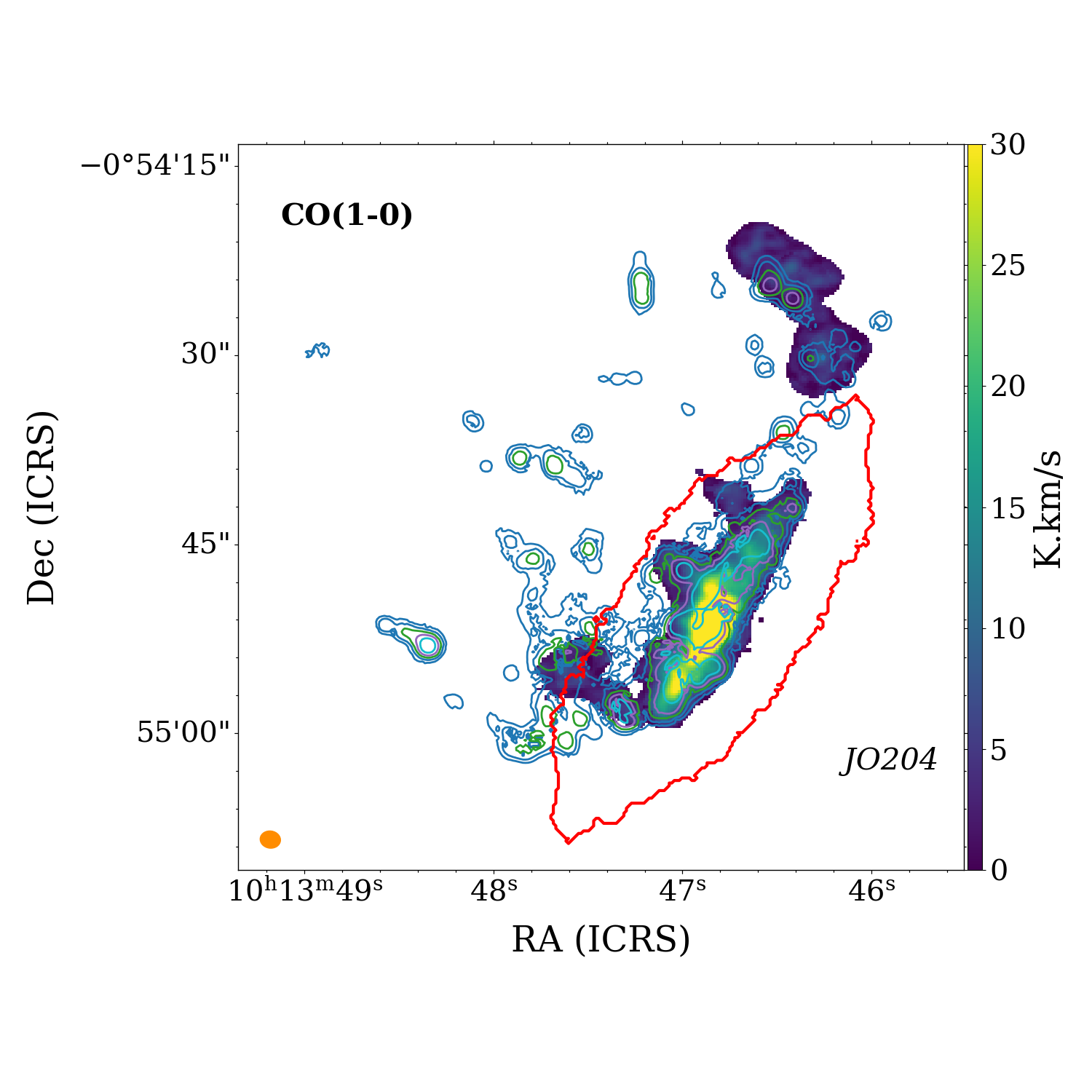}
    \includegraphics[width=0.24\textwidth]{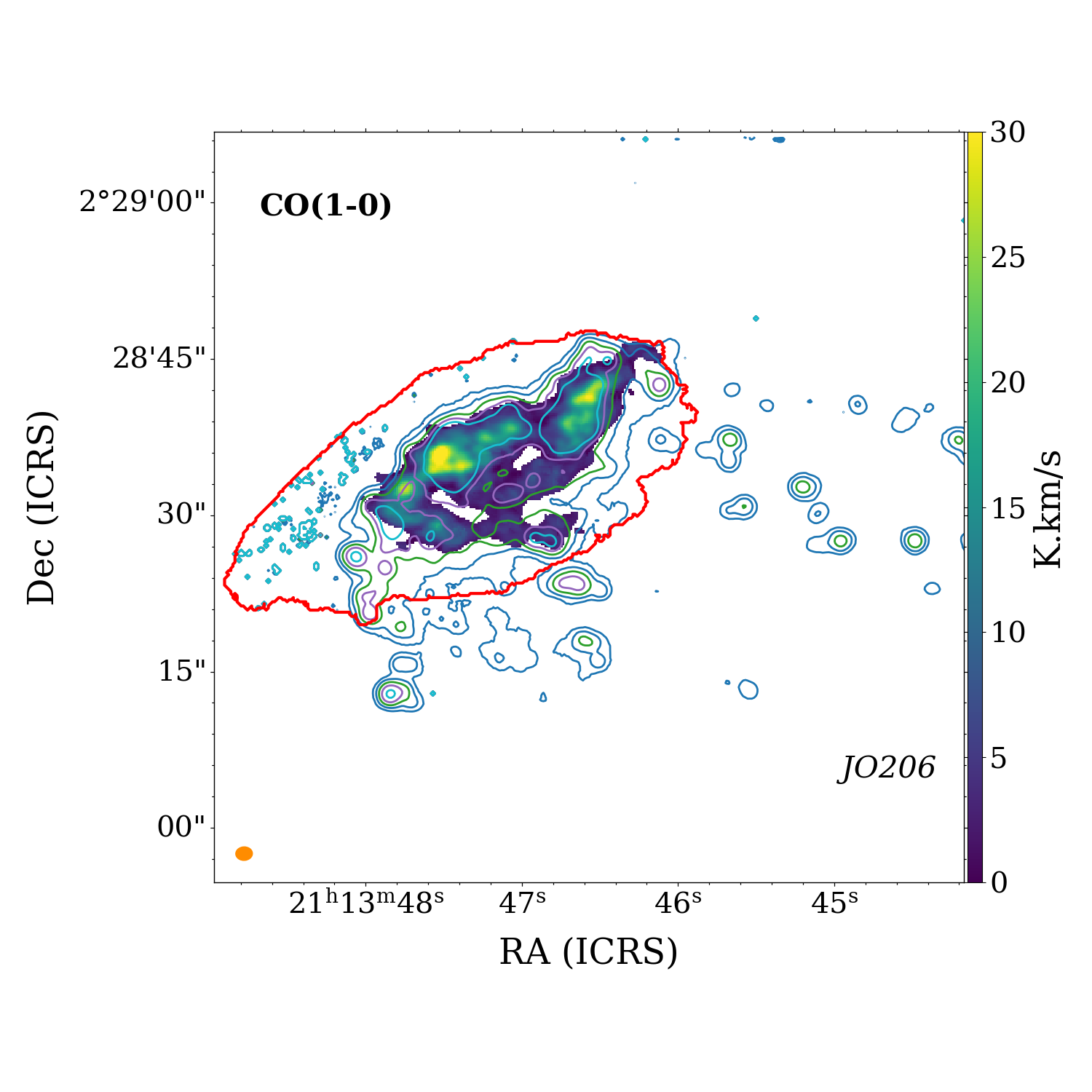}
    \includegraphics[width=0.24\textwidth]{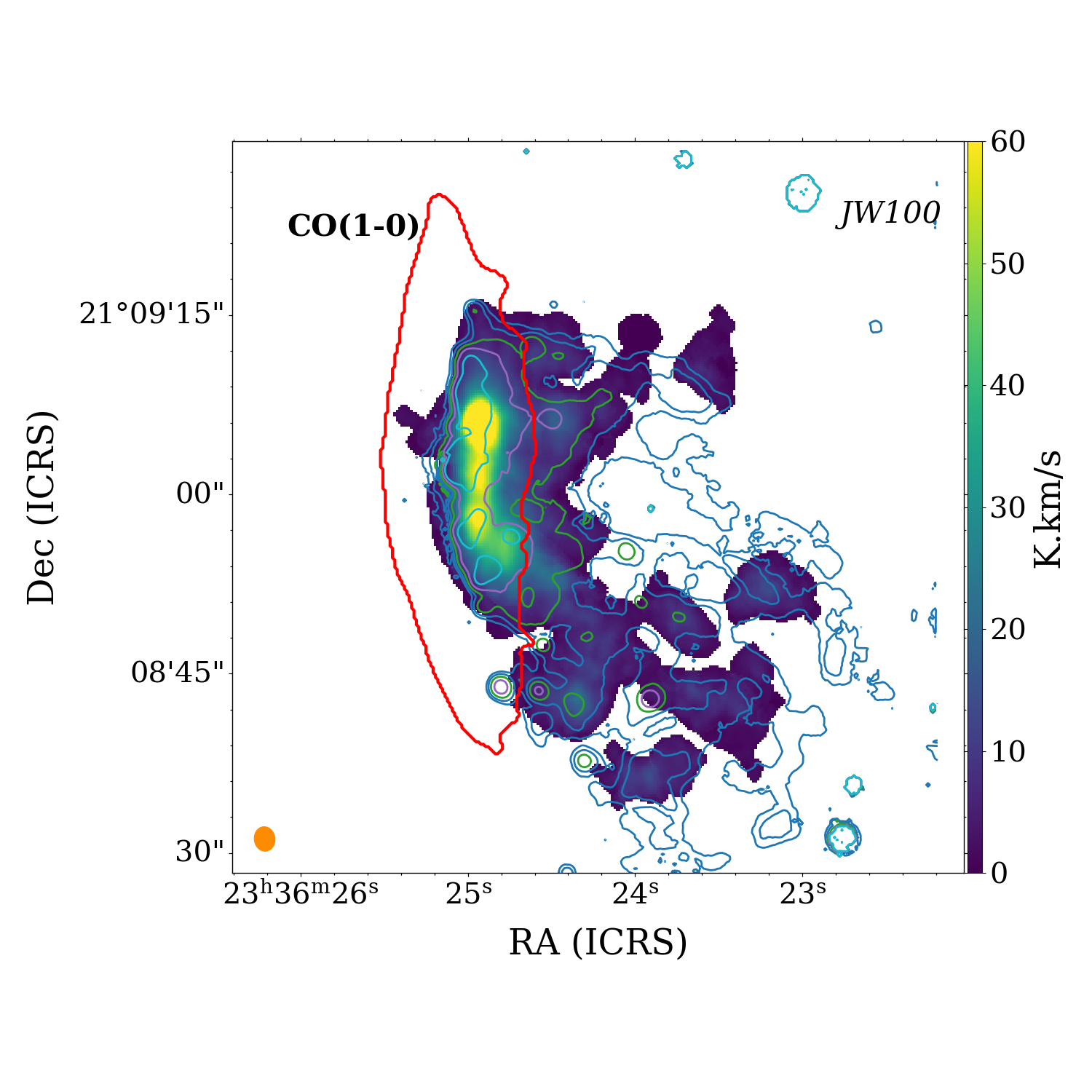}
     \caption{CO(2-1) and CO(1-0) zero-moment maps for JO201, JO204, JO206, and JW100. The red contour shows the extent of the stellar disk, as defined in \citet{GASPXXI}. The beam size is the orange ellipse in the lower left corner of each map.
     Colored contours delineate the \Ha emission derived from MUSE data.}
    \label{fig:co21_maps}
\end{figure*}

Concerning the cold gas properties of the galaxies, while Band 3 data have been analyzed in \citet{GASPXXII}, here we add the analysis of Band 6 observations.
The data were calibrated using the ALMA pipeline (version Pipeline-CASA51-P2-B) and imaged as described in \citet{Moretti+2020}. Datacubes were obtained smoothing the original spectral resolution (of $\sim$ 3 \kms) to 20 \kms to increase the signal-to-noise ratio (S/N), and channel velocities were computed in the kinematic local standard of rest (LSRK) frame, with a reference frequency corresponding to the redshifted frequency of the CO transition (reported in Table~\ref{tab:data} for each target). The continuum-subtracted dirty cubes were cleaned with a natural weighting, resulting in the synthesized beam of $\sim 1$\arcsec, similar to the MUSE point spread function (PSF). 
The line-emitting regions were initially identified with the automatic algorithm available in the CASA task {\it{tclean}} and interactively refined to better account for faint residual emissions. The achieved root mean square (rms) in the final cubes was measured in line free channels. In Table~\ref{tab:data} we report the rms and the properties of both CO(2-1) and CO(1-0) images, including the maximum recoverable scales (MRSs) provided by the arrays used to perform the observations. For the galaxy JW100 ACA observations have been added, as discussed in \citet{GASPXXII}.  
Figure~\ref{fig:co21_maps} shows the CO(2-1) zero-moment maps, and for comparison the corresponding CO(1-0) maps already presented in \citet{Moretti+2020}. Similarly to what we did for the CO(1-0) maps, we used the SOFIA software \citep{sofia} to construct the detection masks for our galaxies.
In two galaxies, JO201 and JO204, the emission detected outside the stellar disk is more prominent in the CO(2-1) line emission map, indicating the presence of regions filled with dense, star-forming gas, with warmer temperatures connected to the ongoing star formation. JW100, on the other hand, has a gaseous tail that indicates a multicomponent gas, where the star formation is less efficient (as is discussed in the following).

These data allowed us to derive the molecular gas surface densities in the disk and in the tails of our jellyfish galaxies on scales of $\sim 1-2$ kpc, using the following equations \citep{WatsonKoda2016} and the CO(2-1) and CO(1-0) line emission, respectively:
\begin{equation}\label{eqn:wk_co21}
\left(\frac{M_{\rm H_2}}{M_{\odot}}\right) = 
3.8 \times 10^3 \left( \frac{\alpha_{CO}}{4.3}\right)
\left(\frac{r_{21}}{0.7}\right)^{-1}
\left(\int S_{21}dv \right)
\left(D_L \right)^2
\end{equation}

and

\begin{equation}
\left(\frac{M_{\rm H_2}}{M_{\odot}}\right) = 
1.1 \times 10^4 \left( \frac{\alpha_{CO}}{4.3}\right)
\left(\int S_{10}dv \right)
\left(D_L \right)^2
,\end{equation}
where $S_{ij}$ is the integrated CO line flux of the i-to-j transition, r$_{21}$ is the CO(2–1)/CO(1–0) flux ratio, and D$_L$ is the luminosity distance in megaparsecs.
This formulation is equivalent to the one from \citet{Solomon2005} at the 5\% level.
To distinguish disk and tail spaxels, we made use of the definition of stellar disk as given in \citet{GASPXXI}.

In order to compare our data with the literature, we generally used as a conversion factor, $\alpha_{CO}$, the Milky Way factor, i.e., 4.3 $M_{\odot} \rm (K \, km \, s^{-1} \, pc^{2})^{-1}$\citep{Bolatto2013}. The calculation of the gas surface densities uses as line ratio r$_{21}$ the latest value found in nearby resolved star-forming galaxies from the PHANGS data \citep{Leroy+2022}, i.e., 0.65, which is slightly lower than the usually adopted value of 0.79 derived from unresolved observations \citep{Leroy2009,Saintonge2017,Brown+2021}. 
The overall distribution of r$_{21}$ in the xCOLDGASS sample \citep{Saintonge2017} and in the VERTICO survey of Virgo galaxies \citep{Brown+2021} shows indeed a quite broad distribution peaked at $\sim$0.8 with a large scatter ($\sim$0.3), which is in agreement with the value adopted here.
We also used our own data to estimate the r$_{21}$ ratio where both emissions were detectable, i.e., mostly within disks, and we obtained values compatible with the ones in the literature \citep[on average 0.59 $\pm$ 0.2, see][]{denbrok+2021,Egusa+2022}.

\begin{figure*}
    \centering
    \includegraphics[width=0.225\textwidth]{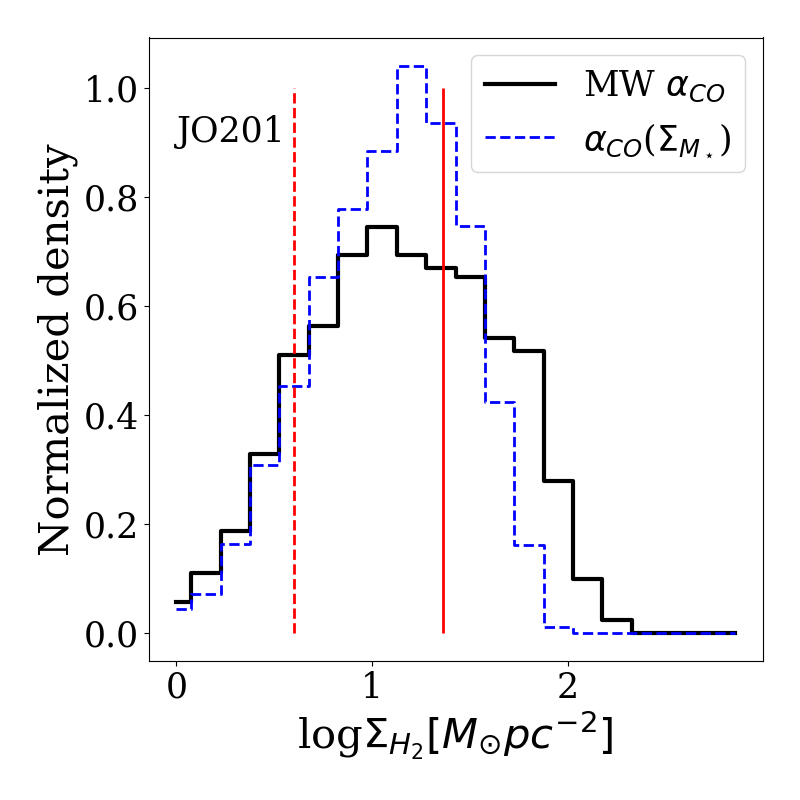}
    \includegraphics[width=0.225\textwidth]{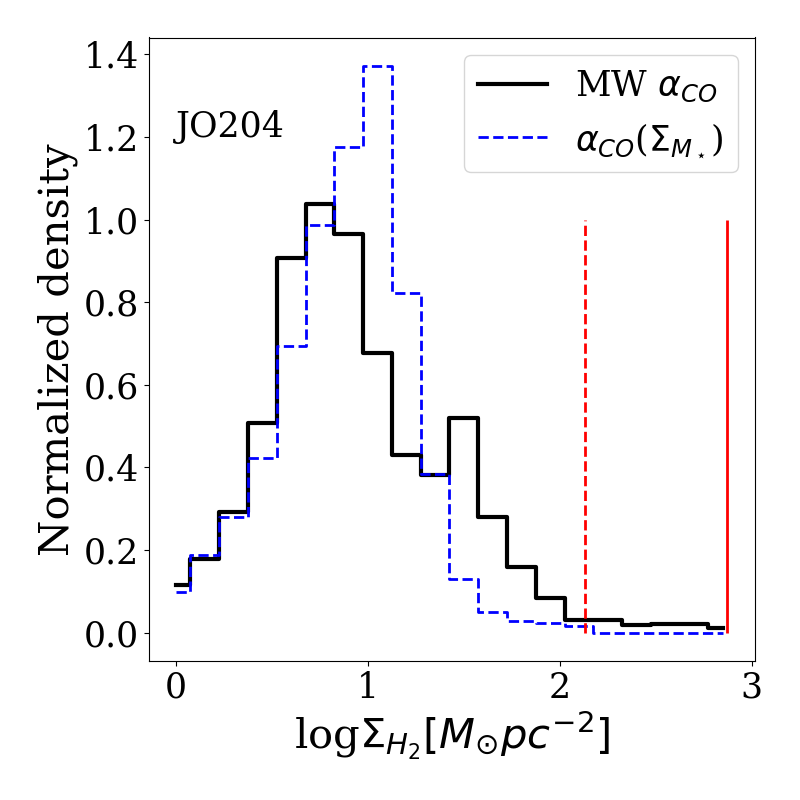}
    \includegraphics[width=0.225\textwidth]{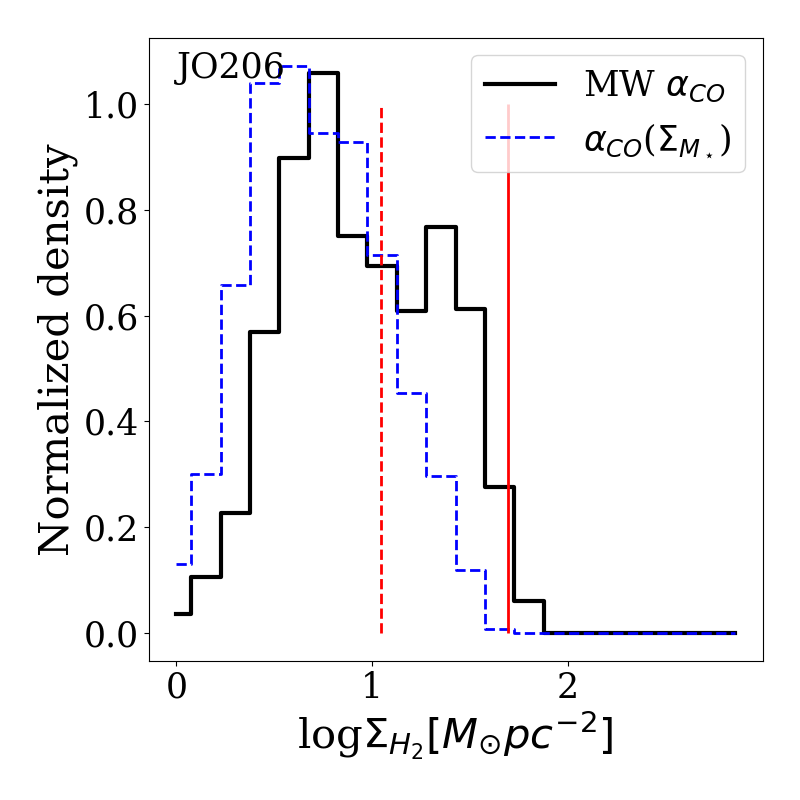}
    \includegraphics[width=0.225\textwidth]{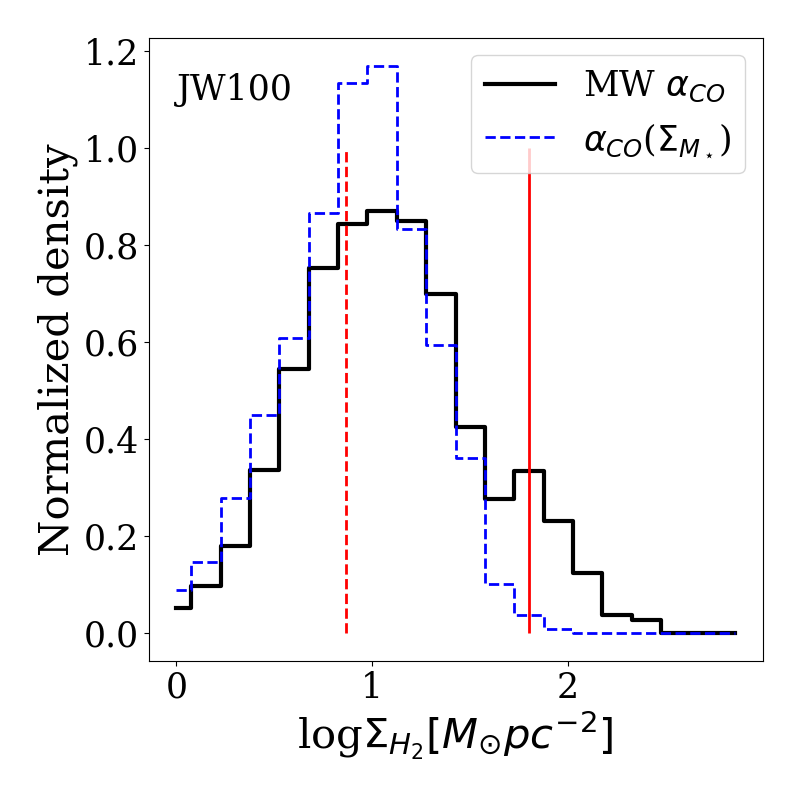}
     \caption{Distribution of the CO(2-1) derived molecular gas mass surface densities using the MW \aco (continuous black histogram) and a varying \aco that follows the stellar mass density distribution (dashed blue histogram), using the relation given in \citealt{Sandstrom+2013}. The vertical dashed and continuous red lines show the average density at the galaxy centers, estimated over the central $\simeq$1 kpc$^2$, assuming the variable \aco and the MW \aco, respectively.}
     
    \label{fig:sigma_co}
\end{figure*}

Molecular gas surface densities in our cluster galaxies subject to RPS range between 1 and $\leq$100 M$_{\odot}$ pc$^{-2}$, as can be seen from Fig.~\ref{fig:sigma_co}, and therefore span the typical range exhibited by normally star-forming galaxies \citep{Bigiel2011,Pessa+2021}.
The two distributions shown in each subpanel of Fig.~\ref{fig:sigma_co} refer to the spaxels' de-projected surface mass densities of molecular gas derived using the inclinations measured by \cite{GASPXXVII} and the Milky Way conversion factor (continuous black histogram) and the variable \aco calculated assuming the relation with the stellar mass surface density described in \citet[dashed blue histogram]{Sandstrom+2013}. In fact, the correlation between the \aco and the stellar mass surface density has been shown to be the strongest among those studied by \cite{Sandstrom+2013}, possibly tracing more closely changes in the ISM pressure.

Stellar mass surface densities were derived using the optical data from MUSE and the continuum fitting from SINOPSIS, as described in \cite{GASPI,GASPIII}.                                                      
In all galaxies assuming a variable \aco results in a narrower and unimodal \hdue  surface density distribution, and in particular to the disappearing of the secondary peak at high densities.
The vertical red lines in Fig.~\ref{fig:sigma_co} show the values of the molecular gas mass surface density reached averaging the measurements in the central region ($\simeq$1 kpc$^2$) assuming the two different \aco (MW \aco is the continuous line, the variable \aco is the dashed line).
 The highest molecular gas mass densities are reached in the regions closest to the galaxy center in JO204,  whereas the central regions of the other galaxies generally exhibit lower densities. This is at odds with normally star-forming galaxies, which show enhanced central densities, at least at the cloud scale \citep{Sun+2020b}, especially in the case of barred galaxies, such as those presented in this paper \citep[see also][]{GASPIV,Bacchini+2023}.
What causes this central depletion of molecular gas is not yet clear, as the complex interplay between ram-pressure and AGN activity is still puzzling. We notice, however, that JO204 is the galaxy hosting, probably, the least powerful AGN \citep{poggianti2017,GASPXIX}, suggesting a positive correlation between the AGN power and the clearing of the central regions from the molecular gas, while RPS would cause a more general and mild effect.
For the remainder of our analysis, we use the widely used standard \aco derived from Milky Way observations.

To derive the $\Sigma_{SFR}$ of our sample, we used the \Ha emission measured on the MUSE datacubes after the subtraction of the stellar component, and after the correction for internal extinction as traced by the Balmer decrement.
These measurements were performed on the 5x5 smoothed MUSE datacubes \cite[as described in][]{GASPI}.
As a result, optical datacubes are characterized by an effective spatial resolution of $\sim1$\arcsec\,, which corresponds to $\sim$ 1 kpc at the redshift of the target sample.
In order to build a meaningful comparison with ALMA data, we convolved the MUSE \Ha emission with a Gaussian 2D kernel, resulting in the size and position angle of the corresponding ALMA beam appropriate for each galaxy.
We then converted these \Ha luminosities using the following prescription, valid for a  \cite{Chabrier2003} initial mass function (IMF):
\begin{equation}
    \mathrm{SFR (M_{\odot}
\, yr^{-1})} = 4.6 \times 10^{-42} L_{\rm H\alpha} \mathrm{(erg \, s^{-1}})
.\end{equation}
Spaxels characterized by AGN and LINER-like line ratios, according to the BPT diagram \citep{BPT} involving the [NII] lines, were excluded from the calculation.
They represent a small fraction of the total number of spaxels with a measurable \Ha emission (from 3\% in JO206 to 13\% in JW100).
In this way, we are confident that we are analyzing here regions dominated mostly by star formation and not by the AGN.

\section{The Kennicutt-Schmidt relation on a 1 kpc scale}\label{sec:KS1}
Thanks to the combination of ALMA and MUSE data, we can derive the relation between the ongoing $\Sigma_{SFR}$ and the corresponding surface mass density of molecular gas, $\Sigma_{H_2}$, the so-called KS relation \citep{Schmidt1959,Kennicutt1998}, on a scale of $\sim$ 1 kpc, after having convolved both the ALMA and MUSE data to the common resolution given by the largest beam size of each galaxy (the worst resolution).
To build this relation, we used only star-forming and composite spaxels that lie within the SOFIA masks. 

\begin{figure*}
    \centering
    \includegraphics[width=0.45\textwidth]{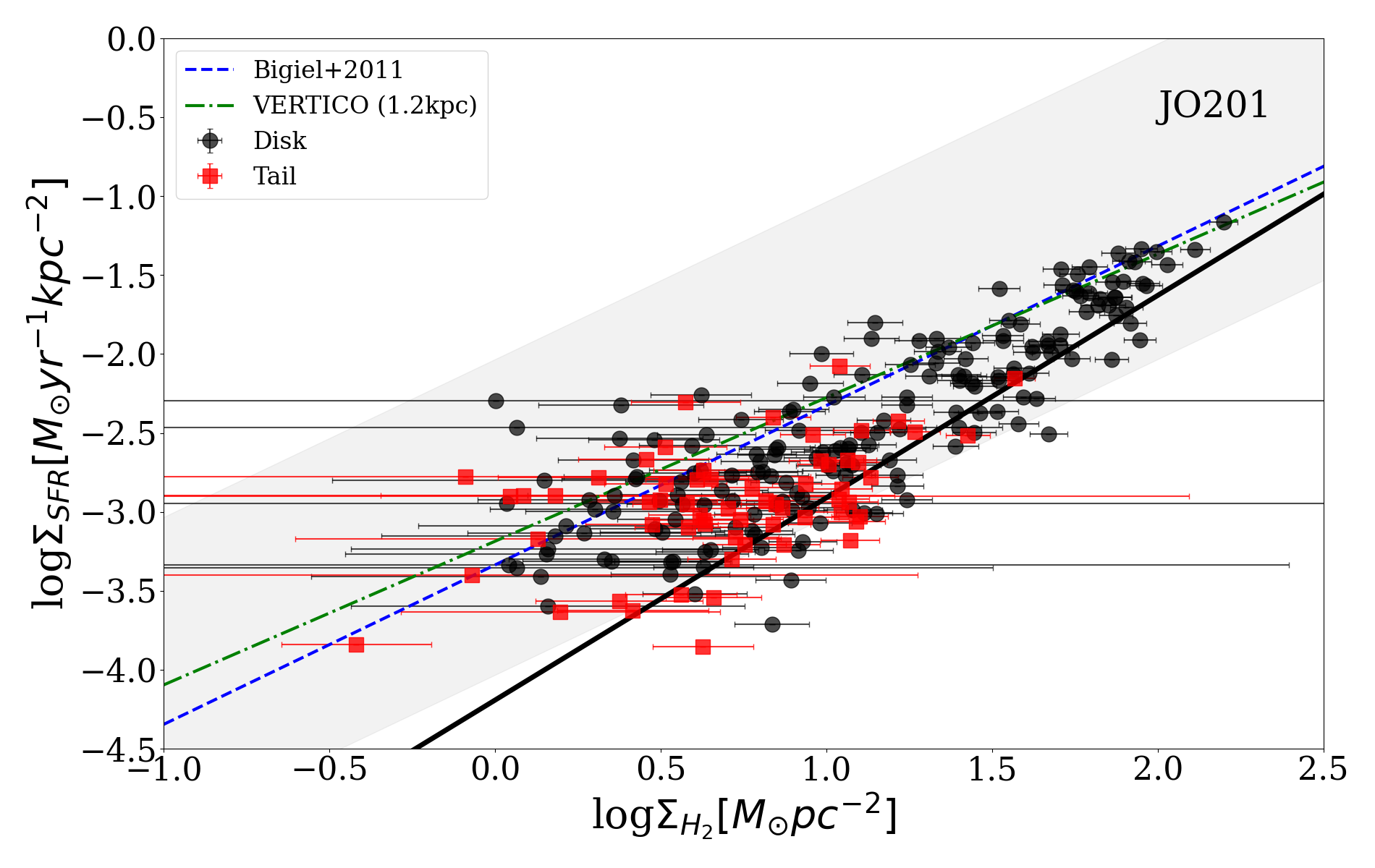}
    \includegraphics[width=0.45\textwidth]{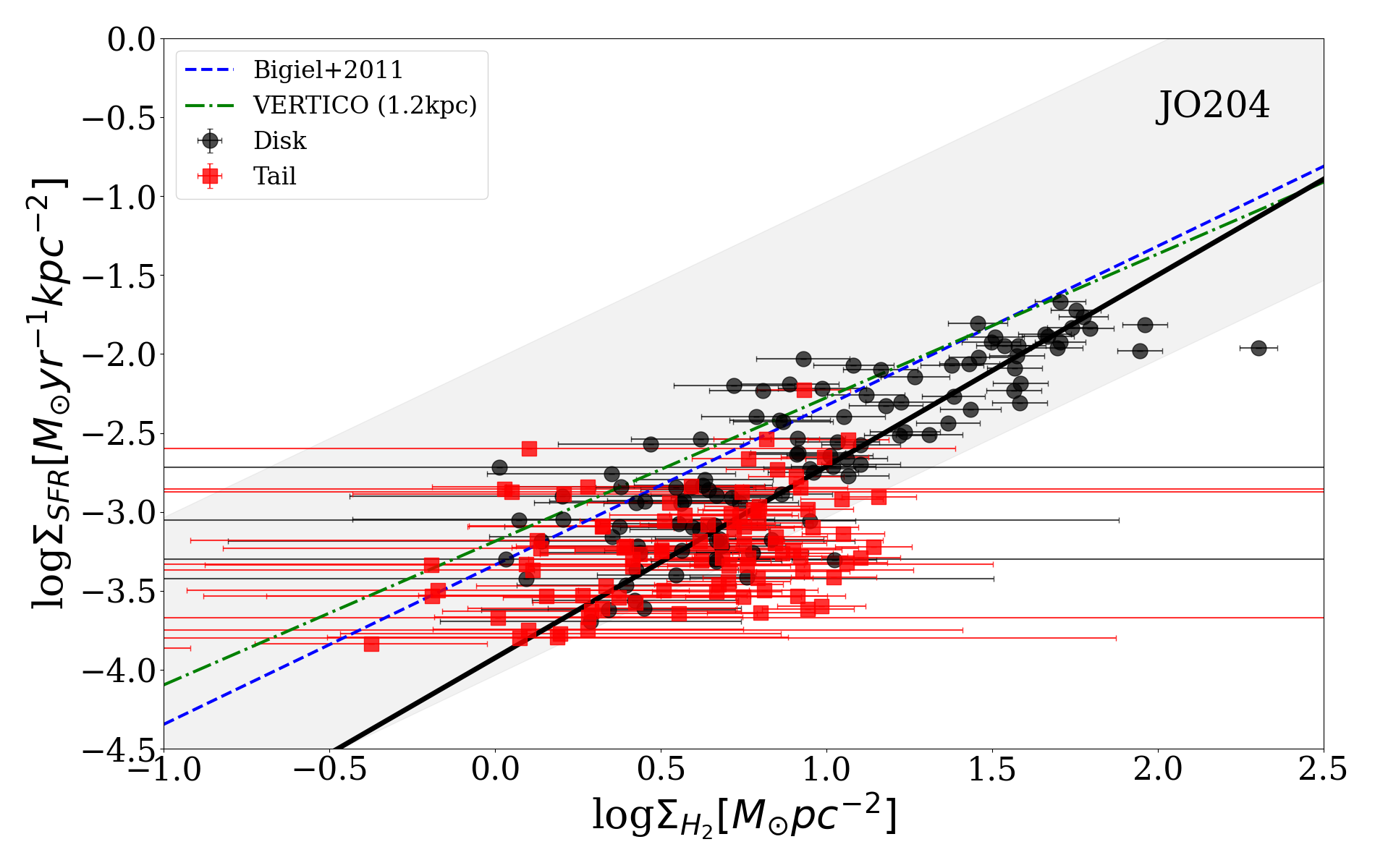}
    \includegraphics[width=0.45\textwidth]{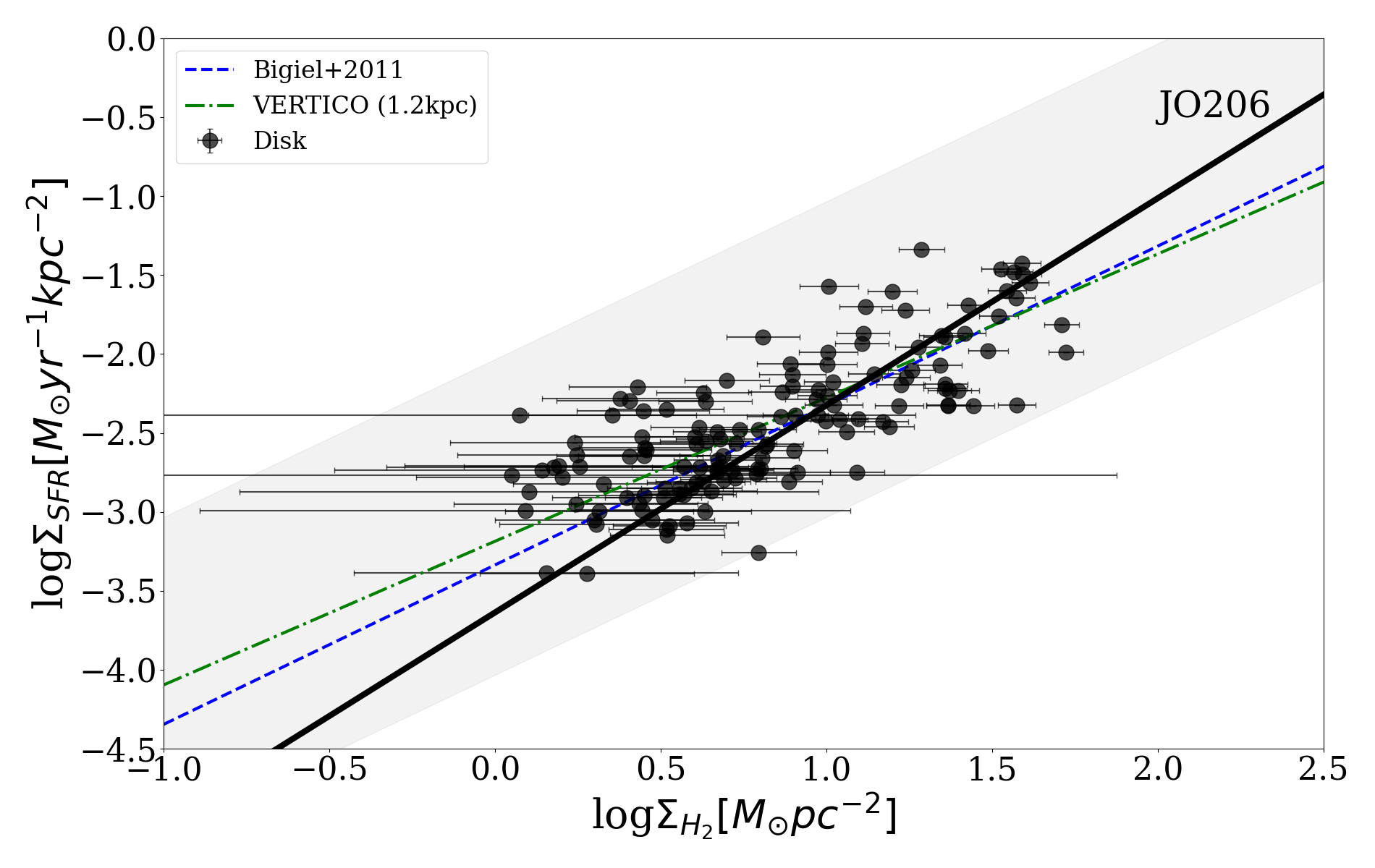}
    \includegraphics[width=0.45\textwidth]{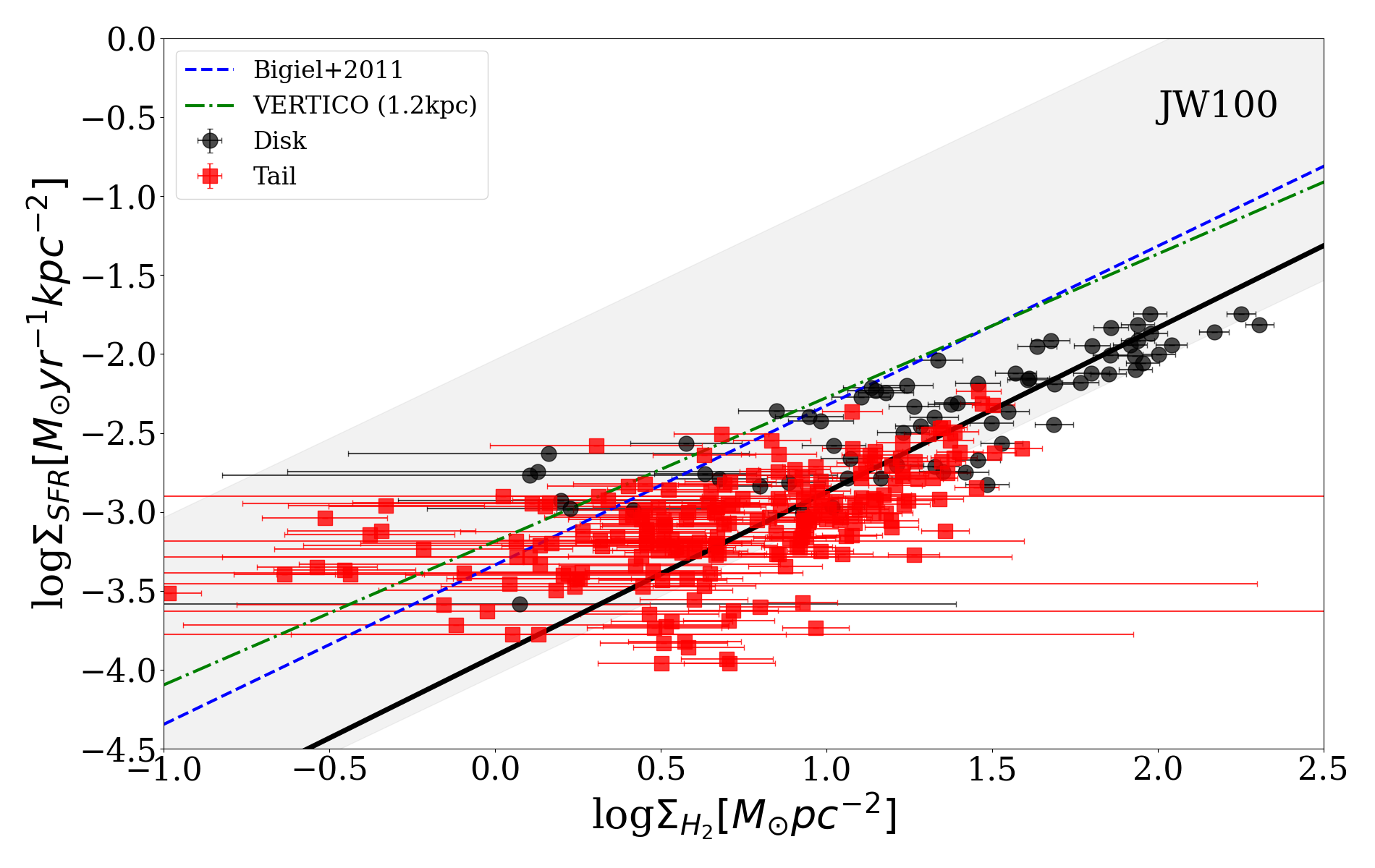}
     \caption{
     Star formation rate density as a function of the \hdue surface mass density from the CO(2-1) measured on 1 kpc scale for the GASP galaxies JO201, JO204, JO206, and JW100 (from top left to bottom right). The shaded region encompasses the zone with depletion times between 0.1 and 10 Gyr. The continuous black line is the fit to the disk regions (in black) that takes into account errors on both coordinates, the dashed blue line is the local relation on 1 kpc scale from \citet{Bigiel2011}, and the dashed green line was derived for VERTICO galaxies on a 1.2 kpc scale from \cite{jimenez+2023}. The red squares show the measurements corresponding to the tails.
     }
    \label{fig:ks_global_21}
\end{figure*}

Figure~\ref{fig:ks_global_21} shows the SFR density as a function of the H$_2$ surface mass density derived from the CO(2-1) line emission for the four galaxies in our sample. 
The shaded gray region shows the region of this plane where the depletion time ranges between 0.1 and 10 Gyr.
Black dots and red squares refer to the regions located within the stellar disk of the galaxies and in the tail, respectively.
In order to fit the data we used the orthogonal distance regression method to model a linear relation, taking into account the errors in both coordinates.
The error in the $\Sigma_{SFR}$ originates from the fit of the \Ha flux per spaxel in the emission-line maps, as described in \citealt{GASPI}, while for the uncertainty on the molecular gas mass surface density we used the rms of the datacubes.
We fit the data of each galaxy with the following equation describing the KS relation:
\begin{equation}
    log\Sigma_{SFR}=m\times log\Sigma_{H_2}+q
,\end{equation}
and show the slope (m) and the intercept (q) of each relation in Table~\ref{tab:ksfit}.
In Fig.~\ref{fig:ks_global_21} the black line is the fit we obtained for the disk regions, which turns out to be superlinear (i.e., m$>$1) for all galaxies (with the lowest slope shown by JW100). For the sake of clarity, we do not show in the plots the fit of the tail regions, but slopes and intercepts are given in Table~\ref{tab:ksfit}.
The dashed blue line shows instead the fit at 1 kpc scale of undisturbed star-forming disks as derived in \citet{Bigiel2011} and converted to the \citealt{Chabrier2003} IMF. The green line shows the relation found for Virgo cluster galaxies using ALMA data \citep{jimenez+2023} at a 1.2 kpc resolution, which is similar to the one we reach for our galaxies.
In all plots the fit is dominated by the high-density regions, where the errors are smaller.
Band 6 data shown in Fig.~\ref{fig:ks_global_21} led us to conclude that the CO(2-1) relation has an average slope of $\sim$ 1.3 for 3/4 of our galaxies, with the only exception being JW100, the most massive galaxy of the sample ($M_\star = 3 \times 10^{11} M_{\odot}$).
In fact, this galaxy shows always low efficiencies (or long depletion times), both in the disk and in the tail, and when using both CO(2-1) and CO(1-0) measurements.
In particular, this is true also for the regions with the highest molecular gas densities.
As already discussed in \citealt{GASPXXII}, JW100 could show such low efficiencies even within the disk because of an increased turbulence related to the formation of the central spheroid \citep{Haywood+2016,khoperskov+2018,Gensior+2020}.
It is worth also noticing that JO206 does not have significant emission in the tail in the masked cube.

In general, though, the fact that our galaxies show a SFE that becomes higher in the regions characterized by higher molecular gas mass surface densities suggests that the star formation is regulated by the gravitational free-fall time of the molecular clouds \citep{Kennicutt1998}. 
This effect is reminiscent of the local variations in the SFE found by \citet{ellison+2020} in a sample of 34 galaxies possessing ALMA and MANGA data.
Variations in the \aco conversion factor have also been invoked to explain the environmental dependence of the gas depletion time at a 1 kpc scale \citep{Leroy2013}.
We note that assuming a \aco variable with the stellar mass surface density \citep{Sandstrom+2013} would only increase the slope of the KS relation, as the regions with highest $\Sigma_{H_2}$ would move toward lower values, conserving their SFR density (see Fig.~\ref{fig:sigma_co}).
None of the fit relations would change significantly assuming our own $r_{21}$ measurements, which do not deviate strongly from the assumed value. Flatter slopes, though, could be recovered if $r_{21}$ becomes smaller in the densest regions, which is not the case in our galaxies.

Tail regions are always characterized by a lower SFE, as most of them always lie close to the $\tau_{dep}\sim10$ Gyr line.
This suggests that the star formation is inefficient in these regions.
These findings show that the KS relation of undisturbed star-forming galaxies is mostly flatter than in these peculiar objects at a scale of $\sim$ 1 kpc, when referring to the disk regions, and when using the CO(2-1) line emission as a dense gas tracer.

Studying both the CO(2-1) and CO(1-0) transitions is essential because they trace different physical conditions of the molecular gas: CO(1-0) is more sensitive to the diffuse, cold component, characterized by critical densities of $10^2-10^3$ cm$^{-3}$ and temperatures in the range of 10-20K, while CO(2-1) predominantly traces denser and warmer regions (with n$_{H_2}=10^3-10^4$ cm$^{-3}$, T$_{kin}$=20-50K), allowing for a more complete understanding of the molecular gas distribution and excitation within galaxies.

\begin{figure*}
    \centering
    \includegraphics[width=0.45\textwidth]{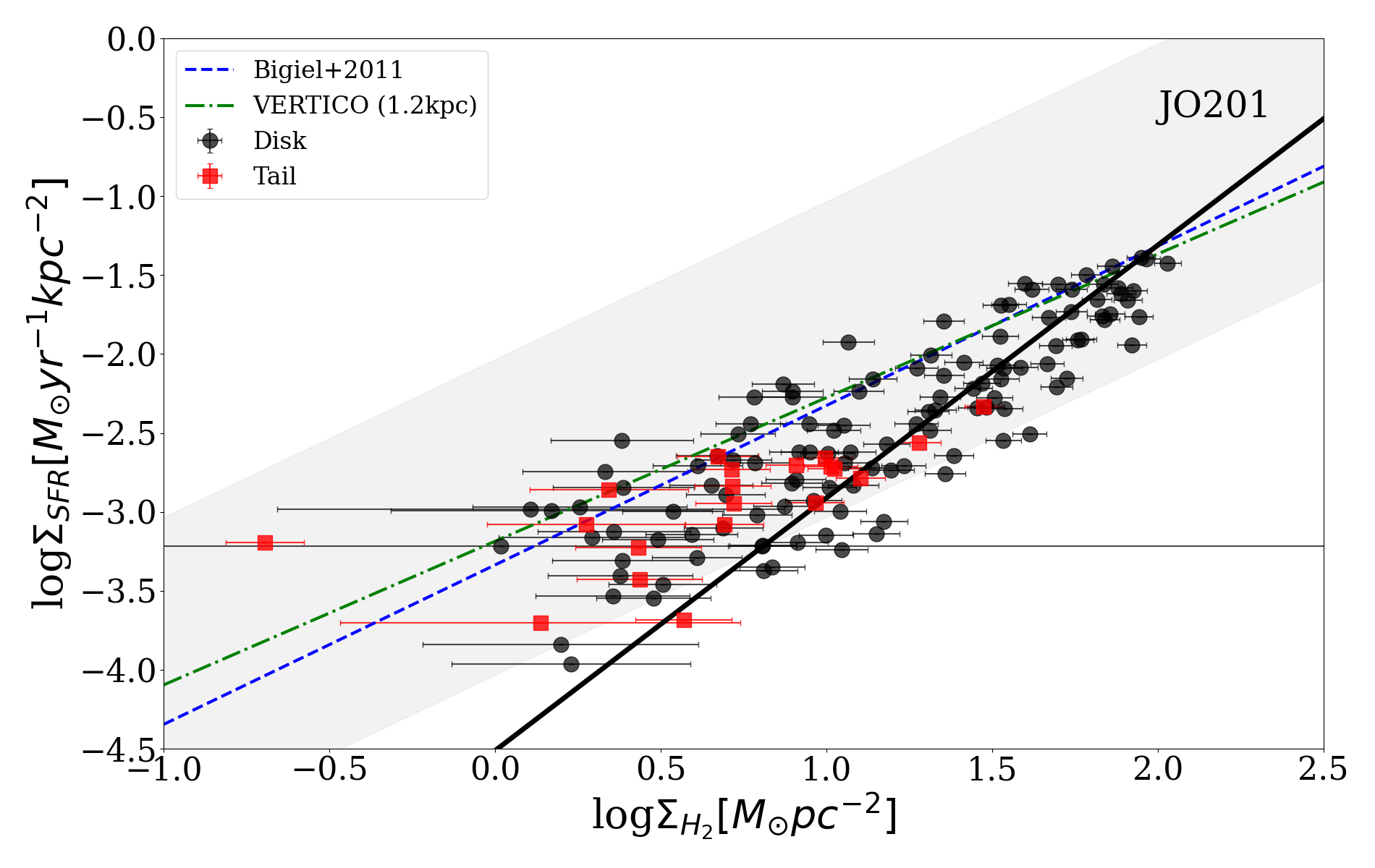}
    \includegraphics[width=0.45\textwidth]{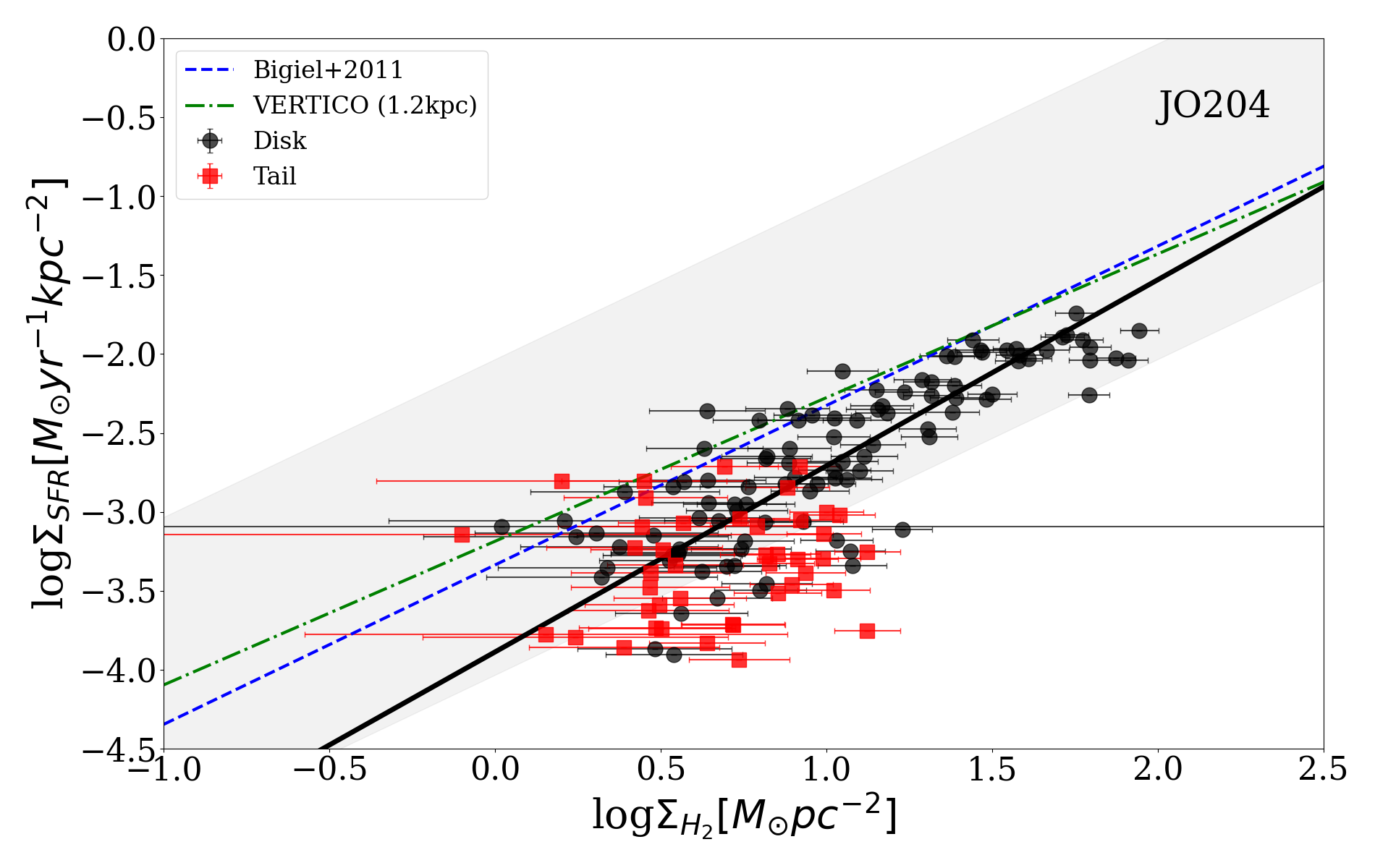}
    \includegraphics[width=0.45\textwidth]{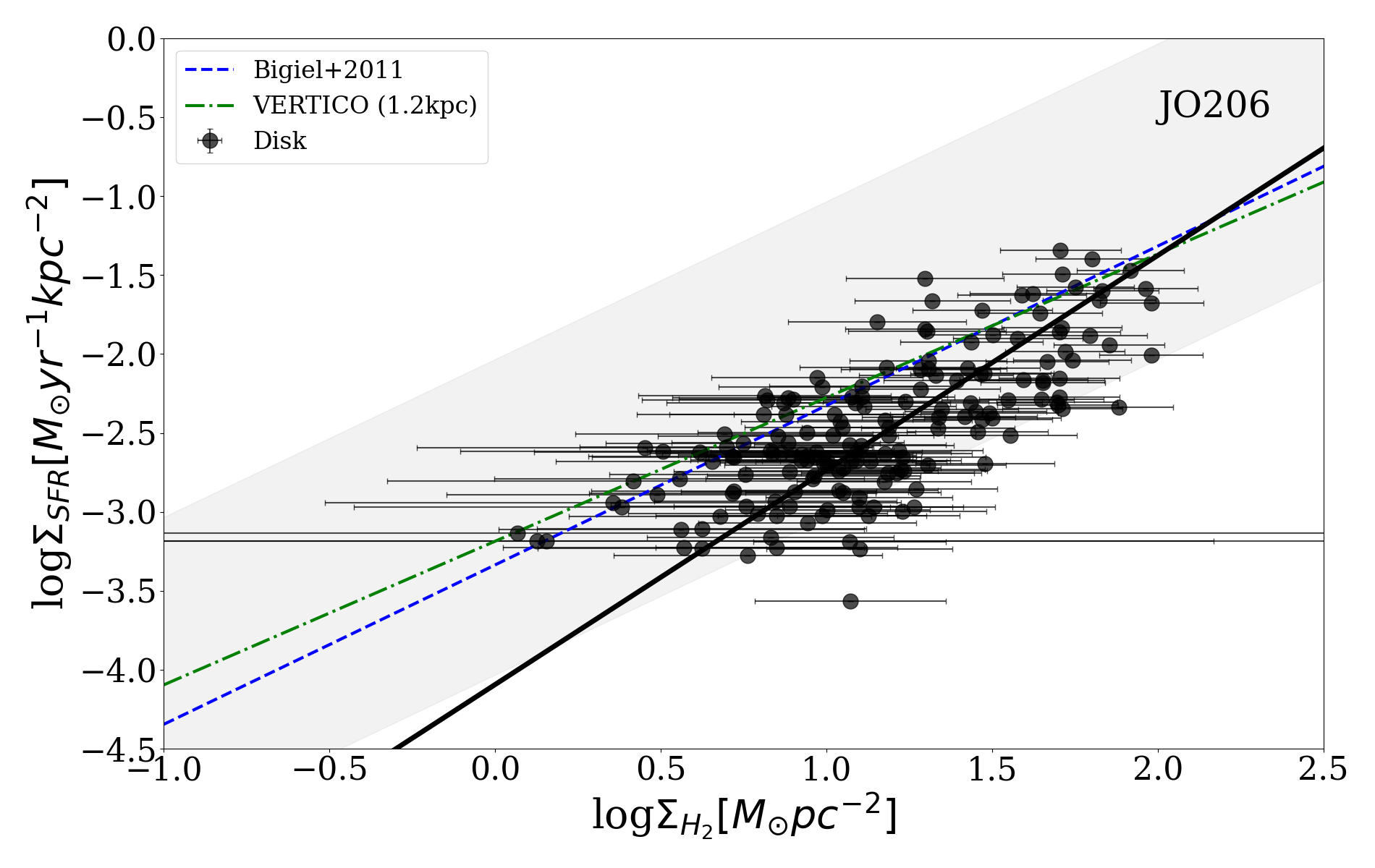}
    \includegraphics[width=0.45\textwidth]{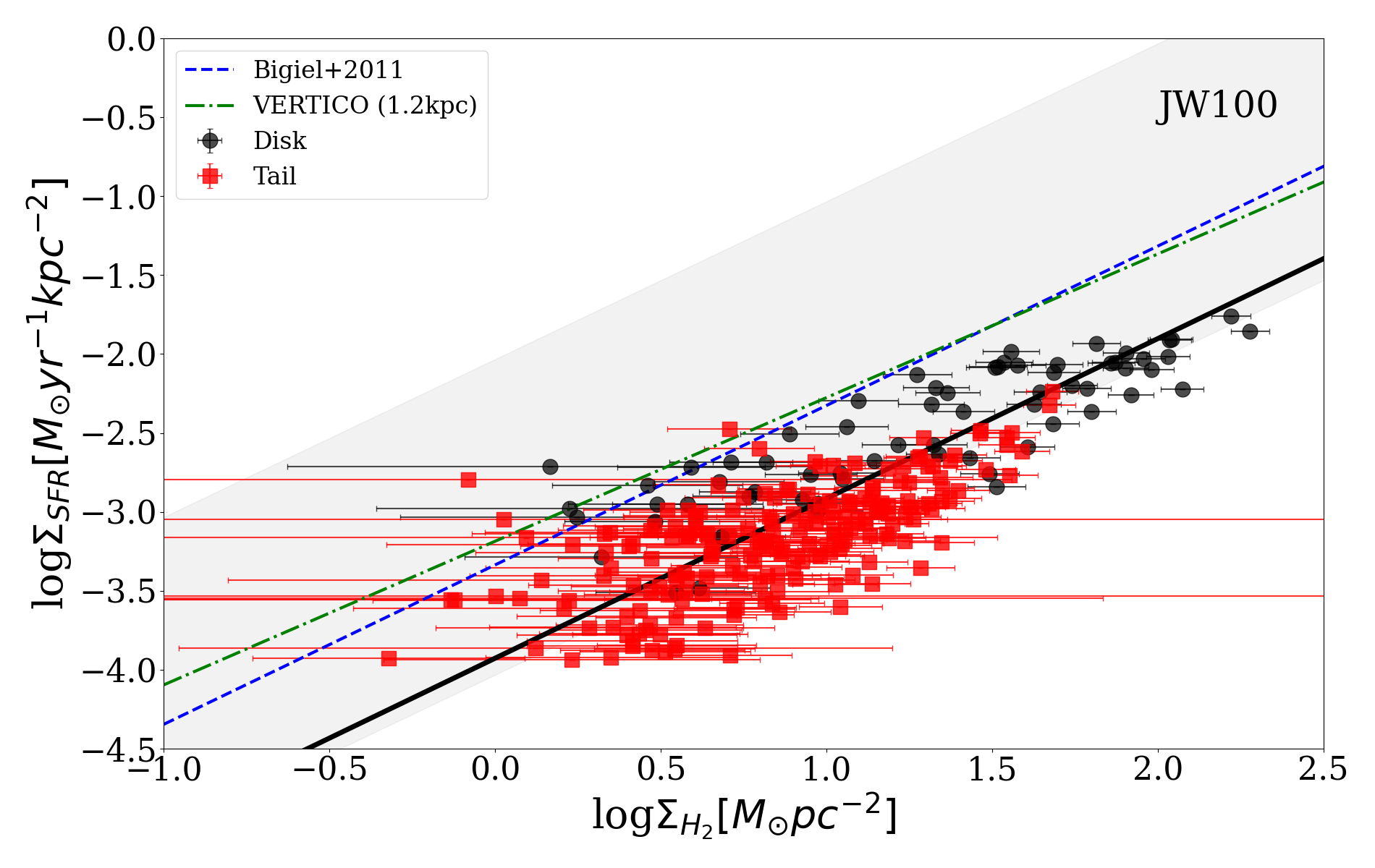}
    \caption{As in Fig.~\ref{fig:ks_global_21} but for the CO(1-0) emission.}
    \label{fig:ks_global_10}
\end{figure*}

In our sample, when using the CO(1-0) line as a molecular gas tracer, as shown in Fig.~\ref{fig:ks_global_10} and Table~\ref{tab:ksfit}, it becomes immediately evident that the relation becomes steeper only for JO201, while it is compatible with the one derived from the CO(2-1) line for the other galaxies.
This implies that the effect of the RPS on these galaxies remains even when using the CO(1-0) as a gas tracer: denser regions have shorter depletion times with respect to less dense regions.
This could be due to the fact that we assumed a constant \aco factor in the whole disk, which can lead to an underestimation of the real molecular gas content in the regions where the star formation is more efficient, steepening the observed slope with respect to the one exhibited by undisturbed star-forming disks.
The same effect could also be due to the presence of CO-dark regions at the lowest densities, which makes the CO line an inefficient tracer of the cold gas.

However, we notice also that all the CO(1-0) based measurements are shifted toward lower efficiencies, i.e., they show lower star formation rate densities for a given cold gas column density, or, in other words, we can measure a given density of star formation only when reaching higher cold gas densities.
We found no significant differences when measuring the CO(2-1) gas densities within the larger CO(1-0) beams, demonstrating that this is not the cause of the shift.
Most probably, as expected, it is the nature of the more diffuse CO(1-0) emission that reflects into longer depletion times.

\begin{figure*}
    
\vspace{1cm} 

\begin{minipage}[c]{\textwidth}
    \begin{minipage}[c]{\textwidth}
         \begin{minipage}[c]{0.5\textwidth}
         \includegraphics[width=\textwidth]{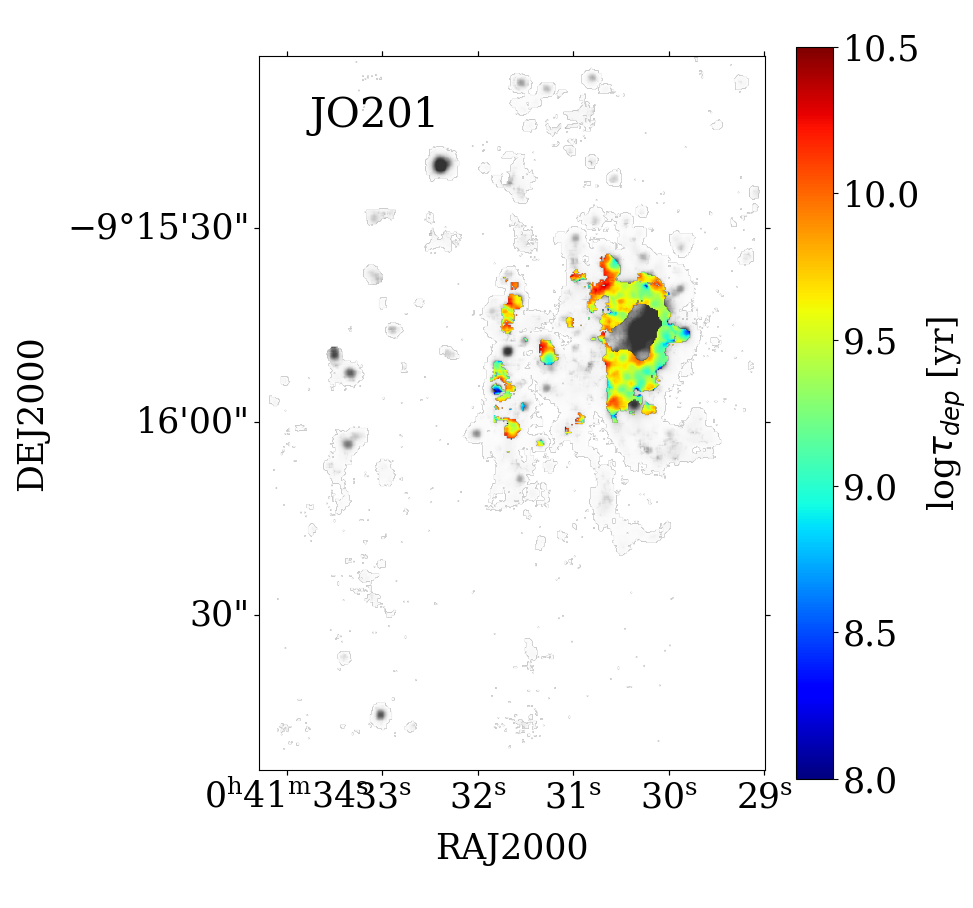}
         \end{minipage}
\hfill
        \begin{minipage}[c]{0.45\textwidth}
        \includegraphics[width=\textwidth]{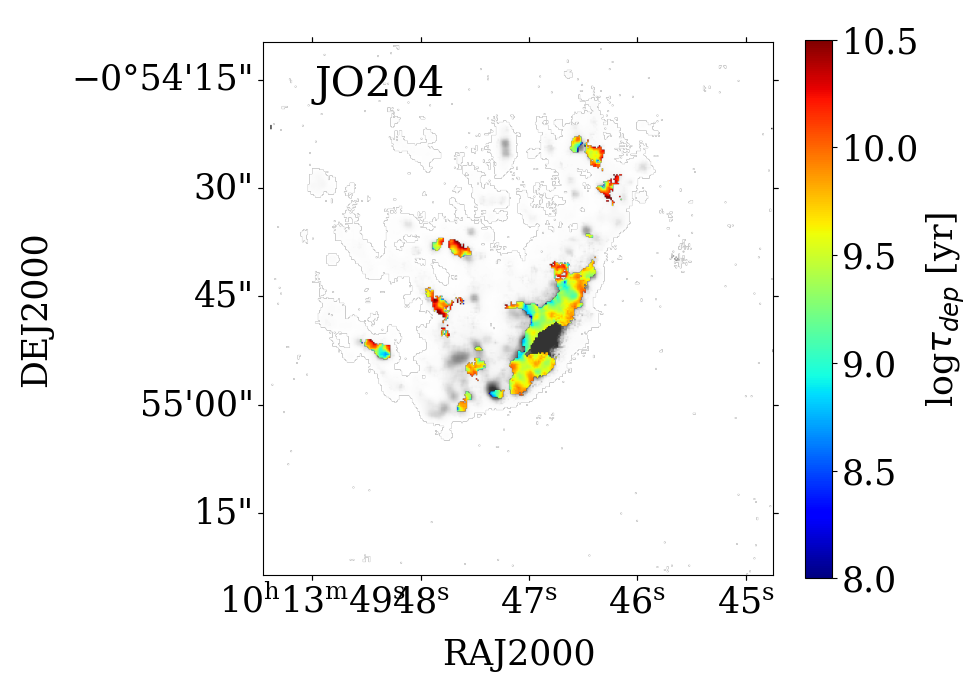}
        \includegraphics[width=\textwidth]{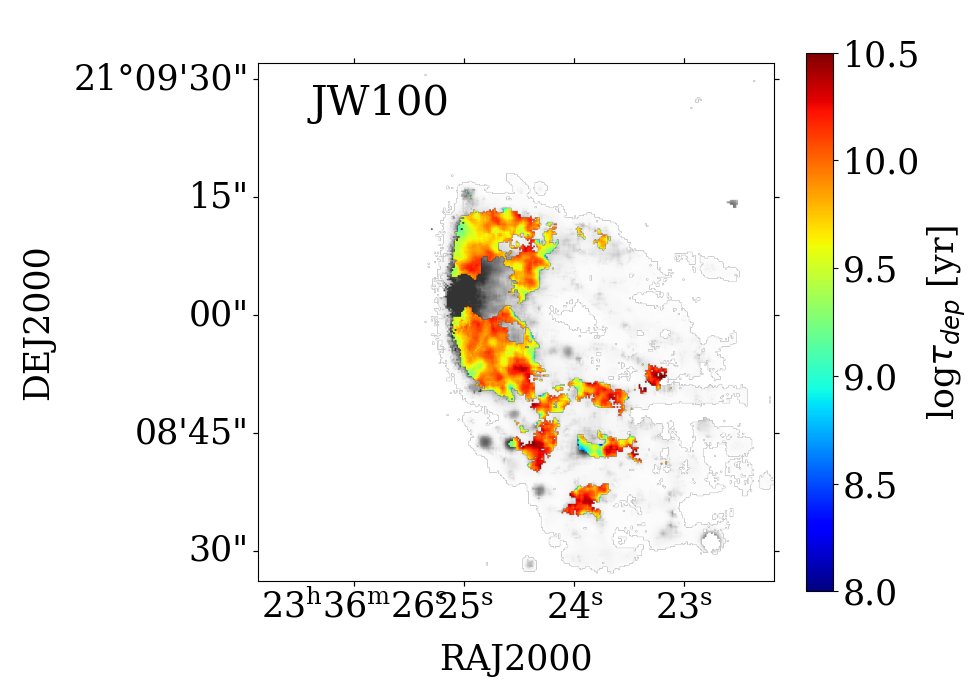}
        \end{minipage}
    \end{minipage}
    \begin{minipage}[c]{\textwidth}
    \centering
    \includegraphics[width=\textwidth]{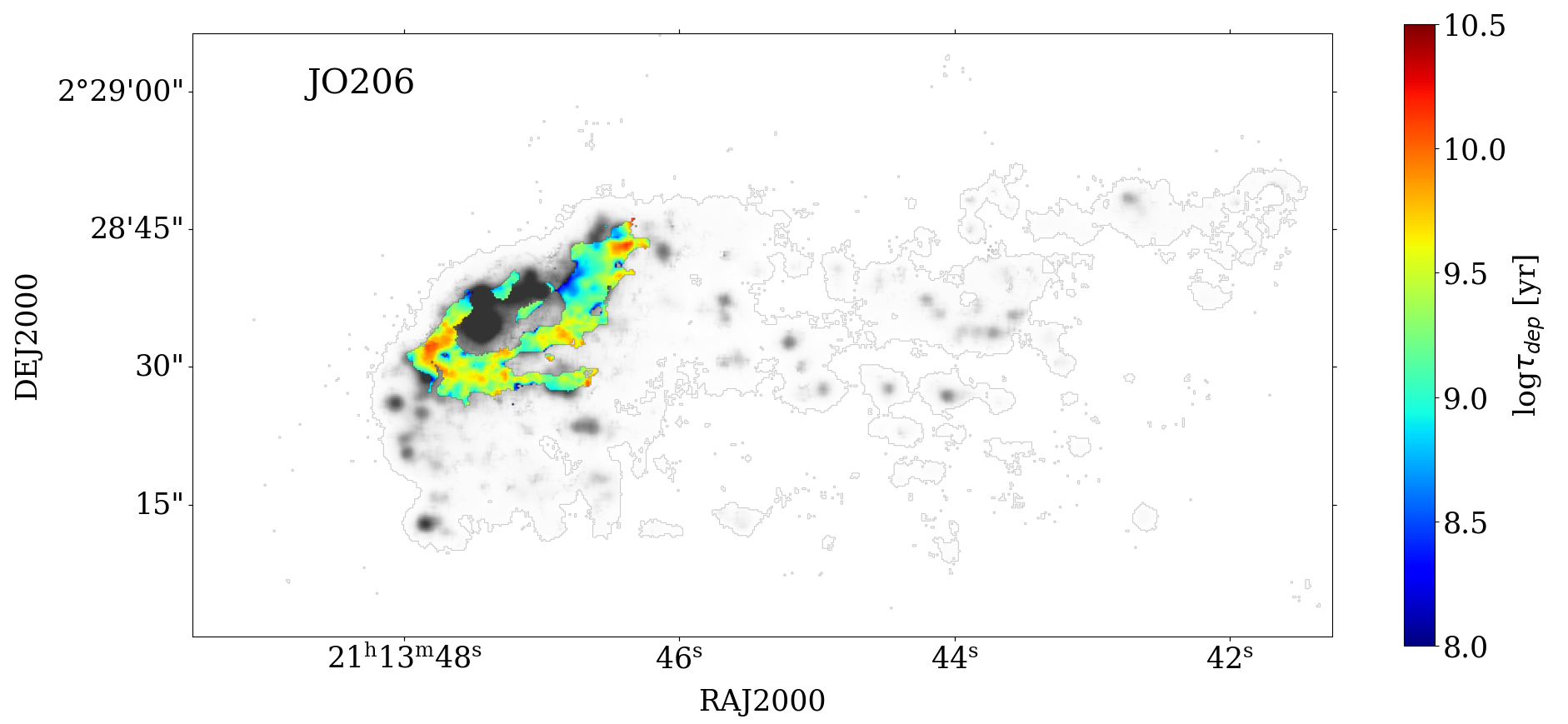}   \end{minipage}
\end{minipage}
\caption{Depletion time maps for the four jellyfish galaxies JO201 (top left), JO204 (top right), JW100 (middle right left), and JO206 (bottom). The gray background layer shows the extent of the \Ha emission from MUSE.}
\label{fig:tau_dep}
\end{figure*}

\begin{table}[]
    \centering
        \caption{KS relation slope and intercept for the disk spaxels in our galaxies, derived from the CO(2-1) and the CO(1-0) data.}
    \begin{tabular}{|c|c|c|c|c|}
    \hline
    Galaxy & Slope & Intercept  \\
    \hline
    CO(2-1)&\multicolumn{2}{|c|}{Disk}\\
    \hline
    JO201     & 1.28 $\pm$ 0.06 & -4.19 $\pm$ 0.10 \\
    JO204     & 1.21 $\pm$ 0.08 & -3.92 $\pm$ 0.12 \\
    JO206     & 1.31 $\pm$ 0.08 & -3.64 $\pm$ 0.10 \\
    JW100     & 1.04 $\pm$ 0.09 & -3.91 $\pm$ 0.15 \\
    \hline
    CO(1-0)&\multicolumn{2}{|c|}{Disk}\\
    \hline
    JO201     & 1.60 $\pm$ 0.09 & -4.51 $\pm$ 0.14 \\
    JO204     & 1.18 $\pm$ 0.07 & -3.89 $\pm$ 0.09 \\
    JO206     & 1.36 $\pm$ 0.08 & -4.099 $\pm$ 0.11   \\
    JW100     & 1.01 $\pm$ 0.09 & -3.93 $\pm$ 0.15  \\
    \hline
        CO(2-1)&\multicolumn{2}{|c|}{Tail}\\
    \hline
    JO201     & 1.89 $\pm$ 0.29 & -4.66 $\pm$ 0.29 \\
    JO204     & 0.87 $\pm$ 0.10 & -3.88 $\pm$ 0.09 \\
    JO206     &  &  \\
    JW100     & 1.31 $\pm$ 0.14 & -4.23 $\pm$ 0.15 \\
    \hline
    CO(1-0)&\multicolumn{2}{|c|}{Tail}\\
    \hline
    JO201     & 0.96 $\pm$ 0.62 & -3.41 $\pm$ 0.47 \\
    JO204     & 0.75 $\pm$ 0.20 & -3.57 $\pm$ 0.12 \\
    JO206     &  &    \\
    JW100     & 1.34 $\pm$ 0.08 & -4.46 $\pm$ 0.09  \\
    \hline
    \end{tabular}
    \label{tab:ksfit}
\end{table}
The trend that we find when fitting the whole set of spaxels within our galaxies could in principle hide the presence of different relations connected to the geometry of the stripping. Nearby Virgo cluster galaxies in fact have a higher SFR along the compression front \citep{KoopmannKenney2004,Cramer+2020}, due to the impact of the ram pressure. Numerical simulations also predict enhanced SFRs at the leading side of cluster galaxies subject to RPS \citep{RoedigerBruggen2007}.
In order to understand whether the geometry of the gas compression is responsible for altering the SFE, we also produced the maps of the depletion times (i.e., the inverse of the SFE) from the CO(2-1) line emission, shown here in Fig.~\ref{fig:tau_dep}, in which the gray underlying map shows the extent of the ionized gas distribution from MUSE.
In all galaxies, the central zones are excluded from the depletion time map, since they are classified as AGN and/or LINERs from the optical line ratios.
Tail regions for which depletion times could not be calculated either are not star-forming or do not possess a significant emission in the CO(2-1) line.

In general, the depletion time increases in the stripped tails, where it is often longer than the Hubble time.
The compression front is not easily distinguishable in JO201, where the stripping has a strong component along the line of sight \citep{GASPII,GASPXXIX}, but some hint of short depletion times are seen on the west side, in agreement with the direction of the compression \citep{GASPII,Bacchini+2023}.
JW100 shows a similar behavior in the northeast, again following the infalling trajectory of the galaxy, while along the tail the depletion time reaches 10-20 Gyr \citep[see also][]{GASPXXII}.
The correlation with the infalling trajectory of galaxies in the cluster is much less clear in JO204 and JO206, where we would expect shorter depletion times in the west side and in the northeast side, respectively, while we only see a hint of this. Obviously, orientation issues as well as projection effects due to the 3D distribution of the gas \citep{Bacchini+2019_MW,Bacchini+2019_vol} may influence these maps.

A comparison with the trend of SFE with the galactocentric distance derived in normally star-forming galaxies with a similar spatial resolution \citep{Villanueva+2021} is not straightforward, as the galaxies that we analyze here are strongly disturbed by the ram pressure, and as such are not in dynamical equilibrium. Our maps clearly show a decline in the efficiency moving toward the external parts of the disk, but a measurement of the gradient would be meaningless, as the morphology of the galaxies is disturbed by the ram pressure.

\section{The Kennicutt-Schmidt relation in \Ha emitting knots}\label{sec:KSc}
The analysis of the optical MUSE data of the GASP survey has revealed a number of star-forming knots both in the disk and in the ionized gas tails of ram-pressure-stripped galaxies.
We have identified these regions on the basis of their \Ha emission \citep{GASPI} at a scale of $\sim 1$ kpc, and characterized their properties thanks to the ionized gas line ratios in \citet{GASPXIII}.
In \citet{GASPXXX} we also characterized the SFR-mass relation in tail clumps, finding that they show a relation similar to the one of disk regions, once the contribution of older stellar populations contaminating the disk is excluded.
 Their diffuse ionized gas (DIG) content was estimated by \cite{GASPXXXII}, based on the analysis of the [SII] emission.

In order to estimate the molecular gas properties of these ionized gas regions, we used the CO(2-1) line in the spectrum extracted from each \Ha identified knot, with the underlying assumption that the molecular gas emission arises from a region with the same extent of the one traced by the ionized gas. We used this gas tracer, as it is strongly connected to the star formation, while we used the information on the CO(1-0) emission only to confirm low-significance detections.
When the area of the MUSE-detected clump was smaller than the ALMA beam, we extracted the spectrum from a region as large as the beam itself.
We defined the S/N of each spectrum as the ratio between the peak of the spectrum and its standard deviation calculated in a line-free portion of the spectrum centered on the \Ha line and extending from -500 to +500 \kms.

For each \Ha knot we therefore derived the corresponding velocity, velocity dispersion, and flux of the molecular gas using a Gaussian fit\footnote{We used the Levenberg-Marquardt algorithm and least squares statistic included in \textit{astropy}}.
As a further constraint, we used as a first guess for the centering of the CO line the ionized gas emission velocity, i.e., 
we focused on a velocity window centered on the \Ha mean velocity and with a width equal to $\pm 5$ times the \Ha velocity dispersion, and left the CO intensity, central velocity, and velocity dispersion as free parameters.

Among the CO detections, we performed a visual inspection, and we considered as good  those  spectra that have a S/N larger than 2 in at least two adjacent channels. For those close to this limit, we also kept in the catalog (with flag=2) the detections with the same velocity of the \Ha and a CO(1-0) line emission, under the assumption that these conditions correspond to a 3D coincidence in space with confirmation from the more diffuse gas. Spectra with S/N larger than 2 in only one channel were  excluded from the final catalog. The total number of disk and tail regions fitted with this procedure is given in Tab.~\ref{tab:knots}.

In most cases a single Gaussian fit was able to describe the CO line emission arising from the star-forming knots in JO201, JO204, and JO206, while in JW100 we often found multiple peaks in the observed spectra (see e.g., Fig.~\ref{fig:multipeaks}), implying the presence of separated components along the line of sight simply superimposed in the integrated sky region.
When possible we therefore attributed to the star-forming region the Gaussian fit of the component that was coincident also with the \Ha emission, and we assigned them a flag 3.

When the Gaussian fit did not reproduce signal in the wings of the line, we preferred to use the integral of the positive flux within $\pm 3\sigma_{H_{\alpha}}$ as a proxy for the total amount of gas to be associated with the \Ha emission, irrespective of the Gaussian fit.
In the catalog given in Table~\ref{tab:catalog} we flagged these spectra as 0. Detections flagged as 2 were kept but not used in the following plots, as they represent detections with a lower significance (defined above). 
Table~\ref{tab:catalog} gives the characteristics of each star-forming region analyzed hereafter, i.e., the sky coordinates, the area in square kiloparsecs, the velocity of the CO(2-1) line relative to the galaxy redshift, and the corresponding velocity dispersion.
We then give the CO flux integrated using the Gaussian fitting and the integral of the positive emission with the relative error. We finally give the S/N of the CO(2-1) line and the visual flag explained above together with the final value of the flux used to extract the \hdue mass given in the last column.

\begin{figure*}
    \centering
    \includegraphics[width=0.225\linewidth]{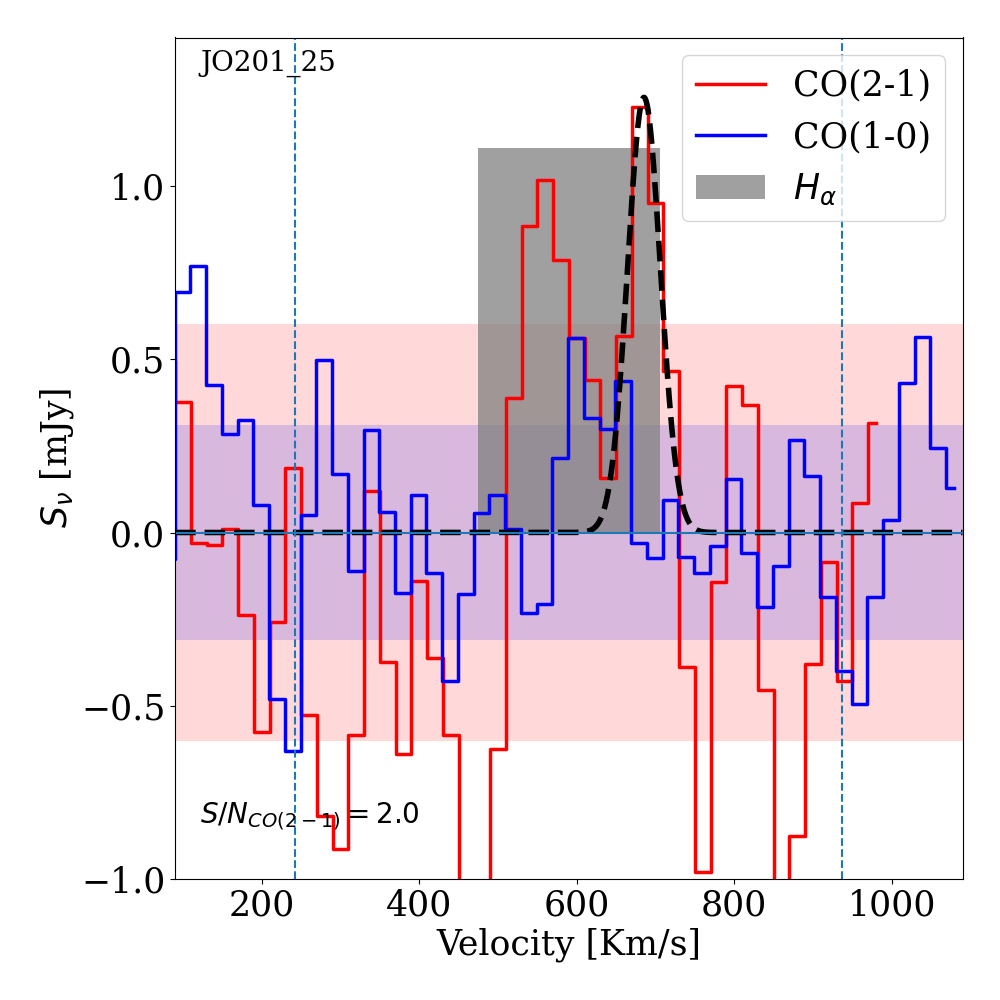}
    \includegraphics[width=0.225\linewidth]{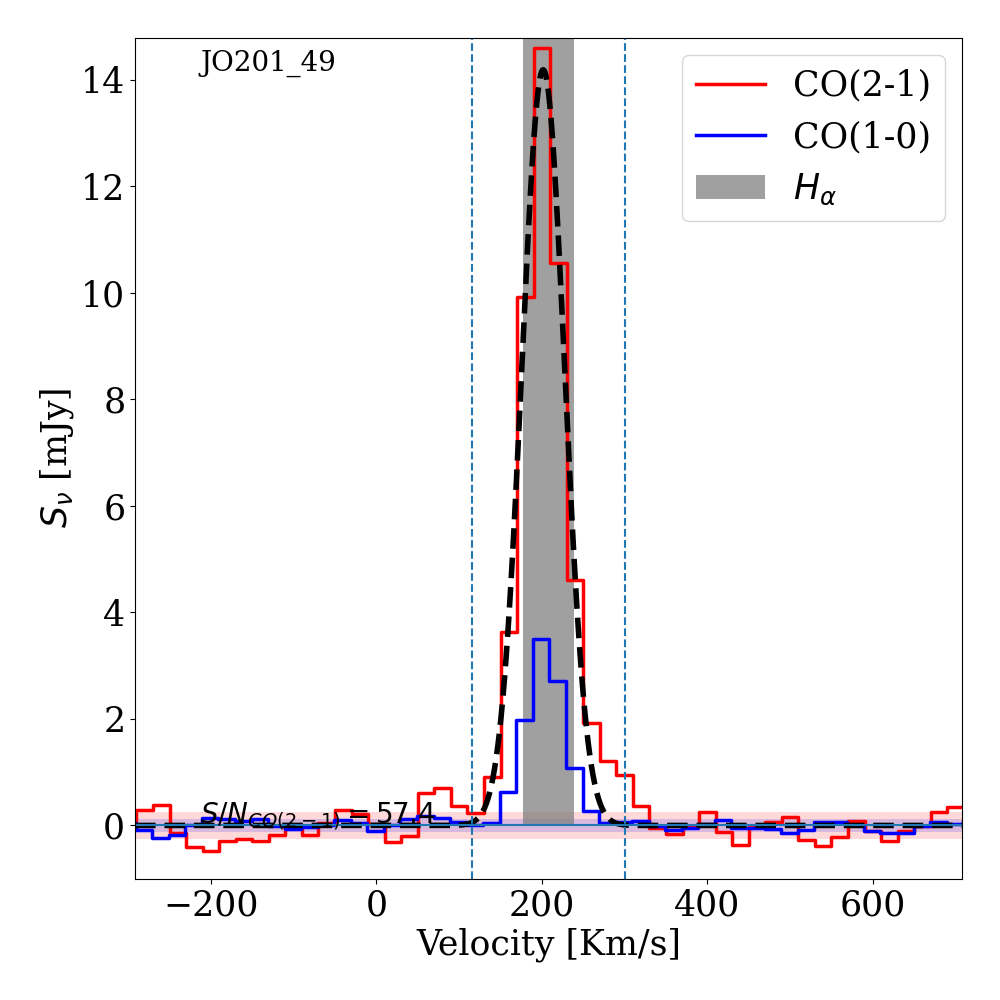}
    \includegraphics[width=0.225\linewidth]{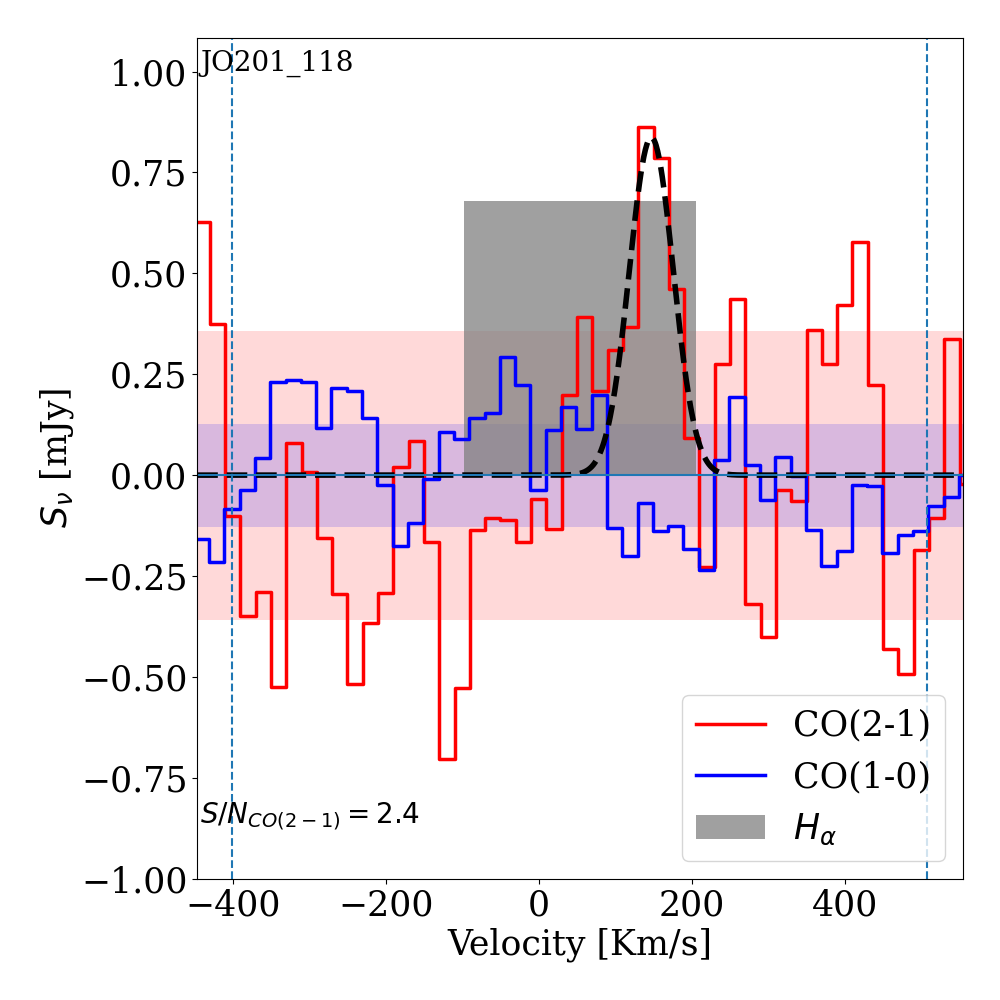}
    \includegraphics[width=0.225\linewidth]{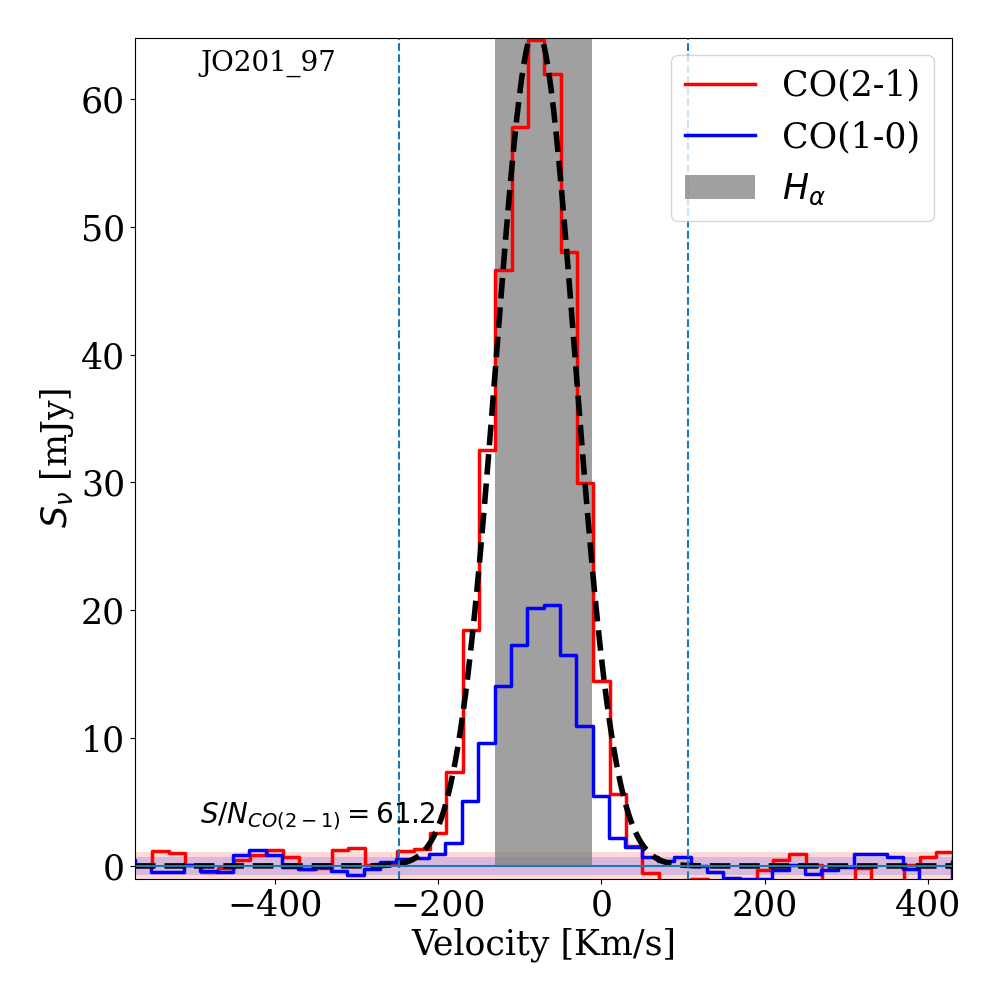}
    \includegraphics[width=0.225\linewidth]
    {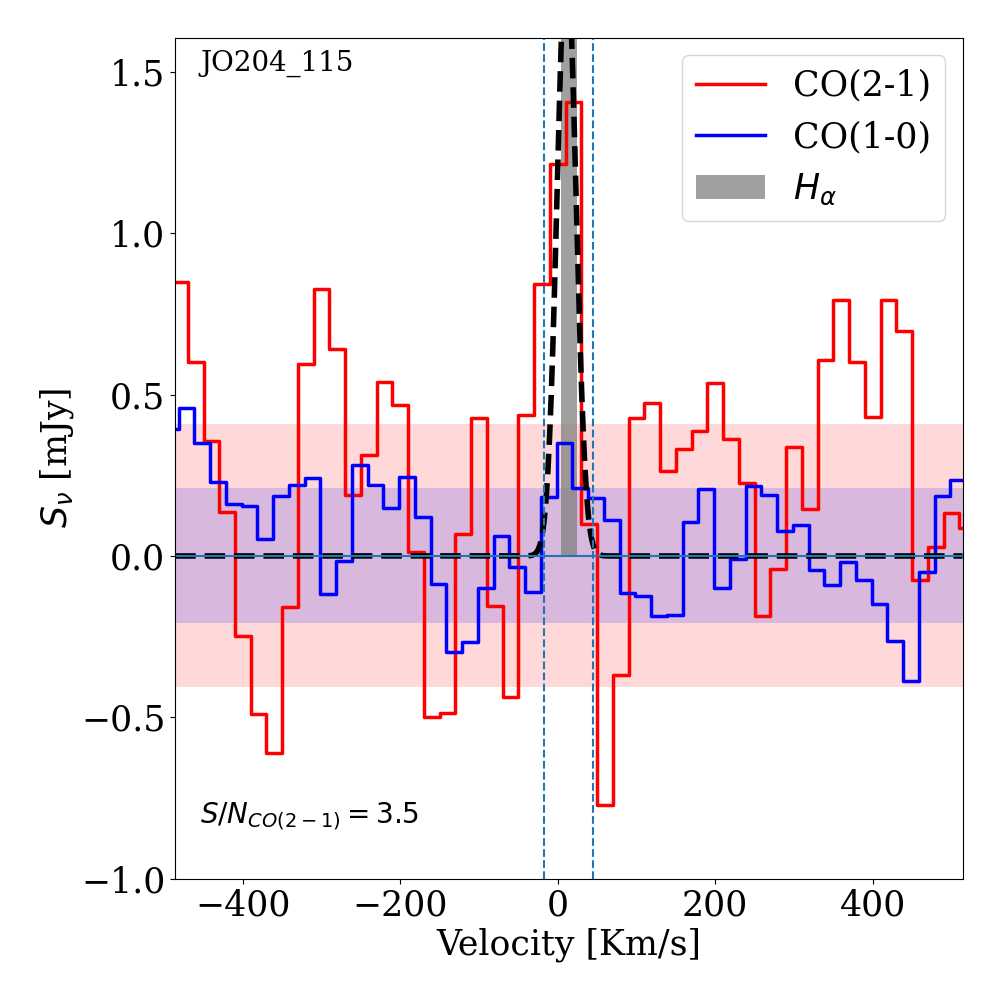}
    \includegraphics[width=0.225\linewidth]{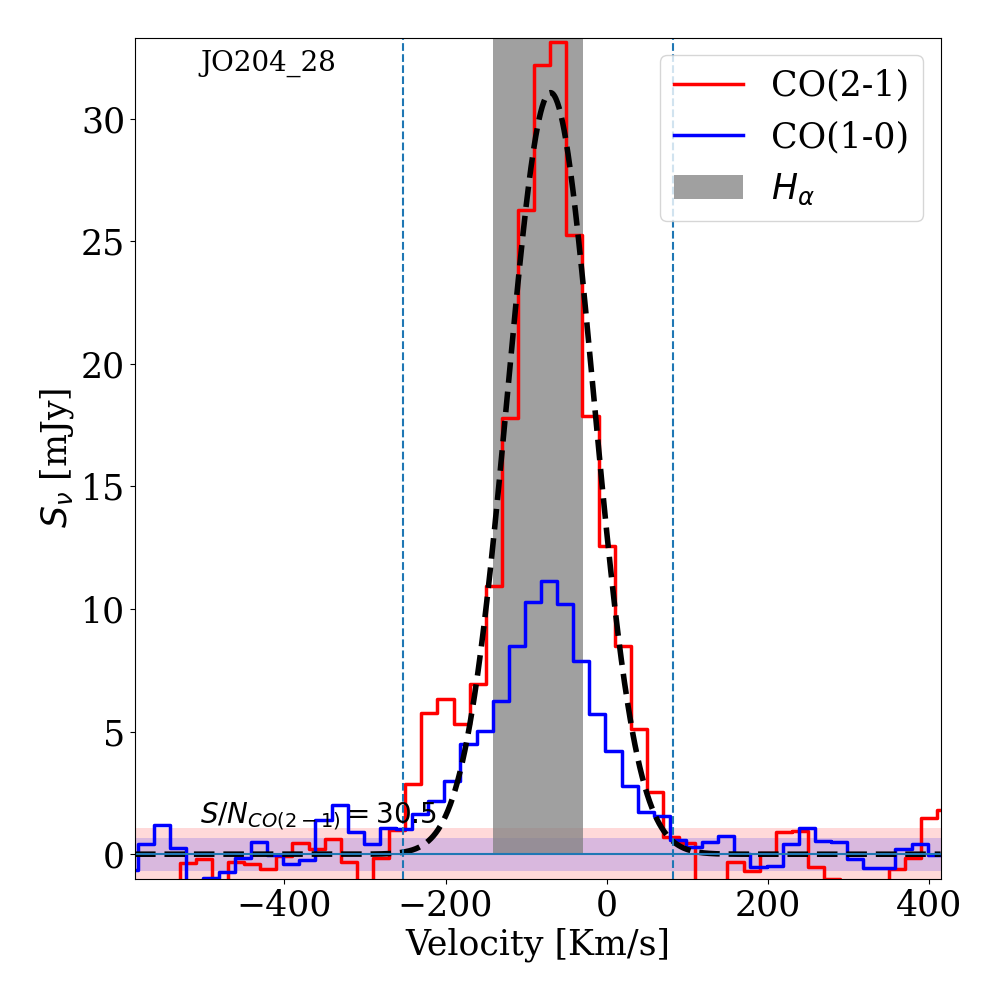}
    \includegraphics[width=0.225\linewidth]{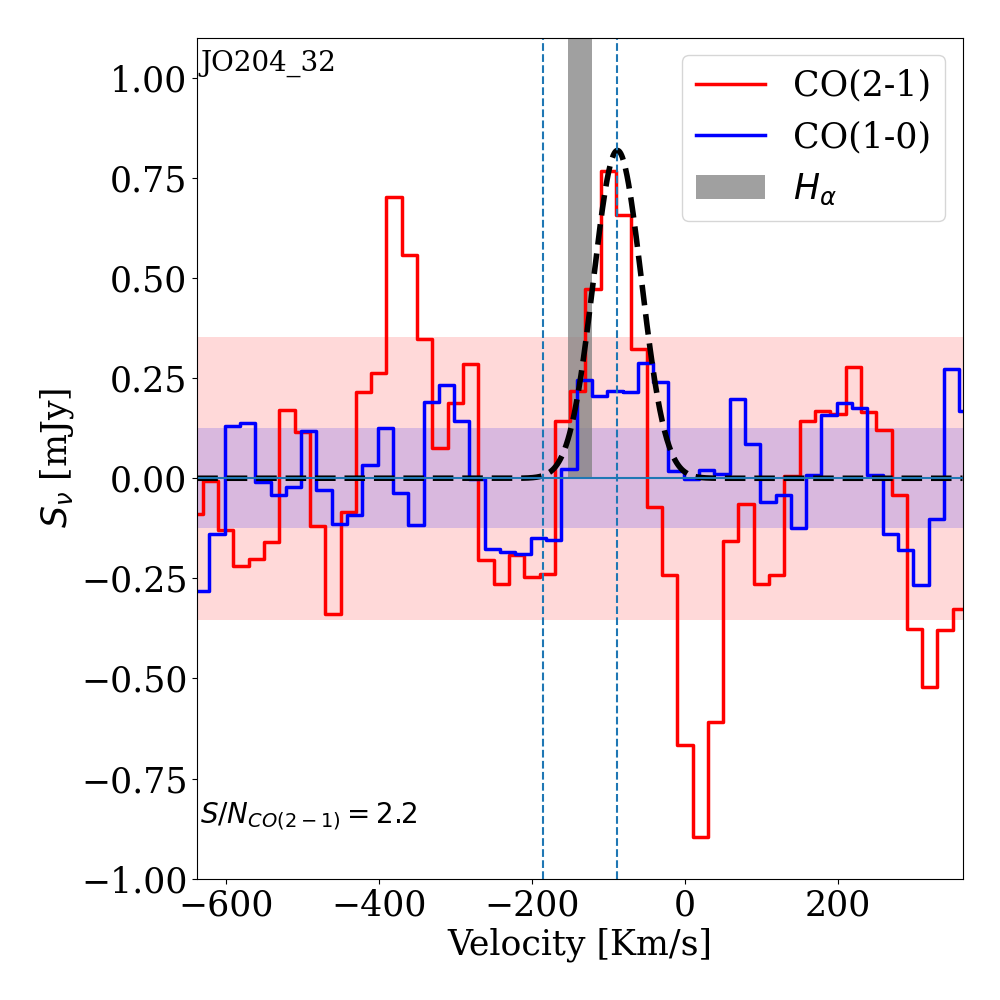}
    \includegraphics[width=0.225\linewidth]{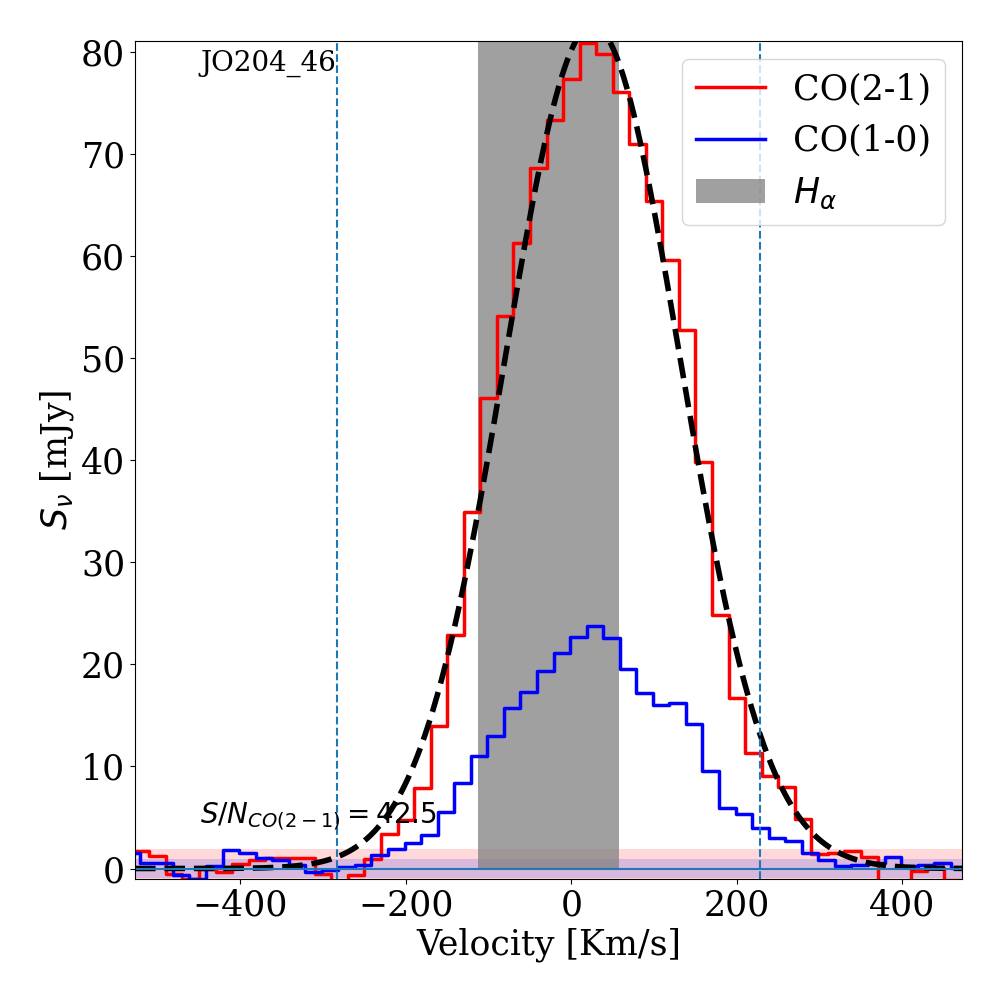}
    \includegraphics[width=0.225\linewidth]
    {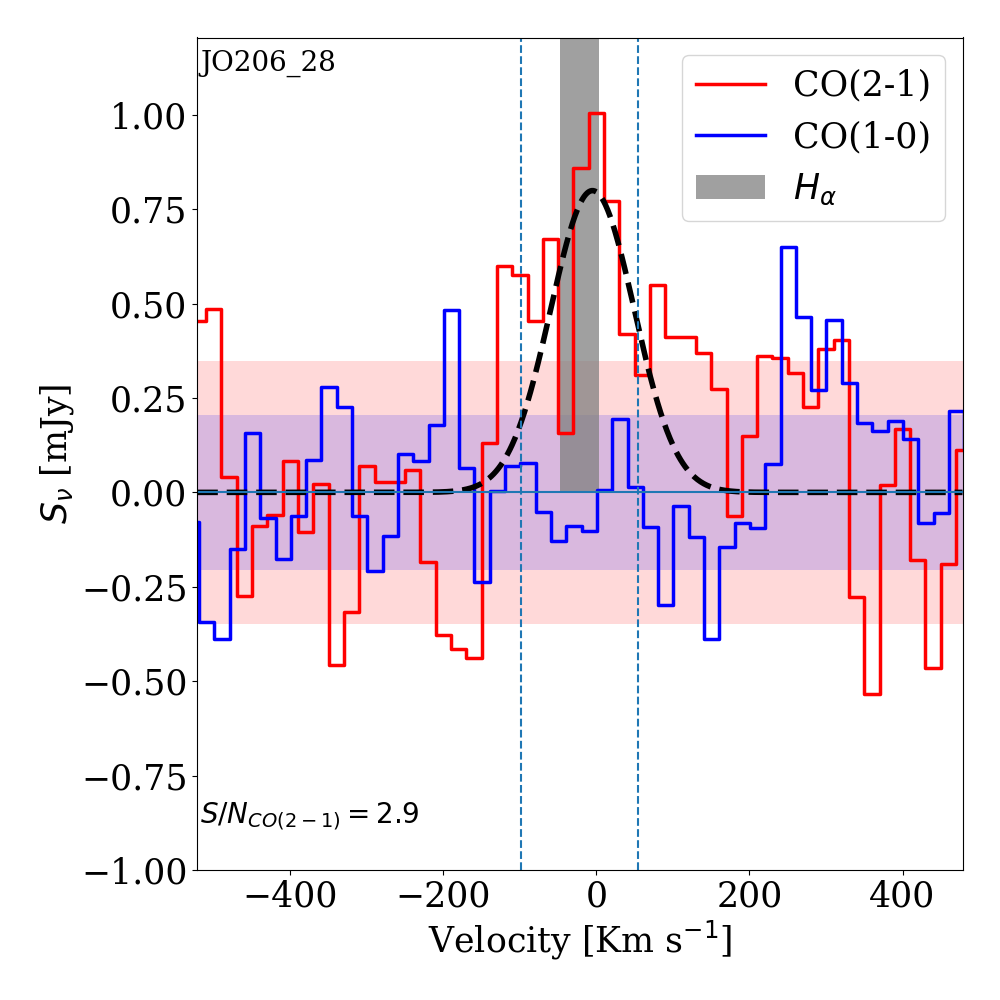}
    \includegraphics[width=0.225\linewidth]{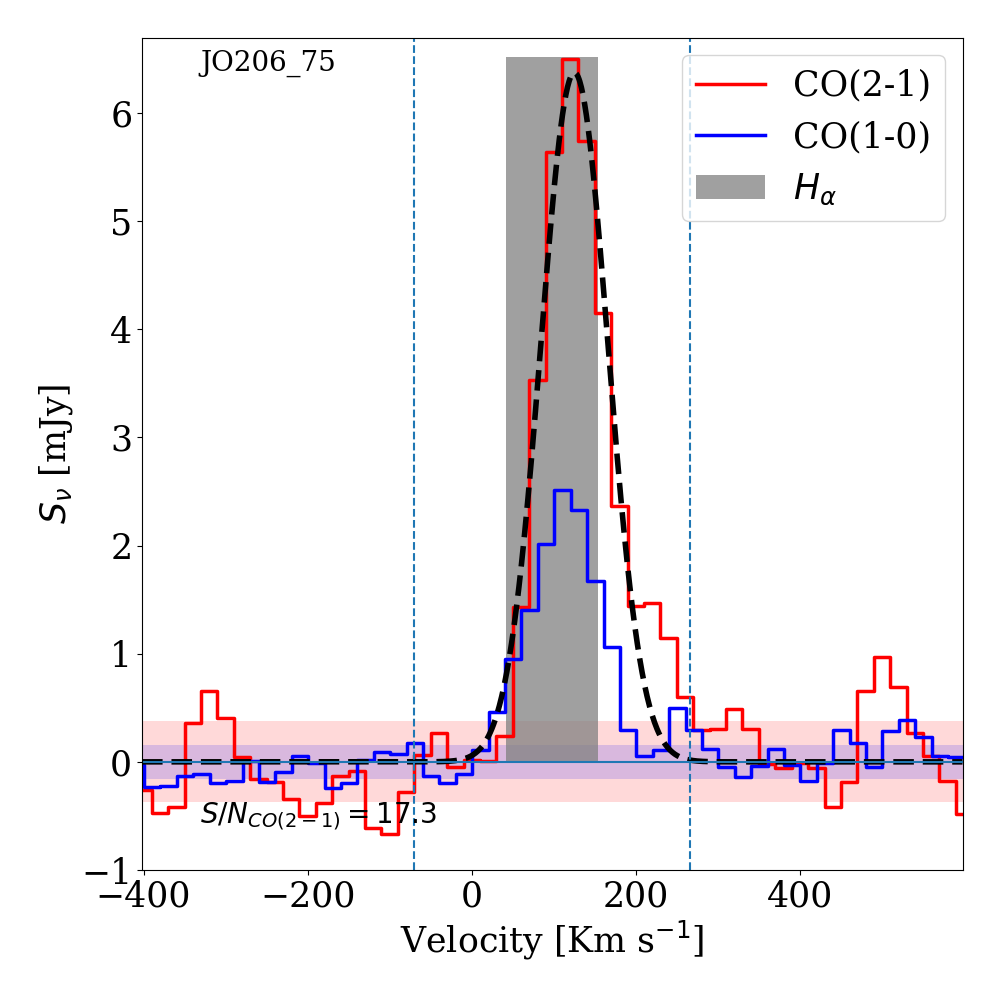}
    \includegraphics[width=0.225\linewidth]{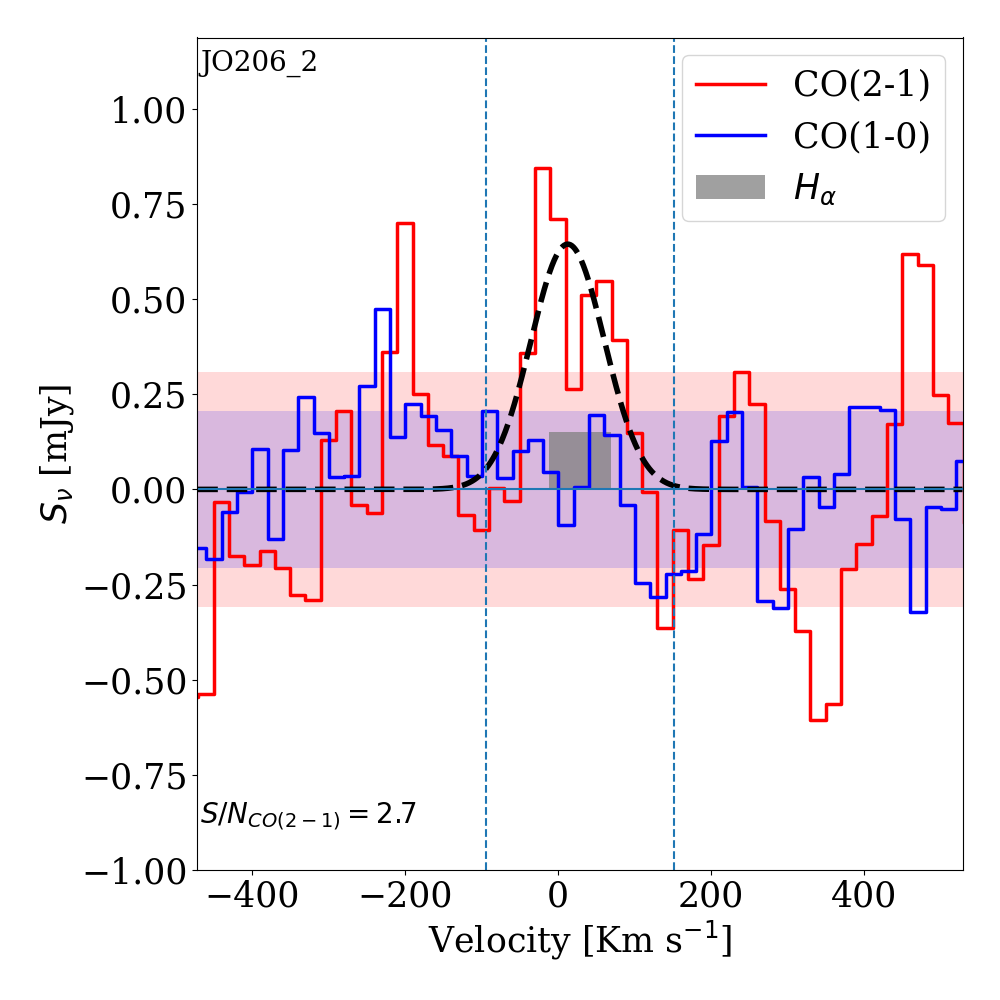}
    \includegraphics[width=0.225\linewidth]{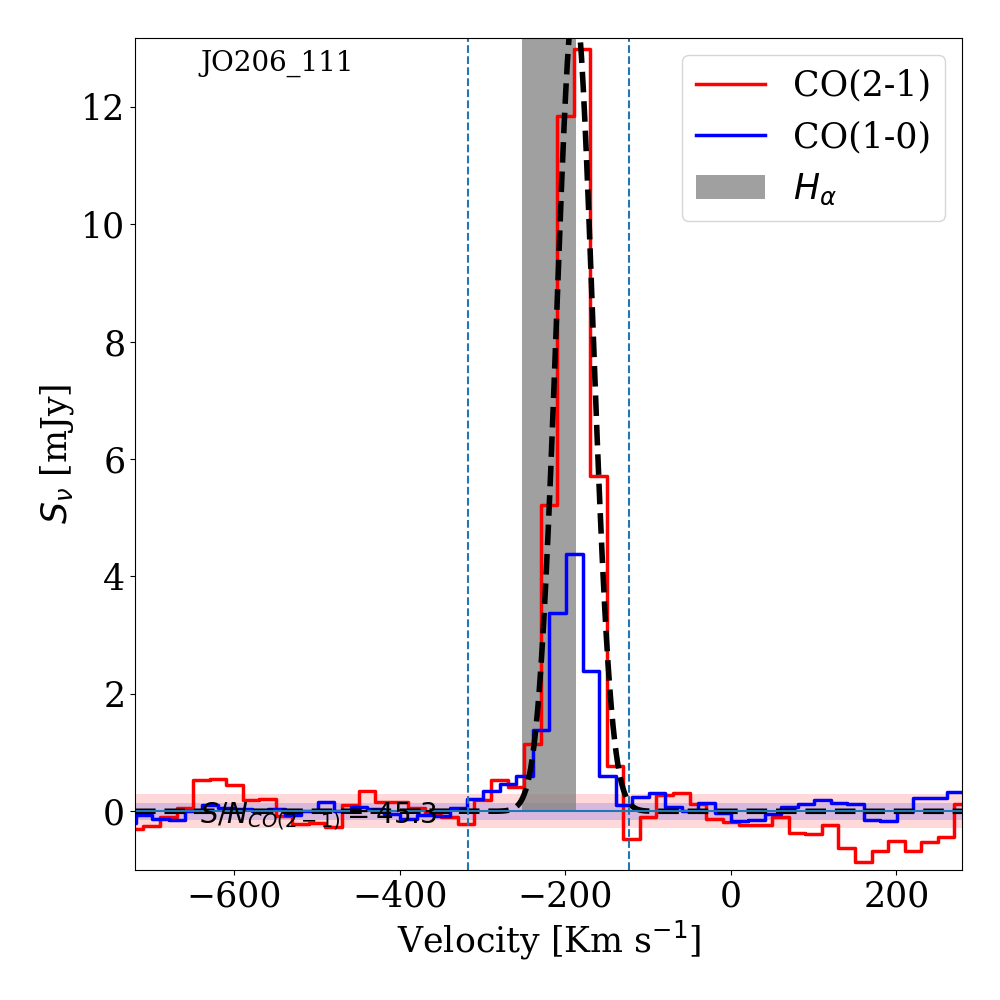}
    \includegraphics[width=0.225\linewidth]{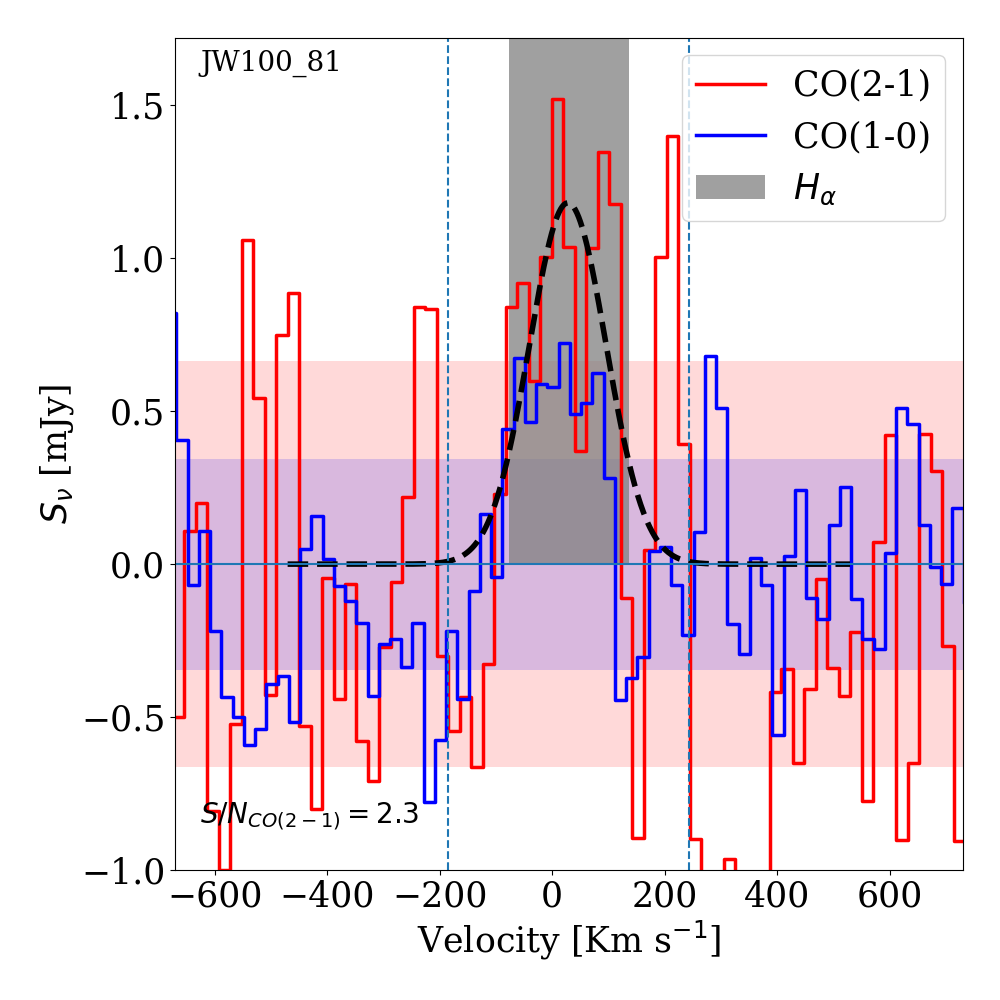}
    \includegraphics[width=0.225\linewidth]{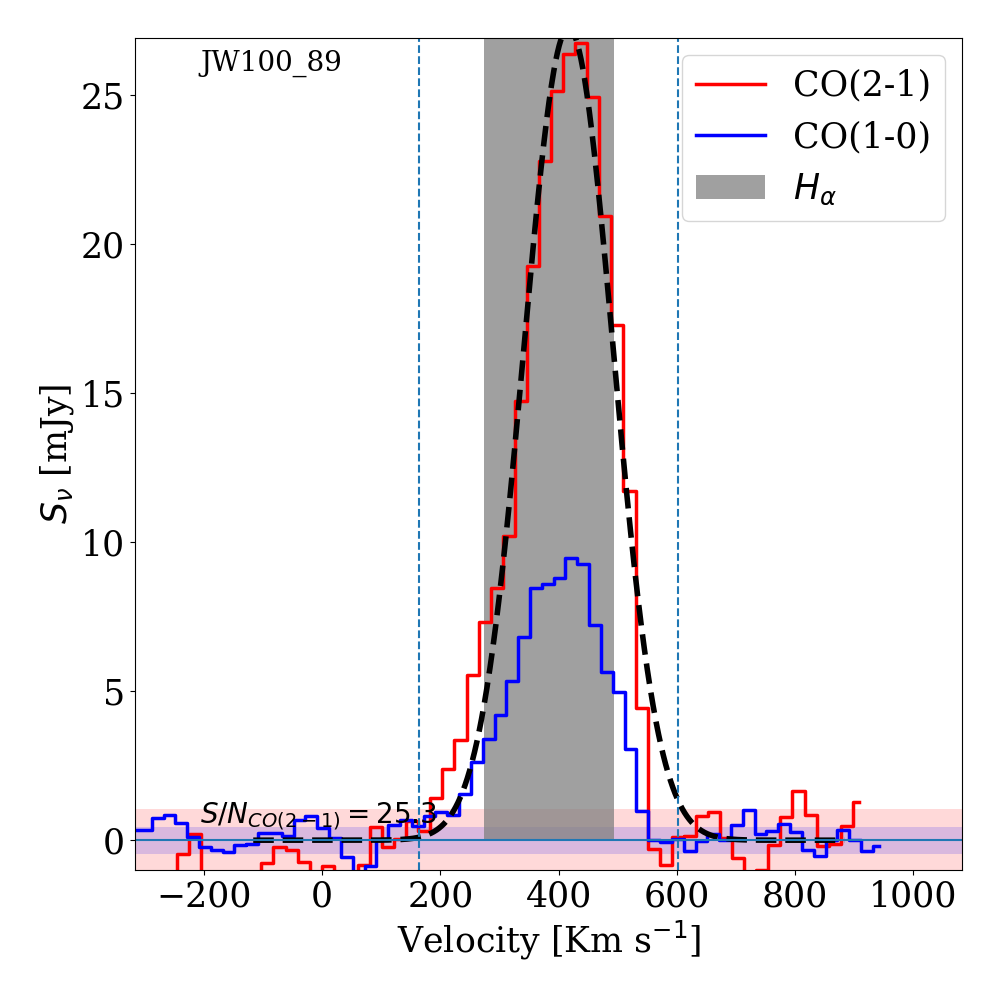}
    \includegraphics[width=0.225\linewidth]{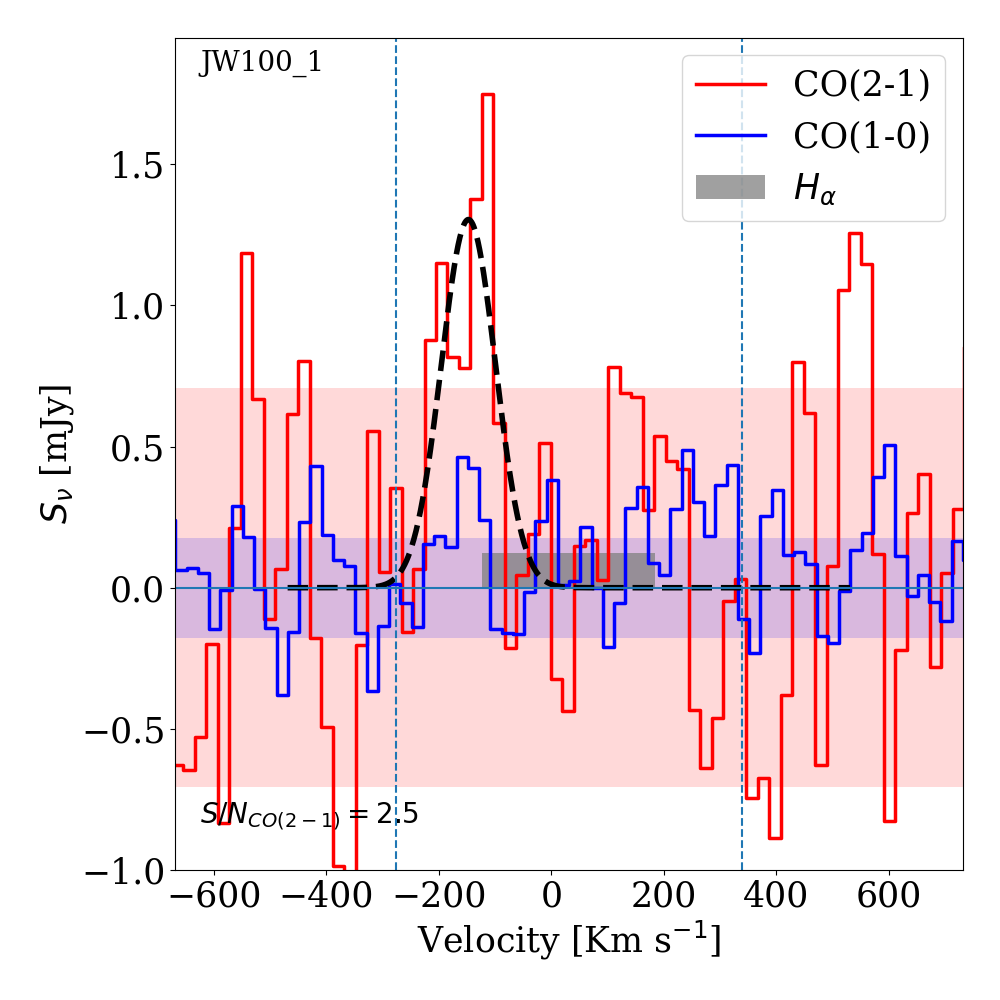}
    \includegraphics[width=0.225\linewidth]{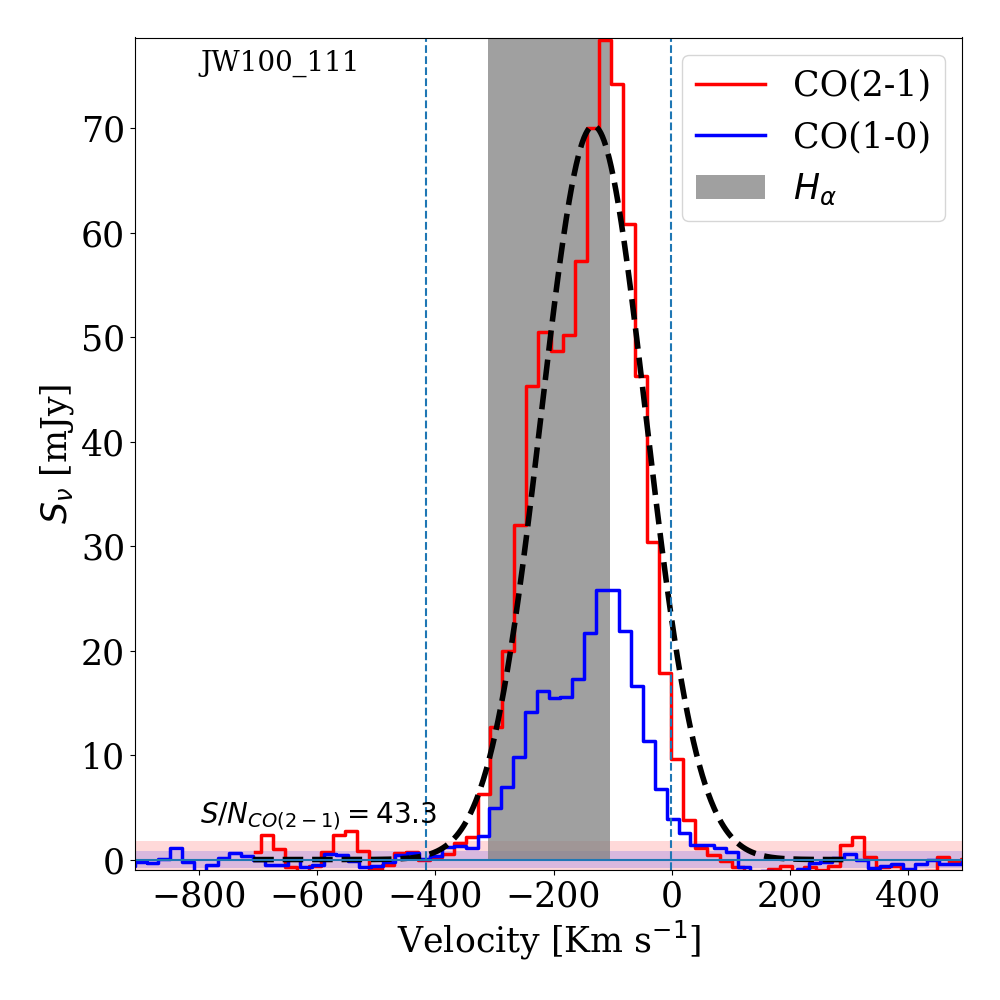}

    \caption{Gallery of spectra extracted from the ALMA cubes within the star-forming regions identified by the \Ha emission from MUSE. The top row refers to JO201 clumps, the second row to JO204, the third row to JO206, and the fourth row to JW100. Red lines show the CO(2-1) line emission, and the dashed black line its Gaussian fit, while blue lines show the CO(1-0) line emission. The gray box indicates the location of the \Ha emission and has a width corresponding to twice the \Ha velocity dispersion within the clump. Its height is proportional to the \Ha flux. The shaded horizontal regions mark the rms of the CO(2-1) and CO(1-0) emission, in red and blue, respectively. These spectra are representative of the lowest (first and third columns) and highest (second and fourth columns) -S/N spectra in the analyzed sample and have been flagged differently according to the preferred method to evaluate the total gas emission: the first two columns show spectra flagged as 0 (flux derived from the integral of the positive emission), while columns 3 and 4 are flagged as 3 (flux derived from a Gaussian fit).}
    \label{fig:spectra}
\end{figure*}

Figure~\ref{fig:spectra} shows a gallery of spectra extracted from the star-forming regions in the four galaxies (JO201, JO204, JO206, and JW100, from top to bottom) and with low (first and third columns) and high (second and fourth columns) S/N in the CO(2-1) line. The first two columns show spectra flagged as 0, while columns 3 and 4 are flagged as 3.

\begin{table}[]
    \centering
       \caption{Number of \Ha emitting regions in the four observed galaxies.}
    \begin{tabular}{|c|c|c|c|c|c|}
     \hline
    Galaxy    & N$_{tot}$ & N$_{disk}$ & N$_{disk,fit}$ & N$_{tail}$ & N$_{tail,fit}$\\
    \hline
     JO201    & 148 & 85   & 59 &  63  & 26 \\
     JO204    & 121 & 69   & 49 &  52  & 13 \\
     JO206    & 139 & 64   & 53 &  75  & 19 \\
     JW100    & 131 & 58   & 53 &  63  & 42 \\     
    \hline
    \end{tabular}
    \tablefoot{The total number is shown in col. 2, disk regions in col. 3, disk regions with a good fit of the CO(2-1) line in col. 4, tail regions in col. 5, tail regions with a good fit of the CO(2-1) line in col. 6 
    .}
    \label{tab:knots}
\end{table}

\begin{table*}[]
    \centering
    \small
    \caption{First ten entries of the clump catalog. }
\begin{tabular}{|l|l|l|l|l|l|l|l|l|l|r|l|l|}
\hline
  \multicolumn{1}{|c|}{ID} &
  \multicolumn{1}{c|}{RA} &
  \multicolumn{1}{c|}{DEC} &
  \multicolumn{1}{c|}{AREA} &
  \multicolumn{1}{c|}{v} &
  \multicolumn{1}{c|}{$\sigma_v$} &
  \multicolumn{1}{c|}{FLUX21g} &
  \multicolumn{1}{c|}{FLUX21} &
  \multicolumn{1}{c|}{$\sigma_f$} &
  \multicolumn{1}{c|}{S/N} &
  \multicolumn{1}{c|}{FLAG} &
  \multicolumn{1}{c|}{F21\_FIN} &
  \multicolumn{1}{c|}{M$_{H_2}$} \\
    \multicolumn{1}{|c|}{} &
  \multicolumn{1}{c|}{deg} &
  \multicolumn{1}{c|}{deg} &
  \multicolumn{1}{c|}{kpc$^2$} &
  \multicolumn{1}{c|}{\kms} &
  \multicolumn{1}{c|}{\kms} &
  \multicolumn{1}{c|}{Jy} &
  \multicolumn{1}{c|}{Jy} &
  \multicolumn{1}{c|}{Jy} &
  \multicolumn{1}{c|}{} &
  \multicolumn{1}{c|}{} &
  \multicolumn{1}{c|}{Jy} &
  \multicolumn{1}{c|}{10$^9$\Msun} \\
\hline
  JO201\_3 & 10.38908 & -9.27676 & 2.35 & 489.02 & 21.87 & 0.07 & 0.09 & 0.05 & 3.3 & 0 & 0.09 & 0.02\\
  JO201\_8 & 10.37775 & -9.26998 & 7.2 & 463.4 & 105.45 & 0.34 & 0.46 & 0.19 & 3.6 & 3 & 0.34 & 0.09\\
  JO201\_9 & 10.38085 & -9.26976 & 2.35 & 313.35 & 11.94 & 0.02 & 0.05 & 0.07 & 2.2 & 3 & 0.02 & 0.01\\
  JO201\_10 & 10.37764 & -9.26934 & 1.32 & 827.46 & 20.53 & 0.04 & 0.06 & 0.07 & 2.7 & 3 & 0.04 & 0.01\\
  JO201\_11 & 10.38142 & -9.26908 & 9.4 & 217.64 & 14.94 & 0.07 & 0.10 & 0.12 & 2.6 & 3 & 0.07 & 0.02\\
  JO201\_13 & 10.38014 & -9.26857 & 2.35 & 338.32 & 16.45 & 0.03 & 0.04 & 0.06 & 2.9 & 0 & 0.04 & 0.01\\
  JO201\_16 & 10.37542 & -9.26832 & 1.32 & 339.3 & 90.91 & 0.10 & 0.12 & 0.09 & 2.6 & 0 & 0.12 & 0.03\\
  JO201\_18 & 10.38191 & -9.26803 & 9.4 & 415.79 & 41.81 & 0.18 & 0.43 & 0.22 & 4.2 & 3 & 0.18 & 0.05\\
  JO201\_20 & 10.37488 & -9.26772 & 2.35 & 363.92 & 19.19 & 0.05 & 0.10 & 0.10 & 3.7 & 0 & 0.1 & 0.02\\
  JO201\_22 & 10.38895 & -9.26759 & 14.69 & 290.82 & 35.68 & 0.33 & 0.35 & 0.16 & 4.1 & 3 & 0.33 & 0.08\\
\hline\end{tabular}
    \caption{For each clump we give: the ID of the clump (col. 1), its coordinates (col. 2 and 3), its size in square kiloparsecs as derived from the MUSE data analysis (col. 4), the line-of-sight velocity with respect to the galaxy center (col. 5), the velocity dispersion of the line from the Gaussian fit (col. 6), the integrated line intensity in Jy both derived using a Gaussian fit (col. 7) and integrating all the positive contribution within $\pm 3$ \Ha velocity dispersion (col. 8), the error on the line flux (col. 9), the S/N of the CO(2-1) line (col. 10), the visual flag used to classify clumps (col. 11), the final integrated line flux (col. 12), and the corresponding \hdue mass in 10$^9$ \Msun (col. 13). }
    \label{tab:catalog}
\end{table*}

\begin{figure*}
    \centering
    \includegraphics[width=0.45\textwidth]{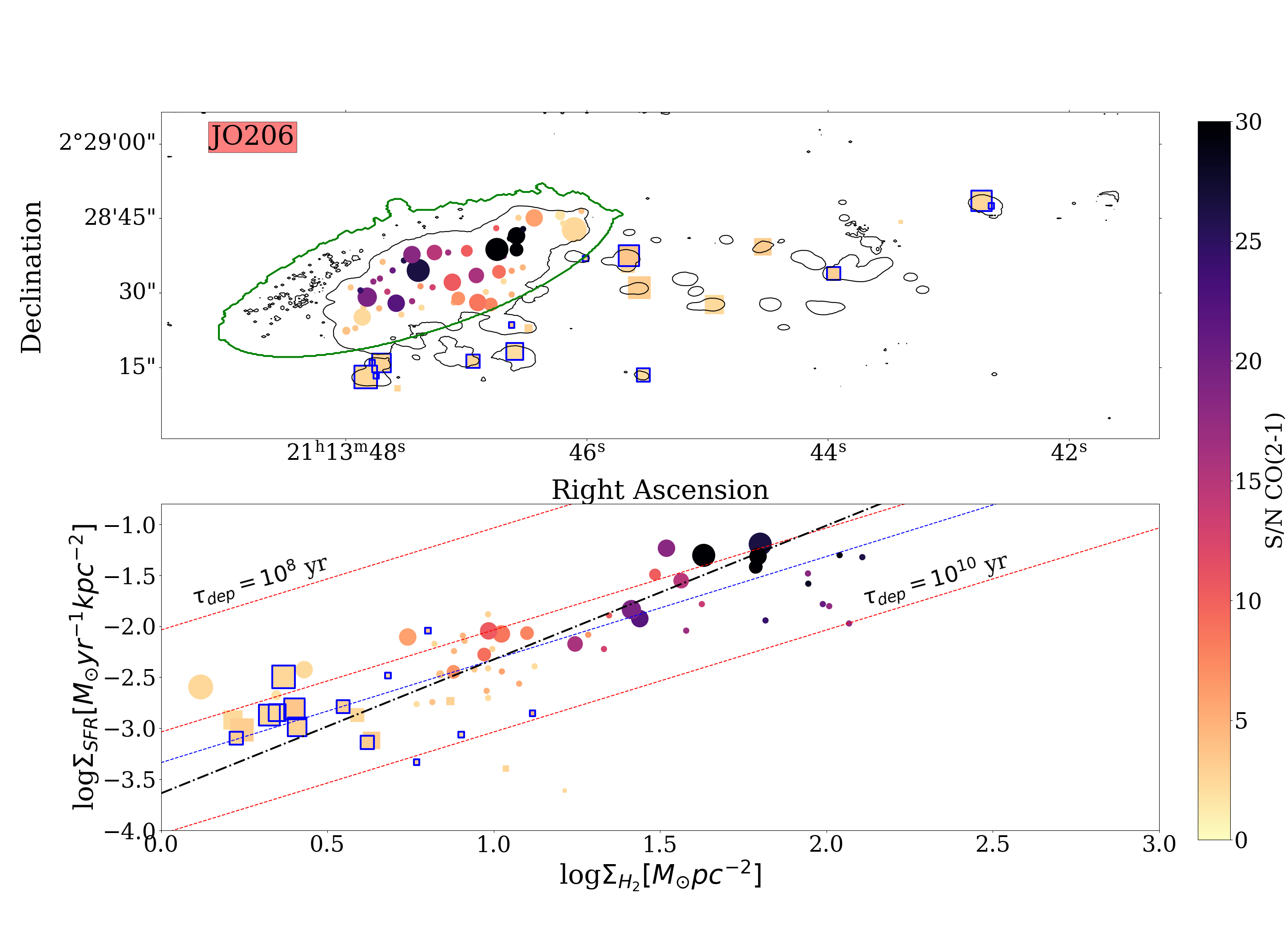}
    \includegraphics[width=0.45\textwidth]{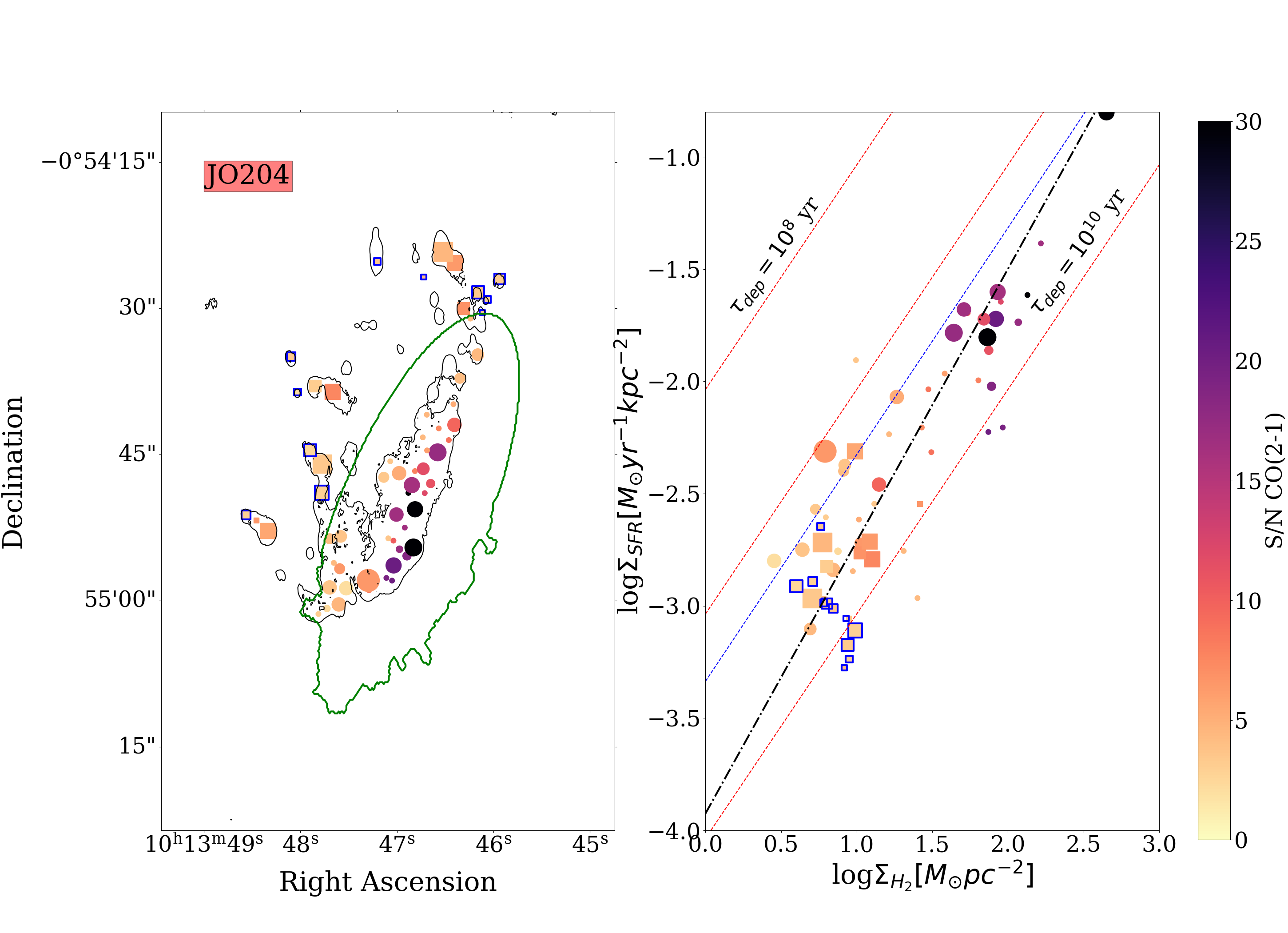}
    \includegraphics[width=0.45\textwidth]{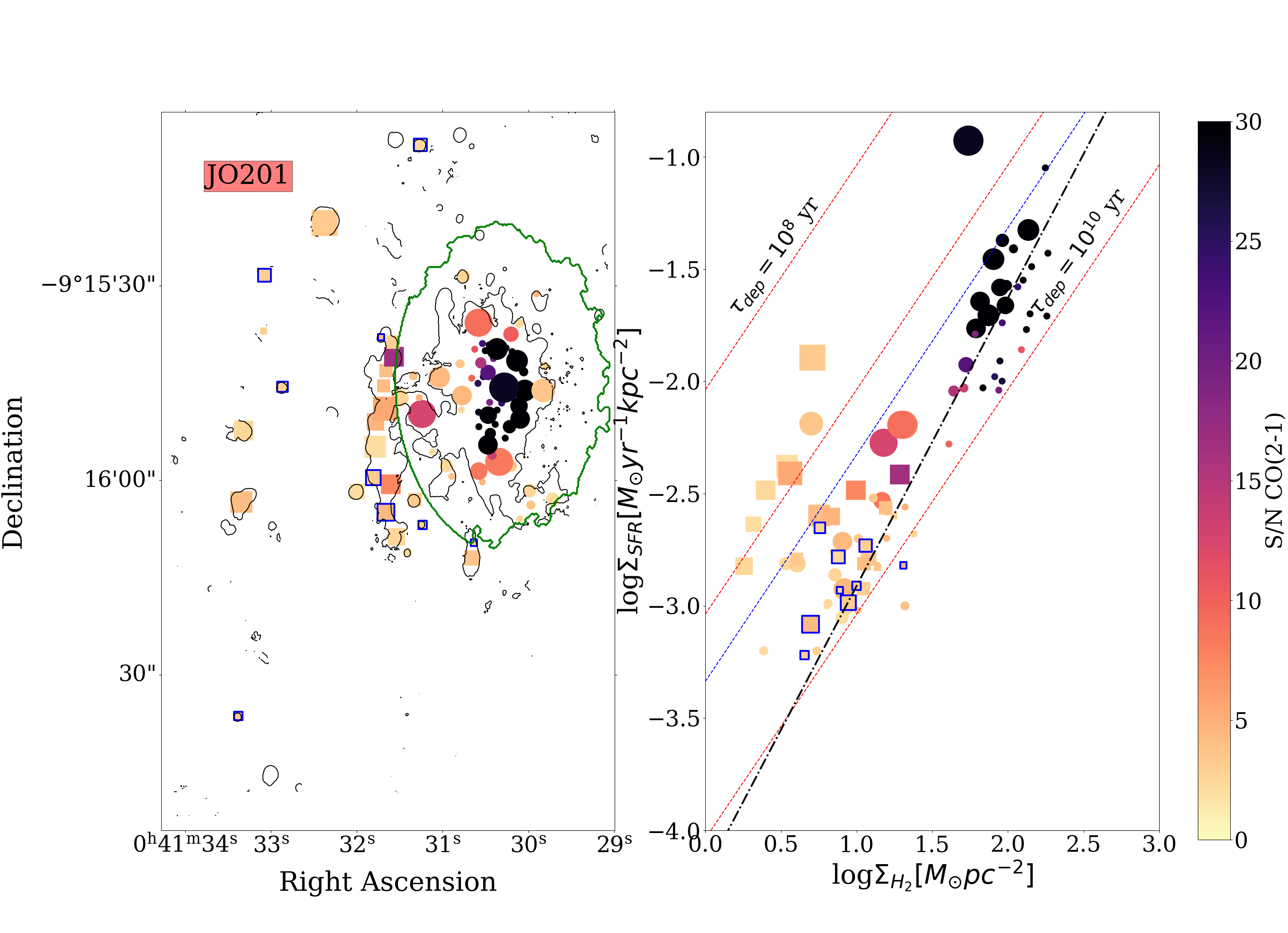}
    \includegraphics[width=0.45\textwidth]{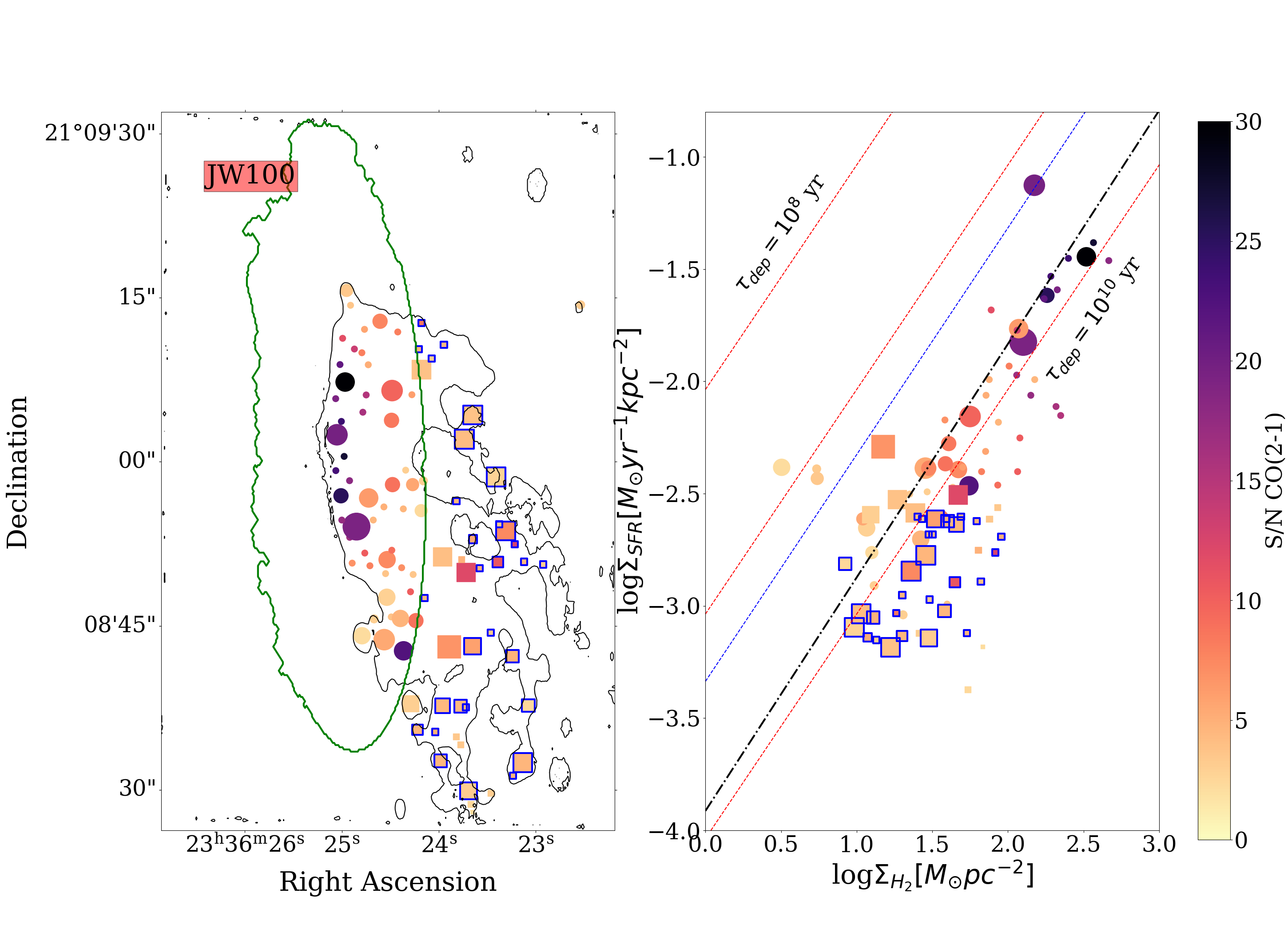}
    \caption{Maps of the \Ha emitting knots and corresponding KS relation in JO206, JO204, JO201, and JW100 from top to bottom, with different symbols for knots in the disk (circles) and in the tail (squares). Blue contours indicate knots dominated by the DIG. The size of each symbol is proportional to the size of the emitting region, and it is color-coded according to the S/N in the CO(2-1) line. The dashed red lines show lines of a constant depletion time (from $10^8$ to $10^{10}$ yr), while the dashed blue shows the relation at a 1 kpc scale from \citealt{Bigiel2011}. The dash-dotted black lines show the relation found analyzing independent beams, the slopes and intercepts of which are given in Table
    ~\ref{tab:ksfit}.}
    \label{fig:ks_knots}
\end{figure*}

For each galaxy, we show in Fig.~\ref{fig:ks_knots} the distribution of the star-forming knots in the sky and the corresponding KS relation.
In both panels star-forming regions in the disk are identified with circles, while squares represent star-forming regions in the tail.
The color coding follows the S/N of the CO(2-1) line emission, where dark magenta regions have the highest S/N and the yellow ones show a lower S/N (albeit always larger than 2).
The size of each symbol is proportional to the size of the star-forming region as derived from MUSE data and corresponds to the spatial extension of the ALMA cube integration.
Finally, whenever a symbol has a blue contour, it means that the region is dominated by the DIG emission (DIG>70\%). In doing this, we used the DIG estimation from \citealt{GASPXXXII}.
The dash-dotted black line shows the relation found when analyzing independent beams (see Table~\ref{tab:ksfit})

In all galaxies star-forming regions located in the stripped gas tails have a lower surface mass density of CO when compared with those in the disk.
They also have on average a lower S/N, as well as a strong contribution from the DIG.

The analysis of their \hdue content confirms that JW100 is characterized by a low overall efficiency, both within and outside the disk. 
Its star-forming knots in the tail, though, are generally bigger than those found in the other galaxies. This evidence, coupled with the presence of double peaks in the extracted molecular gas emission, as shown in Fig.~\ref{fig:multipeaks}, suggests that more knots are superimposed along the line of sight, as expected given that JW100 is almost  edge-on, and therefore its stripped tail can be formed by different filaments superimposed in space but departing from the galaxy disk at different velocities.
\begin{figure}
    \centering
    \includegraphics[width=0.85\linewidth]{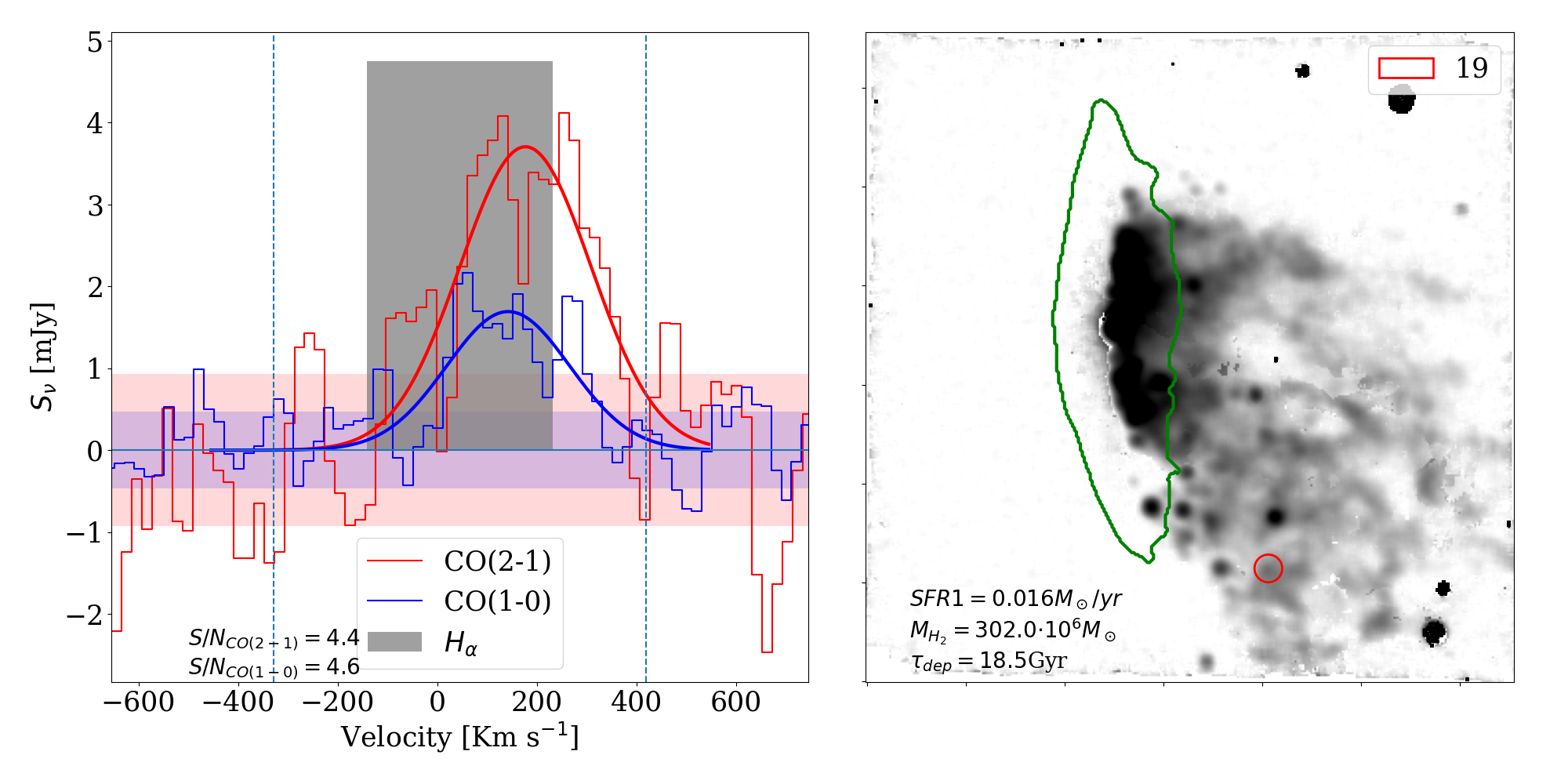}
    \caption{Evidence of multiple peaks in the CO(2-1) spectrum extracted from the region 19 in JW100, located in the stripped gas tail (as seen from the rightmost panel). The meaning of the lines in the left panel is the same as in Fig.~\ref{fig:spectra}.}
    \label{fig:multipeaks}
\end{figure}

The remaining galaxies host star-forming knots that are normally efficient at converting gas into stars (i.e., they follow the KS relation of normally star-forming disks), and this is true also in the stripped gas tail, as the spectacular tail of JO201 demonstrates (where they appear even to be super-efficient).
In JO201 and JO206 the star-forming regions in the tails have a lower surface mass density of molecular gas but a relatively high SFR, so that they lie often above the KS relation derived in normally star-forming galaxies.
The clumps dominated by the DIG, present both in the disk and in the tail, instead show a generally lower SFE, as expected.

These findings indicate that the star formation process, despite happening in an hostile environment that produces low-mass-density regions, is universal. 
This also implies, interestingly, that within the identified knots the cold gas is efficiently converted into stars even when clumps are not embedded in the stellar disks. 

In nearby galaxies, the ability of clumps with typical sizes of $\sim 100$ pc to remain gravitationally bound has been verified using the virial parameter $\alpha_{vir}$ \citep{Sun+2020b}, which is  the ratio between the kinetic and the gravitational energy of a clump. When $\alpha_{vir}\sim1$ the gravitational energy and the kinetic energy of the clump are comparable, while $\alpha_{vir}\gg1$ indicates that the gravitational energy is lower than the kinetic one. In both cases, though, the clumps may be in a steady state, provided that the external pressure is balanced by the internal one. In this case, the clumps are confined by the external pressure, which dominates over the self-gravity \citep{Keto+1986,Elmegreen1989,Bertoldi+1992}. 
The virial parameter was also estimated for the nearby jellyfish galaxy ESO137-001, for which 350 pc resolution ALMA data exist \citep{Jachym+2019}, where the measurement of $\alpha_{vir}\gg1$ has been interpreted as a probe that the emitting regions will probably dissolve with time.

Motivated by the fact that star formation is present and efficient both in the disk and in the tail clumps of our ram-pressure-stripped galaxies, we decided to estimate the virial parameter of our $\sim 1$ kpc clumps, even though they likely host a complex, unresolved substructure of clouds, clumps, and cores with nonuniform density profiles. The virial parameter derived in this way can be interpreted as a diagnostic tool to assess whether the observed kiloparsec-scale gas structures can be compatible with star formation, under reasonable assumptions about their internal structure, while a direct comparison with high-spatial-resolution clumps is not feasible.

Following the original formulation in \citet{Bertoldi+1992}, we calculated the virial parameter for our 1 kpc resolution clumps as
\begin{equation}\label{eqn:alfavir}
    \alpha_{vir}=\frac{5\sigma^2 R}{(fGM)}
\end{equation}
using the sizes of the \Ha regions from MUSE (R) and the molecular gas velocity dispersion ($\sigma$) and \hdue mass (M). 
The factor, \textit{f}, includes the effect of the radial density distribution within the cloud and the ellipticity of the cloud itself.
In particular, neglecting the ellipticity dependence, $ f=(1-\gamma/3)/(1-2\gamma/5)$, where $\gamma$ is the power-law exponent of the density profile, i.e., $\rho(r) \propto r^{-\gamma}$.
If we assume a constant density within the clouds ($\gamma=0$), then f=1 and the $\alpha_{vir}$ that we derive from our low-resolution data is much larger (by about 1 order of magnitude) than the one calculated on cloud scales.
If we assume, instead, a steeper distribution (with $\gamma=2.3$, i.e., $f=3$), corresponding to a more centrally concentrated gas distribution, we can reach more standard values for the virial parameter, which will range between 0 and $\sim$100.

 \begin{figure}
    \centering
    \includegraphics[width=0.45\textwidth]{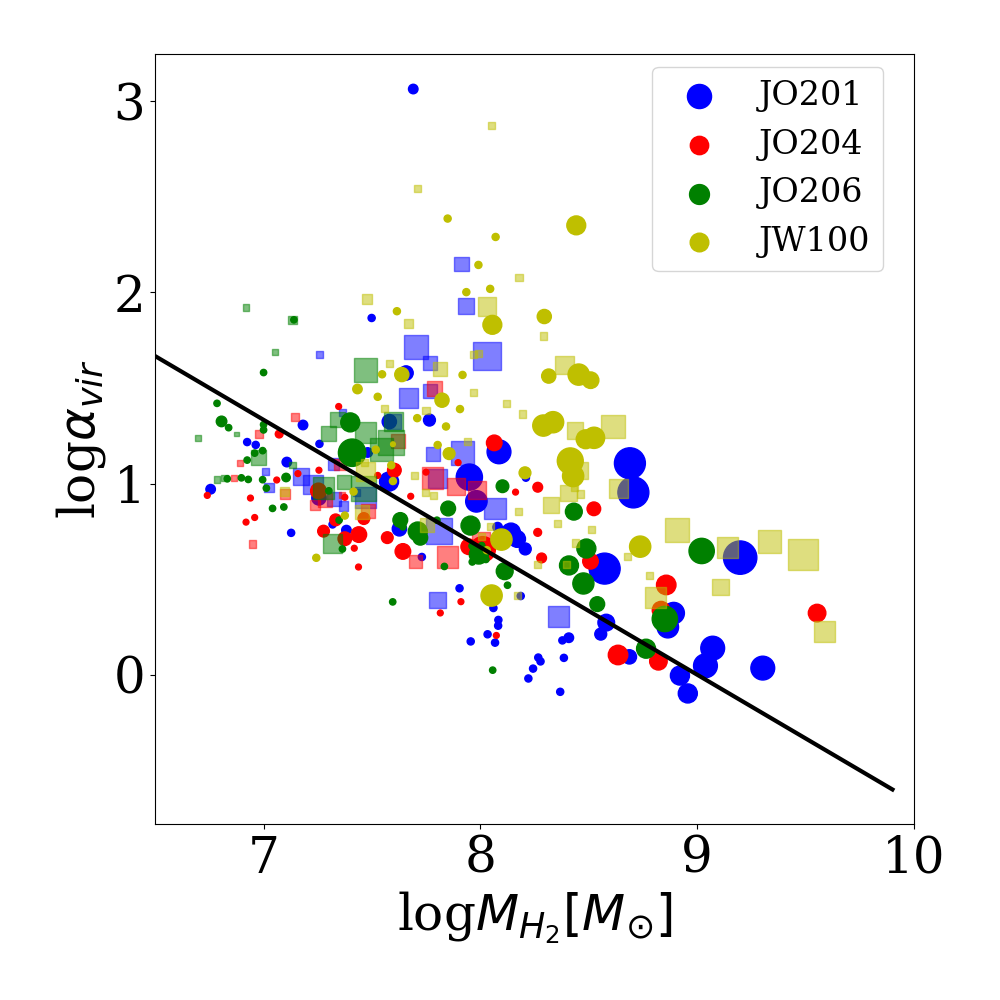}
    \caption{Virial parameter of the star-forming regions against their \hdue mass derived from the gas traced by the CO(2-1) line emission for the four galaxies. Circles refer to disk clumps, squares to tail clumps. The black line shows the theoretical relation for pressure-confined clumps, arbitrarily scaled. Sizes are proportional to the clump sizes. The virial parameter assumes $f=3$.}
    \label{fig:alfavir}
\end{figure}

Figure~\ref{fig:alfavir} shows the virial parameter that we estimated using $\gamma=2.3$, i.e., $f=3$ in Eqn.~\ref{eqn:alfavir} as a function of the \hdue mass in our star-forming complexes, with different colors referring to the different galaxies.
Disk clumps (circles in Fig.~\ref{fig:alfavir}) show on average smaller $\alpha_{vir}$ than their tail counterparts (squares), but always larger than the value of 1 that is typically interpreted as being due to star-forming clumps in gravitational equilibrium.
Interestingly, disk clumps, and to some extent also the ones in the tails, show a correlation between the virial parameter and the mass that is compatible with the theoretical one derived for pressure-confined clumps \citep{Bertoldi+1992}, i.e., $\alpha_{vir} \propto M^{-2/3}$ that is shown as a continuous black line in the figure.
In other words, clumps might be in equilibrium, and therefore survive, even though at face value the measured $\alpha_{vir}$ values are high.
Larger and massive clumps, especially in the disk, show a larger displacement with respect to this relation, possibly due to an overestimation of the clump size due to the underlying disk emission.
Our interpretation is that within our identified clumps the gas density profile is not constant, but follows a steep profile. Most probably, each region hosts one (or more than one) embedded clump that is star-forming as confined by the pressure of the surrounding medium.

\section{Overcoming the limited spatial resolution}\label{sec:hst}

The results presented so far make use of the best observational sample available to study the star-forming properties of clumps born in the gaseous tail produced by the RPS, and therefore represent a valuable tool with which to understand how the star formation proceeds in regions that are not the usual galaxy disks.
The major drawback of the analyzed sample is certainly the spatial resolution, which is one order of magnitude larger than the one attainable for nearby galaxy disks that can actually resolve the star-forming complexes.

\begin{figure*}
    \centering
    \includegraphics[width=\textwidth]{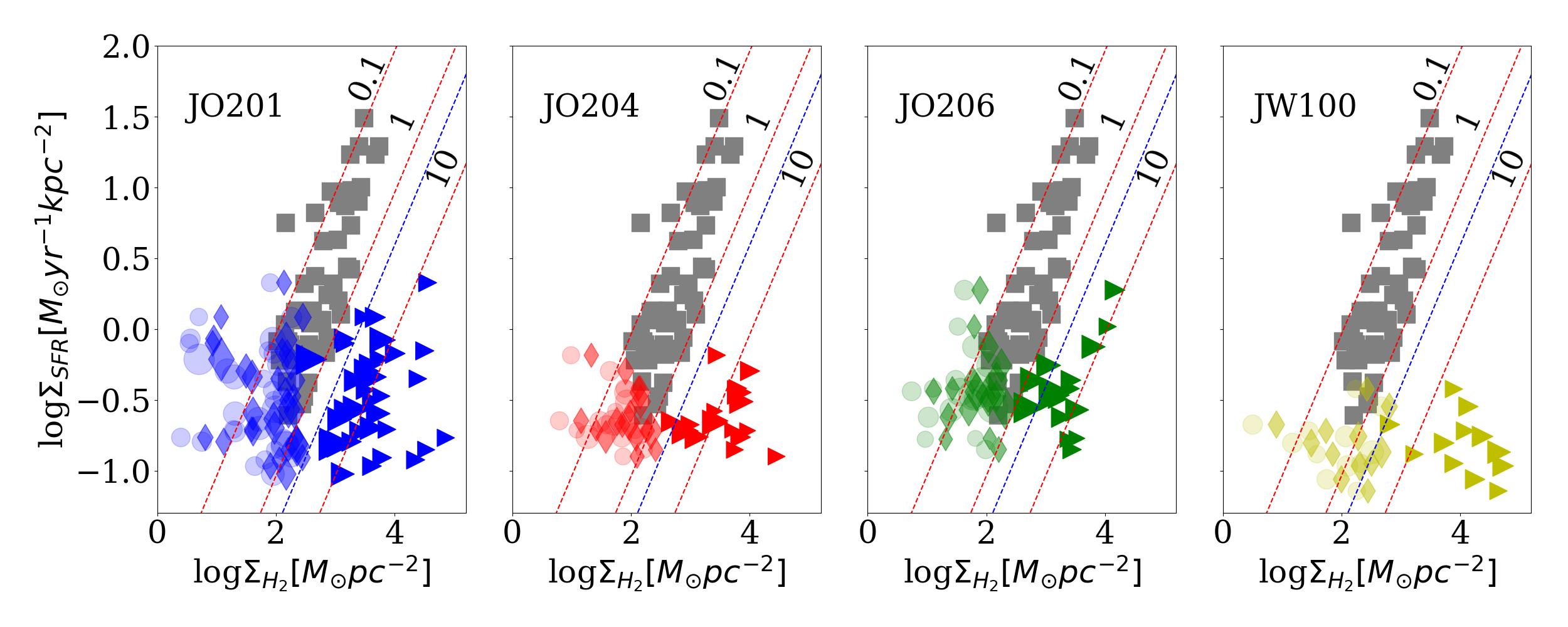}
    \caption{Star formation rate density of HST star-forming and resolved (disk and tail) clumps \citep{Giunchi+2023} as a function of the CO(2-1) surface mass density derived from the ALMA data and converted to small scales assuming that the CO is distributed: (a) over the entire MUSE clump (transparent circles), (b) as in local galaxies using the 1 kpc-150pc scaling relation \citep[filled diamonds;][]{Sun+2022}, and (c) over the HST clump (filled triangles), as detailed in the text. The four panels from left to right refer to the JO201 (blue), JO204 (red), JO206 (green), and JW100 (yellow) clumps. 
    The size of symbols is proportional to the HST clump size. Gray squares refer to the disks of starburst galaxies \citep{Kennicutt+2021}. Dashed red lines show constant depletion times from 0.1 to 10 gigayears, while the blue one refers to the relation at 1 kpc scale \citep{Bigiel2011}.}
    \label{fig:ks_hst}
\end{figure*}

For the four galaxies analyzed here, though, we possess HST photometry \citep{Gullieuszik+2023,Giunchi+2023} with both the broadband optical and the narrow-band \Ha filters that has confirmed the presence of star-forming clumps in the whole observed field encompassing the disk, the extra-planar and the tail regions around each galaxy.
About 15\% of these clumps are spatially resolved and have size measurements down to a scale of 140 pc (the HST resolution limit at the average distance of our galaxies).
The \Ha luminosity of each clump was measured using the narrow-band F680N filter, and the contribution from the [NII] lines falling into the observed wavelength range was taken into account with the procedure described in \citealt{Giunchi+2023}, while we could not establish at these scales the amount of dust attenuation.
Any correction for internal extinction would only make the \Ha fluxes (and therefore SFRs) larger.
By comparing the clumps' \Ha luminosity with the resolved sizes, \citealt{Giunchi+2023} have shown that the HST clumps generally possess a higher \Ha luminosity for a fixed size with respect to their counterparts in undisturbed star-forming galaxies \citep{Wisnioski+2012}. Their position in the luminosity-size plane is closer, therefore, to the region occupied by their counterparts in starburst galaxies \citep{Fisher+2017}.
The measured \Ha luminosities, corrected for the [NII] contribution, have been then converted in SFRs as we did in GASP \citep{GASPI}, and are shown in Fig.~\ref{fig:ks_hst} as transparent circles.

In an attempt to overcome the limited spatial resolution of ALMA data, we propose in this section two different approaches: the first one consists of using the scaling relation between 150 pc size regions and 1 kpc size region found in nearby PHANGS galaxies \citep{Schinnerer+Leroy2024} to calculate the mass density of molecular gas present on small scales.
As has been demonstrated, in fact, by \citealt{Sun+2022}, the 150 pc resolved molecular gas surface mass density is well correlated (with a scatter of $\sigma=0.14$ dex) with the one measured at 1 kpc scale through the following equation:
\begin{equation}\label{eqn:sun22}
log\Sigma_{mol}^{150pc}=0.88log\Sigma_{mol}^{1kpc}+0.46
.\end{equation}
Adopting this relation would increase our molecular gas mass surface densities by 0.4 (for the lowest densities) to 0.2 dex (for the highest densities) in the measured interval. These points are shown as filled diamonds in Fig.~\ref{fig:ks_hst}.

The second assumption that we make is that the fraction of the \hdue mass contained in the HST clumps follows the fraction of the MUSE \Ha luminosity recovered by HST in the corresponding clumps, i.e., on average $70\%$ \citep[see Fig. 4.5 in ][]{Giunchi_PhD}. These points are shown in Fig.~\ref{fig:ks_hst} as filled triangles.

Each star-forming region (either in the disk or in the tail) with a significant fit of the CO(2-1) line has therefore been matched with the corresponding nearest HST clump and is shown in Fig.~\ref{fig:ks_hst} with different colors following the different host galaxy. The size of each point is proportional to the size of the HST clump, while the symbols meaning are described above.
In many clumps within the region identified by MUSE at 1 kpc resolution, a few small clumps coexist. For the sake of this analysis, we used the closest clump identified both by MUSE and by HST. Including in the SFR densities the other nearby matched clumps would only move the points at higher SFR densities (and only for the big MUSE clumps).
The distribution of the clump properties in the KS plane shown in Fig.~\ref{fig:ks_hst} is nearly flat, as the range in SFR densities exhibited by our star-forming regions is limited (1 dex).
In fact, as is demonstrated in \citealt{Giunchi+2023}, resolved clumps' \Ha luminosities correlate with the size of the star-forming clumps, and in particular the disk ones follow a power law relation, $L_{H_{\alpha}} \propto r^{2}$, so that the derived SFR densities are almost constant, reaching very high values, due to the concentrated \Ha emission that fills the MUSE clumps.

MUSE regions with a matched HST clump show a very short depletion time if we assume that the star formation is concentrated in the HST clumps while the CO is distributed uniformly within the MUSE clumps (transparent circles in all the plots). 
Adopting instead the relation found in local galaxies (Eqn.~\ref{eqn:sun22}), we find only slightly longer depletion times (0.01 Gyr$<\tau_{dep}<$1 Gyr), as is demonstrated by the location of the colored diamonds in each of the subplots in Fig.~\ref{fig:ks_hst}.
As a comparison, we also show in Fig.~\ref{fig:ks_hst} the location of starburst disks (gray squares, from \citealt{Kennicutt+2021}), following a linear slope KS relation, albeit with a higher efficiency, suggesting that our star-forming clumps indeed follow a super-efficient star formation mode.
Finally, assuming that 70\% of the CO flux comes from the same small region that is responsible for the \Ha emission detected by HST produces very high molecular gas mass densities (filled triangles in all plots) that would then be mostly gas-rich but unable to turn this gas into stars in less than 10 Gyr, implying a very inefficient process.

We notice that an invariant KS relation is found in the literature at all scales from 1 kpc \citep{Bigiel2011} down to 750 pc \citep{Bigiel2008}, suggesting that any change in the relation could happen only below 300 pc \citep{Bigiel2008,Casasola+2022}.
However, the resolved relation found in PHANGS galaxies seems to confirm the trend even at a 100 pc scale \citep{Pessa+2021}.

Star-forming clumps in our sample of jellyfish galaxies therefore exhibit a highly efficient KS relation, but also demonstrate the existence of very low-gas-mass-density clumps where the SFR density is enhanced (especially in JO201, as is demonstrated by the big blue dots in Fig.~\ref{fig:ks_hst}).
Taken at face value, our data seem to imply that there is a sort of threshold in the SFR density below which clumps in our disturbed galaxies cannot form stars (or, better, they cannot be detected as resolved clumps emitting in the \Ha NB HST filter).
On the other hand, when they do host active star formation, this is happening in a quasi-starburst regime, i.e., the SFR density is particularly high, as is also suggested by the $L_{H_{\alpha}}-size$ relation shown in \citealt{Giunchi+2023}.
The same trend has been demonstrated to hold also in the Antennae galaxies \citep{Saravia+2025}, as well as within the star clusters in tidal dwarf galaxies \citep{Kovakkuni+2023}.

To recover the SFE normally observed in star-forming galaxies, we would need to invoke either a much lower SFR density in our clumps or a ($\sim 1$ dex) higher molecular gas mass surface density.
In the first case, as the SFR density directly comes from the \Ha emission, we would need to assume that a stochastic IMF is shaping the SF in every clump, a hypothesis that has already been put forward but never proved, since it would require one to isolate single stars.
On the other hand, moving our clumps to higher surface gas mass densities is easily obtained by assuming that the molecular gas is much more concentrated than in galaxy disks, albeit assuming that the ionized gas emission and the cold gas one come from the same region leads to unrealistically low star formation efficiencies.

Taken at face value, our analysis suggests that assuming that the cold molecular gas fills the whole extent of the MUSE emitting regions leads to star-forming efficiencies similar to those exhibited by starbursts, and this is true also if we adopt the scaling relation of normally star-forming galaxy disks from PHANGS (i.e., a clumpy medium). Alternatively, assuming that the cold molecular gas is distributed as the \Ha emission detected by HST produces unrealistic $H_2$ densities and much lower star formation efficiencies.
Higher-spatial-resolution interferometric ALMA data mapping the dense gas emission for our clumps are needed to shed light on this issue.

\section{Summary and conclusions}\label{sec:discussion}
We took advantage of MUSE and ALMA data tracing the ionized and the molecular gas at 1 kpc scale in a sample of four jellyfish galaxies extracted from the GASP sample.
On this scale, a monotonic KS relation between $\Sigma_{SFR}$ and $\Sigma_{H_2}$ has been found to hold in undisturbed nearby galaxies, as well as in high-redshift counterparts \citep{KE2012}, and it is best approximated by a power law with a slope that varies according to the gas phase that is traced: it is thought to be linear when considering the dense gas involved in star formation (mostly CO or HCN; \citealt{Wong+Blitz2002,Bigiel2008,Leroy2008}).
The slope becomes superlinear when including the total amount of gas in the disks of galaxies.
It is, instead, sublinear when the amount of diffuse gas that does not contribute directly to the star formation is included \citep{Shetty+2014}, or when the density profile within the molecular clouds is shallow.
Variations in the adopted \aco factor can also impact the slope of the KS relation \citep{Narayanan+2012}.

Our analysis on the scale probed by the beam size reveals the presence of a superlinear KS relation in the disk of our four ram-pressure-disturbed galaxies, suggesting that in the regions with a high column density the star formation process is more efficient than in regions where the gas density is lower.
Instead, in the tails the process is less efficient at all densities.
Likely, the ram pressure responsible for the increased turbulence in the galaxy disks is therefore altering the cold gas properties so that the star formation becomes particularly efficient on the beam scales.
No plausible variation in the \aco or $r_{21}$ values can reconcile our findings with those characterizing undisturbed star-forming disks at the same spatial resolution, both in the field \citep{Bigiel2011} and in the Virgo cluster \citep{jimenez+2023}.
Interestingly, HI-deficient galaxies in Virgo influenced by the ram pressure also show a steeper KS relation, when compared with undisturbed galaxies in the cluster itself \citep{jimenez+2023}.
Normally star-forming disks in Fornax, when not visibly affected by the ram pressure, instead show a linear relation very similar to the one exhibited by isolated disks, albeit with a large scatter \citep{Zabel+2020}. 

The conclusion based on the beam scale is reversed when looking at single star-forming regions identified by their \Ha emission from MUSE.
These clumpy regions appear actually as large star-forming complexes, due to the limited spatial resolution that can be achieved at the redshift of our galaxies.
We have shown that within the MUSE clumps the star formation is normally efficient, with clumps showing normal depletion times even in the tails.

For the subsample of clumps that have a spatially resolved detection in HST, we confirm a starburst-like SFE, provided that the molecular gas is distributed within the clumps as in normally star-forming disks.
If we assume instead a more concentrated distribution, i.e., that the whole cold molecular gas content is spatially coincident with the region emitting in \Ha, then the SF becomes largely inefficient.

The net effect of the RPS on the molecular gas content of the affected galaxies is therefore a global increase \citep{Moretti+2020} on a galactic scale, possibly due to the efficient conversion of the atomic phase into the molecular one. This wealth of cold gas forms stars very efficiently in the regions where it reaches high densities, while the SFE becomes on average low in the tails.
However, in the big complexes where MUSE can track the ionized gas emission and we have good S/N data from ALMA the SFE follows a universal KS relation, with depletion times comparable to the ones exhibited by undisturbed star-forming disks. 
Our data also suggest that on smaller scales the SFE is greatly enhanced, with cold gas in dense clouds forming stars in a starburst-like way. 
Higher-spatial-resolution data and theoretical simulations would be needed to confirm this and assess whether this follows the geometrical compression front. In fact the larger virial parameter exhibited by our big complexes is expected in the case of higher external pressure exerted on the clouds \citep{Meidt+2025}.

\section*{Data availability}
Table \ref{tab:catalog} is only available in electronic form at the CDS via anonymous ftp to cdsarc.u-strasbg.fr (130.79.128.5) or via http://cdsweb.u-strasbg.fr/cgi-bin/qcat?J/A+A/.
The complete set of spectra extracted from the ALMA cubes is available in Zenodo (\url{https://zenodo.org/records/18678559}).

\begin{acknowledgements}
The authors thank the anonymous referee for their insightful comments and constructive suggestions, which significantly improved the quality of this manuscript.
    This project has received funding from the European Research Council (ERC) under the European Union's Horizon 2020 research and innovation programme (grant agreement No. 833824, GASP project). 
This publication has received funding from the European Union’s Horizon 2020 research and innovation programme under grant agreement No 101004719 (ORP). B. Vulcani and A. E. Lassen acknowledge support from the INAF GO grant 2023
“Identifying ram pressure induced unwinding arms in cluster spirals” (P.I. Vulcani).
\end{acknowledgements}

   \bibliographystyle{aa} 
   \bibliography{references_abbr} 

\end{document}